\documentclass[a4paper,11pt]{article}
\usepackage{jheppub} 
\usepackage{lineno}
\usepackage{natbib}
\usepackage{multicol,multirow}

\def\nn{\nonumber\\}

\title{\boldmath Precision calculations of the $D_{(s)}D_{(s)}V$ and $B_{(s)}B_{(s)}V$ couplings from light-cone sum rules}

\author[a]{Hua-Yu Jiang}
\author[a]{and Su-Ping Jin}
\affiliation[a]{School of Physics, Nankai University, Weijin Road 94, 300071 Tianjin, China}

\emailAdd{huayujiang@nankai.edu.cn}
\emailAdd{jinsuping@nankai.edu.cn}

\abstract{We present an improved calculation of the $HHV$ ($H=D_{(s)},\, B_{(s)}$, $V= \rho$, $K^\ast$, $\omega$, and $\phi$) coupling constants $g_{HHV}$ beyond leading order in $\alpha_s$ from QCD light-cone sum rules (LCSRs) by means of the light-cone distribution amplitudes (LCDAs) of light vector mesons.
Near the light-cone, the next-to-leading order QCD corrections for the vacuum-to-vector-meson correlation function are included at leading power in $\delta_V = m_V/m_Q$ ($Q=b,c$) within the framework of hard-collinear factorization.
The higher-twist corrections from two-particle and three-particle vector meson LCDAs are systematically incorporated at leading order in $\alpha_s$ by applying the method of background field in LCSRs. Based on these improvements, we perform a systematic computation of the strong coupling constants $g_{H HV}$ and extract the effective coupling $\beta$ of the heavy meson chiral perturbation theory (HM$\chi$PT).
Furthermore, we accomplish the analysis for the relation between the coupling $g_{HHV}$ and the residue of the $H\to V$ transition form factor $A_0$ at heavy pseudoscalar pole.
Additionally, we provide a detailed investigation of the $SU(3)$ flavour symmetry breaking effects and conduct a comparative analysis with results from previous studies.}

\begin{document}
\maketitle
\flushbottom
\section{Introduction} \label{sec:intro}

The strong couplings between heavy pseudoscalar mesons ($D_{(s)}$ or $B_{(s)}$) and light vector mesons ($V$ involving $\rho,\,K^\ast,\,\omega$ and $\phi$), denoted as $g_{HHV}$, play an important role for understanding the long-distance dynamics of strong interactions.
On the one hand, the effective coupling $\beta$ that governs the interactions between heavy pseudoscalar and light vector mesons at low energy in the HM$\chi$PT Lagrangian \cite{Casalbuoni:1996pg,Cheng:2004ru} can be directly determined from the $HHV$ couplings in the heavy quark limit.
On the other hand, the coupling constants $g_{HHV}$ enter the residue of the $H\to V$ transition form factor $A_0$ at the heavy pseudoscalar pole.
The $H\to V$ form factor $A_0$ in the large momentum transfer ($q^2$) region completely contains the soft overlap effects of the initial and final hadron states, which can be determined by the method of Lattice QCD or analytical continuation from its small $q^2$ value obtained with the approach of QCD light-cone sum rules (LCSRs) \cite{Ball:2004rg}.
The second approach is well known as the BCL parameterization \cite{Bourrely:2008za}; the analytical continuation strongly depends on the first heavy pseudoscalar pole structure \cite{Ball:2004rg}, and the residue of the form factor $A_0$ at the first pole gives the corresponding coupling $g_{HHV}$ derived from the dispersion relation \cite{Wang:2020yvi,Khodjamirian:2020mlb,Jin:2024zyy}.
Therefore, the $HHV$ couplings can provide deep insights into the analytical structure and the nonperturbative nature of the $H\to V$ form factor $A_0$ in the large $q^2$ region.
Furthermore, the $D_{(s)}D_{(s)}V$ couplings are also important for understanding the direct CP violation in $B$ meson non-leptonic decays \cite{Cheng:2004ru} with the assumption of the final state interactions (FSIs) which can make visible contributions to the long-distance dynamics and strong phases.

The $H\to HV$ transitions can not be detected in experiments due to obviously lack of phase space. Therefore there are no experimental measurements for the $HHV$ couplings and they can only be extracted from theoretical side.
As the aforementioned importance of these couplings, there are a series of calculations and analysis with the employing of different nonperturbative QCD techniques, such as Lattice QCD\cite{Can:2012tx}, light-cone sum rules (LCSRs) \cite{Li:2002pp,Wang:2007mc,Li:2007dv} and SVZ sum rules (QCDSRs)\cite{Holanda:2006xq,Janbazi:2017mpb,Khosravi:2013ad}.
The first $2+1$ flavour Lattice QCD calculation for the $DD\rho$ coupling is performed in \cite{Can:2012tx} using the simulated electromagnetic form factors of $D$ meson and the vector meson dominance model (VMD), the precision of their results is advanced than the calculation in LCSRs but obviously affected by the adopted fitting model.
Thanks to the important progress in the study of light-cone distribution amplitudes (LCDAs) of vector mesons \cite{Ball:1998ff,Ball:1998sk}, the coupling $g_{DD\rho}$ and $g_{BB\rho}$ are firstly calculated with the method of LCSRs in \cite{Li:2002pp} and modified in \cite{Li:2007dv}, their analysis are accomplished at leading order in $\alpha_s$ with incomplete hard-collinear factorization and power expansion in $\delta_V=m_V/m_Q$ ($Q=b,c$, $m_V$ is the mass of vector meson).
In \cite{Wang:2007mc}, by utilizing the same approach as in \cite{Li:2002pp}, the author extend the calculation to include all the $D_{(s)}D_{(s)}V$ coupling constants except for $V=\omega$.
In \cite{Holanda:2006xq,Janbazi:2017mpb,Khosravi:2013ad}, the strong coupling constants $g_{DD\omega}$, $g_{D_sDK^\ast}$ and $g_{D_sD_s\phi}$ are calculated in sequence with the framework of three-point SVZ sum rules, which introduce the $Q^2$ ($Q$ is the momentum of interpolating current for $D$ or $V$ meson) dependence and then additional uncertainties.

In this work, we prefer to perform a systematic calculation and analysis for the $HHV$ couplings with the framework of LCSRs, and we extend the computation of the vacuum-to-vector-meson correlation function in the light-cone operator product expansion (OPE) to next-to-leading order in $\alpha_s$.
Our calculation is based on the solid foundation of the progress in computing the correlation function near the light-cone with the technique of QCD (SCET) factorization \cite{Beneke:1999br,Beneke:2000ry,Beneke:2003pa,Khodjamirian:2023wol} which is responsible for infrared safely separating the physics corresponding to multiple energy scales.
In heavy flavour physics, many essential non-perturbative quantities have been determined with the method of LCSRs combing with QCD (SCET) factorization, for instance, the heavy-to-light ($B\to \pi$, etc.) and heavy-to-heavy ($B \to D$, etc.) transition form factors.
The OPE near the light-cone is carried out from two-distinct aspects \cite{Beneke:2003pa} with the approach of QCD (SCET) factorization: the hard-collinear factorization is used to dealing with the vacuum-to-light-hadron correlation function, such as in \cite{Khodjamirian:1999hb,Ball:2004ye,Duplancic:2008ix,Ball:2004rg,Bharucha:2015bzk,Li:2020rcg}, and the SCET factorization (for separating the hard, hard-collinear, collinear and soft scales) is utilizing to calculate the vacuum-to-heavy-hadron correlation function, e.g., the calculations in \cite{Wang:2015vgv,Cui:2022zwm,Gao:2019lta,Feldmann:2011xf,Wang:2015ndk,Li:2009wq,Gao:2021sav,Cui:2023jiw,Wang:2018wfj,Cui:2023yuq}.
In \cite{Khodjamirian:1999hb,Li:2020rcg,Wang:2020yvi,Khodjamirian:2020mlb,Jin:2024zyy}, the systematic calculation of the $H^\ast H\gamma$, $H^\ast HV$ and $H^\ast H\pi$ couplings up to NLO in the QCD (SCET) factorization-based LCSRs provides us with crucial experience for the computation of the $HHV$ couplings in this work.

In this paper, we aim to perform a comprehensive investigation for the $D_{(s)}D_{(s)}V$ and $B_{(s)}B_{(s)}V$ strong coupling constants with the aforementioned framework. The essential new ingredients are summarized as follows:
\begin{itemize}
\item Near the light-cone, we build up rigorous factorization formula for the vacuum-to-vector-meson correlation function at leading power in $\delta_V$\footnote{The power counting parameters are defined as $\delta_V\equiv m_V/m_Q,\,Q=(b,c)$ and $\delta_s\equiv m_s/m_V$, where $m_V$ is the mass of vector mesons, $m_Q$ is the heavy quark mass, and $m_s$ is the strange quark mass. The effectiveness of our power expansion is shown in Table \ref{tab:power}.}
and next-to-leading order in $\alpha_s$ with the framework of hard-collinear factorization, where the one-loop integrals are calculated by using the method of regions.
As we adopt the physical interpolating currents for heavy pseudoscalar mesons, the factorization scale and renormalization scale at NLO are naturally unified as the same one, we show the exact cancellation of the factorization scale for the correlation function up to $O(\alpha_s^2)$ using the evolution equation of twist-two DAs.
The derivation of the double spectral density for the NLO corrections in the dispersion relation is following the prescriptions in \cite{Khodjamirian:1999hb}, after that we verify the factorization scale independence of the leading power LCSRs for the strong coupling constants $g_{HHV}$ up to $O(\alpha_s^2)$. To achieve next-to-leading-logarithmic (NLL) accuracy, we perform QCD resummation for the enhanced logarithms $m_Q/\Lambda_{\rm QCD}$ in the hard-collinear factorization formula, which involves solving the one-loop renormalization-group (RG) equation for $\phi_{2;V}^{\|}(u)$.

\item We include comprehensive subleading power corrections (up to $\delta_V^4$, $\delta_s^1$, and $(\delta_V\delta_s)^1$) and higher-twist corrections (up to two-particle and three-particle twist-4) of the LCDAs of light vector mesons at leading order. With these improvements, we present the most optimized and detailed computations for the $D_{(s)}D_{(s)}V$ and $B_{(s)}B_{(s)}V$ couplings accomplished so far. This allows us to obtain a more precise prediction for the effective coupling $\beta$ of HM$\chi$PT, which still shows a significant deviation from the vector meson dominance (VMD) model estimation \cite{Isola:2003fh}.

\item We establish the dispersion relation to connect  the residue of various $H \to V$ transition form factors $A_0$ at the $H$ pole with the corresponding $H H V$ strong coupling. The couplings $g_{B_{(s)} B_{(s)} V}$, including systematic uncertainties, are computed from this dispersion relation using form factors obtained from two distinct LCSRs methods \cite{Bharucha:2015bzk,Gao:2019lta}. Additionally, we include the $g_{D_{(s)} D_{(s)} \rho}$ coupling using a similar approach but with $D \to \rho$ transition form factors in Ref.\cite{Melikhov:2000yu}. Our study shows the common characteristics and conclusions as comparing the $H^\ast H V$ couplings calculated from LCSRs and obtained from the $H\to V$ transition form factors $V$ and $T_1$ in \cite{Jin:2024zyy}.

\item We investigate $SU(3)$ flavour symmetry breaking effects arising from changes in light quark mass, decay constants, and the masses of vector and heavy mesons, as well as vector meson LCDA parameters. Our study reveals that these effects significantly influence the coupling constants $g_{H H K^\ast}$, $g_{H H \omega}$, and $g_{HH \phi}$. The coupling $g_{H H \phi}$ shows the largest breaking effects due to the  maximum number of strange quarks in this interaction.
\end{itemize}

This article is structured as follows: In section \ref{sec-2}, we introduce the conventions and definitions for the $HHV$ couplings and the relations among various couplings under $SU(3)$ flavour symmetry. In section \ref{sec-3}, we establish the double dispersion for computing the coupling and the relation between the couplings and the $H\to V$ form factors $A_0$. In section \ref{sec-4}, we build up the factorization formula at leading power and present the calculation details for the LO and NLO double spectral density of the correlation function. Section \ref{sec-5} is dedicated to calculating the subleading power corrections from the two-particle and three-particle vector meson LCDAs up to twist-4. We focus on determining the various input parameters and analysing the numerical results in section \ref{sec-6}. The last section is the summary of this work. Essential details regarding the master integrals are listed in Appendix \ref{app:MI}, while the definitions and parameterizations of the vector meson LCDAs are provided in Appendices \ref{app:def} and \ref{app:lcda}.

\section{The effective Lagrangian for the $D_{(s)}D_{(s)}V$ and $B_{(s)}B_{(s)}V$ couplings} \label{sec-2}
The study begins with the effective Lagrangian \cite{Yan:1992gz,Cheng:2004ru,Casalbuoni:1996pg}, which outlines the strong interactions between heavy pseudoscalar and light vector mesons. This framework incorporates the crucial terms and symmetries of the interactions, enabling accurate predictions of meson dynamics
\begin{align}
\mathcal{L}_{eff} = i g_{HHV} \Big( H_i^\dagger\stackrel{\leftrightarrow}{\partial}_\mu H^j \Big) \big( V^\mu \big)_j^i \,,
\end{align}
where $\stackrel{\leftrightarrow}{\partial}_\mu \equiv \stackrel{\rightarrow}{\partial}_\mu-\stackrel{\leftarrow}{\partial}_\mu $, $g_{HHV}$ is the corresponding strong coupling constants, and the symbol $V$ here stands for the nonet matrix of vector mesons,
\begin{eqnarray}\setlength{\tabcolsep}{3pt}
V=\left(\begin{array}{ccc}
\frac{\rho^{0}}{\sqrt{2}}+\frac{\omega}{\sqrt{2}} & \rho^{+} & K^{*+}\\
\rho^{-} & ~~ -\frac{\rho^{0}}{\sqrt{2}}+\frac{\omega}{\sqrt{2}} ~~~~~& K^{*0}\\
K^{*-} &\bar{K}^{*0} &\phi
\end{array}\right),\,
\end{eqnarray}
with $H = \big( D^{0},D^{+},D_s^{+} \big)$ or $\big( B^{-},\bar{B}^{0},\bar{B}_s^{0} \big)$. The hadronic matrix element of the effective Lagrangian for a specific channel, with defined initial and final states, determines the strong coupling constant $g_{H_1H_2V}$
\begin{equation} \label{eq:coupling_definition}
\langle\, V(p,\varepsilon^\ast)\,H_2(p+q)\,|{\cal L}_{eff}|\,H_1(q)\, \rangle \equiv g_{H_1H_2V}\,(q \cdot \varepsilon^{*}) \,,
\end{equation}
where the vector meson $V(p,\varepsilon)$ is characterized by its mass $m_V$, momentum $p$, and polarization vector $\varepsilon$.

This study focuses on the couplings
$g_{D^{+}D^0\rho^+}$, $g_{D^{+}D^+\omega}$, $g_{D^{+}_s D^0  K^{*+}}$, $g_{D^{+} D^+_s {\bar K}^{*0}}$ and $g_{D^{+}_s D^{+}_s \phi}$ in the charm sector, as well as $ g_{\bar{B}^{0}{B}^-\rho^+}$, $g_{B^{0}D^0\omega}$, $g_{\bar{B}^{0}_s {B}^-  K^{*+}}$, $g_{\bar{B}^{0} \bar{B}^0_s \bar{K}^{*0}}$ and $g_{\bar{B}^{0}_s {B}^{0}_s \phi}$  in the bottom sector.
Other couplings are related to these through isospin symmetry, which dictates relations between couplings involving different isospin states
\begin{eqnarray}
{\rm D~ case}:\quad && g_{DD\rho} \equiv g_{D^{+}D^0\rho^+}=
-\sqrt{2}g_{D^{+}D^{+}\rho^0}=\sqrt{2}g_{D^{0}D^0\rho^0}
=-g_{D^{0}D^+\rho^-}~,\nn
&& g_{D_s D K^{*}} \equiv g_{D^{+}_s D^0  K^{*+}}=
-g_{D^{+}_s D^{+} K^{*0}}~, \nn
&& \,g_{D D_s  K^{*}} \equiv g_{D^{+} D^+_s {\bar K}^{*0}}=
g_{D^{0} D^+_s K^{*-}}~;\label{eq:coupling-def1} \\
\nn
{\rm B~ case}:\quad && g_{BB\rho} \equiv g_{\bar{B}^{0}{B}^-\rho^+}=
-\sqrt{2}g_{\bar{B}^{0}\bar{B}^{0}\rho^0}=\sqrt{2}g_{{B}^{-}{B}^-\rho^0}
=-g_{{B}^{-}\bar{B}^0 \rho^-}~,  \nn
&& g_{B_s B K^{*}} \equiv g_{\bar{B}^{0}_s {B}^-  K^{*+}}=
-g_{\bar{B}^{0}_s \bar{B}^{0} K^{*0}}~ \,, \nn
&& g_{BB_s K^{*}} \equiv g_{\bar{B}^{0} \bar{B}^0_s  \bar{K}^{*0}}=
g_{{B}^{-}_s \bar{B}^{0}_s K^{*-}}~ \,.\label{eq:coupling-def2}
\end{eqnarray}
Through charge conjugation transformation, where initial and final states are swapped, we derive:
\begin{eqnarray}
{\rm D~ case}:\quad
&&  g_{D^{+}_s D^0  K^{*+}}=
-g_{D^{+}_s D^{+} K^{*0}} = g_{D^{+} D^+_s {\bar K}^{*0}}=
g_{D^{0} D^+_s K^{*-}};
\nn
{\rm B~ case}:\quad
&&  g_{\bar{B}^{0}_s {B}^-  K^{*+}}=
-g_{\bar{B}^{0}_s \bar{B}^{0} K^{*0}}= g_{\bar{B}^{0} \bar{B}^0_s  \bar{K}^{*0}}=
g_{{B}^{-}_s \bar{B}^{0}_s K^{*-}}~,\,
\end{eqnarray}
SU(3) flavour symmetry allows us to relate the coupling constants listed above, yielding the following relation:
\begin{eqnarray}
{\rm D~ case}:\quad && g_{DD\rho}=g_{D_s D K^{*}}=g_{D D_s  K^{*}}=g_{D_s D_s \phi}=\sqrt{2} g_{DD\omega} \,;\\
{\rm B~ case}:\quad && g_{BB\rho} =g_{B_s B K^{*}}=g_{BB_s K^{*}} =g_{B_s B_s \phi}=\sqrt{2} g_{BB\omega} \,.
\end{eqnarray}

\section{Double dispersion relations for the $D_{(s)}D_{(s)}V$ and $B_{(s)}B_{(s)}V$ couplings}  \label{sec-3}
To derive sum rules for the coupling $g_{H_1 H_2 V}$, we follow the outlined approach,  which involves examining the correlation function between the vacuum and vector meson,
\begin{align}\label{eq:corr}
\Pi((p+q)^2,q^2) = i \int d^4 x \, e^{i q \cdot x} \big\langle V(p, \varepsilon^\ast) \big| T\big\{ j_5^\dagger(x),\, j_5(0) \big\} \big| 0 \big\rangle,
\end{align}
where $p_B = p+q$, and the interpolating current for $D_{(s)}$ and $B_{(s)}$ meson is given by $j_5 = (m_Q+m_q) \Bar{Q} i\gamma_5 q $, with $Q=(b,c)$ and $q=(u,d,s)$.
This analysis utilizes two identical local interpolating currents for the pseudoscalar heavy mesons which serves as effective probes for understanding the relevant properties and dynamics within the heavy meson sector.

For establishing the light-cone sum rules, we need to employ the analytic properties of the correlation function in two invariant variables $(p+q)^2$ and $q^2$ corresponding to the momentum square of interpolating currents in the two channels of heavy pseudoscalar meson. Starting with the double dispersion representation,  by applying the unitarity relation and including complete sets of intermediate states with quantum numbers of $H$ into these channels, we obtain
\begin{eqnarray}
\Pi((p+q)^2,q^2) = \frac{\langle\, V(p,\varepsilon^\ast)H_2(p\!+\!q)\,|H_1(q)  \rangle\langle H_1(q)|\,j_{5}^\dagger|0\rangle\langle H_2(p\!+\!q)|j_5|0\rangle }{\big[q^{2}-m_{H_1}^{2}\big]\big[(p+q)^{2}-m_{H_2}^{2}\big]} + \cdots \,,
\end{eqnarray}
where $\ldots$ represents the excited states and continuum part. After substituting the definition of the strong coupling in Eq.\eqref{eq:coupling_definition} and expressing the matrix elements of interpolating currents in terms of the decay constants $f_{H_1}$ and $f_{H_2}$, i.e., $\left\langle 0\left|j_5\right| H(q)\right\rangle = f_{H} {m_{H}^{2}}$, we proceed to derive the double dispersion relation,
\begin{align}\label{eq:double_disp}
\Pi((p+q)^2,q^2) & = \frac{(q \cdot \varepsilon^\ast) g_{H_1 H_2 V} f_{H_1}f_{H_2} m_{H_1}^{2} m_{H_2}^{2} }{\big[ m_{H_1}^{2}-q^{2} \big]\big[ m_{H_2}^{2}-(p+q)^{2} \big] } \, \nn
& \quad + \iint\limits_{\hspace{4mm}{\Sigma}} \frac{ \rho_{cont}\left(s_1, s_2\right) d s_1 d s_2}{\left[s_1-q^{2}\right]\left[s_2-(p+q)^{2}\right]} + \cdots \nn
& \equiv (q \cdot \varepsilon^\ast) \, F((p+q)^2,q^2) \,,
\end{align}
where $\Sigma$ stands for the double duality region, covering both excited and continuum states.

The leading term in Eq.\eqref{eq:double_disp} signifies the contribution from the ground-state double-pole, involving the product of the desired strong coupling $g_{H_1H_2V}$ and the corresponding decay constants. The term $\rho_{cont}{(s_{1},s_{2})}$ captures the combined spectral density of excited and continuum states, with $\Sigma$ representing the duality region occupy in the $(s_{1},s_{2})$-plane. Additional terms arising from subtractions, denoted by ellipses, vanish upon the application of the double Borel transformation.

On the other hand, the correlation function $\Pi((p+q)^2,q^2)$ can be computed at momentum transfers $q^2$ and $(p+q)^2$ much smaller than the heavy quark mass $m_Q^2$ using the operator product expansion (OPE) near the light-cone $x^2\sim 0$. The result can be factorized as the convolution of the hard kernel and the LCDAs of vector mesons, categorized based on their twist. The former captures the short-distance perturbative contributions, while the latter describesthe long-distance non-perturbative effects. Conveniently, the OPE computations can be expressed in the form of a double dispersion relation:
\begin{align}
\Pi^{\rm OPE}((p+q)^2,q^2)= (q \cdot \varepsilon^\ast) \iint ds_{1}\,ds_{2} \frac{\rho^{\rm OPE}(s_{1}, s_{2})}{(s_{1}-(p+q)^2)(s_{2}-q^2)}\,,
\label{eq:ddispOPE}
\end{align}
where the involved dual spectral density $\rho^{\rm OPE} (s_{1}, s_{2})$ refers to,
\begin{equation}
\rho^{\rm OPE} (s_{1}, s_{2}) \equiv \frac{1}{\pi^2} \mbox{Im}_{s_{1}}\mbox{Im}_{s_{2}} F^{\rm OPE}(s_{1},s_{2}) \,,
\label{eq:rhoope}
\end{equation}
where ${\rm Im}_{s_{1}}$ and ${\rm Im}_{s_{2}}$ correspond to the sequential extraction of the imaginary part of the corrleation function $F^{\rm (OPE)}(s_{1},s_{2})$ with respect to the variables $s_{1}$ and $s_{2}$.

Following the application of the double Borel transformation  with respect to the variables  $ q^2 \to M_1^2$ and $(p+q)^2 \to M_2^2$, a technique aimed at improving the convergence and suppress the contributions from higher-dimensional operators in the OPE, we obtain the sum rules for the coupling constant $g_{H_1H_2V}$,
\begin{eqnarray}
f_{H_1} f_{H_2} g_{H_1H_2V}= \frac{1}{m_{H_1}^2 m_{H_2}^2}\iint \limits^{\hspace{4mm}\widetilde{\Sigma}} ds_{1} ds_{2} \exp\bigg[\frac{m^2_{H_1}-s_{1}}{M_1^2} + \frac{m^2_{H_2}-s_{2}}{M_2^2}\bigg]
\rho^{\rm OPE} (s_{1}, s_{2})\,.
\label{eq:SR2}
\end{eqnarray}
where the integration boundary $\widetilde{\Sigma}$, dual to the ground-state contribution to \eqref{eq:double_disp}, is determined by subtracting the continuum contributions with the parton-hadron duality ansatz.

In this study, we base our approach to the integration boundary $\widetilde{\Sigma}$ on the discussion presented in Ref.\cite{Khodjamirian:2020mlb} and employ a specific parameterization for this boundary:
\begin{eqnarray}
\left ( {s_{1} \over s_{\ast}} \right )^{\alpha} +
\left ( {s_{2} \over s_{\ast}} \right )^{\alpha}   \leq 1 \,,
\qquad s_{1}, \, s_{2} \geq  m_Q^2\,.
\label{eq:alpha}
\end{eqnarray}
Currently, our investigation is confined to the triangular duality region with $\alpha=1$ and $s_\ast=2s_0$,  wherein  $s_{2} \leq 2 s_0 - s_{1} $. This restriction provides a more definite form for the LCSRs \eqref{eq:SR2}:
\begin{align}
f_{H_1} f_{H_2} g_{H_1H_2V} = \frac{1}{m_{H_1}^2 m_{H_2}^2} \int^{+\infty}_{-\infty}ds_{1} \int^{2s_0-s_{2}}_{-\infty}ds_{2}  \exp\bigg[\frac{m^2_{H_1}-s_{1}}{M_1^2} + \frac{m^2_{H_2}-s_{2}}{M_2^2}\bigg] \rho^{\rm OPE} (s_{1}, s_{2})\,.
\end{align}

In the following, we build up the relation between the $D_{(s)}D_{(s)}V$ and $B_{(s)}B_{(s)}V$ couplings and the $D \to V$ and $B \to V$ transition form factors by using dispersion relation.
To acheive this goal, we need to relate the pseudoscalar current to the axial-vector current $j_{\mu 5} = \Bar{Q} \gamma_\mu \gamma_5 q$ by utilizing the QCD equation of motion for quark field, that is
\begin{align}
\partial^\mu j_{\mu5}(x) = j_5(x) \,,
\end{align}
as a result,
\begin{align} \label{eq:PScAVc}
\Pi((p+q)^2,q^2) & = i \int d^4 x \, e^{i q \cdot x} \big\langle V(p, \varepsilon^\ast) \big| T\big\{ \big( \partial^\mu j_{\mu5}(x) \big)^\dagger, j_5(0) \big\}\big| 0 \big\rangle \nn
& = q^\mu \int d^4 x \, e^{i q \cdot x}  \big\langle V(p, \varepsilon^\ast) \big| T\big\{ j_{\mu5}^\dagger(x), j_{5}(0) \big\} \big| 0 \big\rangle \nn
& \equiv q^\mu \Pi_{\mu}^A ((p+q)^2,q^2) \,,
\end{align}
where the correlation function $\Pi_\mu^A ((p+q)^2,q^2)$ serves as the basis for determining the the $B\to V$ axial-vector form factors $A_{0,1,2}$, defined as
\begin{align}
&  c_V\left\langle V\left(p, \varepsilon^*\right)\left|\bar{q} \gamma_\mu \gamma_5 b\right| H_2(p+q)\right\rangle
= \frac{2 m_V  (q\cdot\varepsilon^*) }{q^2} q_\mu A_0\left(q^2\right)  \nonumber \\
& \hspace{4.8cm} \quad + \left(m_B+m_V\right)\left[\varepsilon_\mu^*-\frac{q\cdot\varepsilon^*}{q^2} q_\mu\right] A_1\left(q^2\right) \nonumber \\
& \hspace{4.8cm} \quad -\frac{q \cdot \varepsilon^* }{m_B+m_V}\left[(2 p+q)_\mu-\frac{m_B^2-m_V^2}{q^2} q_\mu\right] A_2\left(q^2\right) \,,
\end{align}
where we introduce the factor $c_V = \pm \sqrt{2}$ to accommodate the flavor structure of vector mesons for $\rho^0$ and $\omega$. With this adjustment, we construct the single dispersion relation:
\begin{align}\label{eq:HVdisp}
q^\mu \Pi_{\mu}^A ((p+q)^2,q^2) & = { f_{H_2} m_{H_2}^2 q^\mu \left\langle V\left(p, \varepsilon^*\right)\left| \bar{q} \gamma_\mu \gamma_5 b \right| H_2(p+q) \right\rangle \over m_{H_2}^2 - (p+q)^2 } + q^\mu \int_{s_h}^{\infty} ds { \rho_\mu(s, q^2) \over s - (p+q)^2 } \,, \nn
& = \frac{2 m_V}{c_V} \frac{ (q\cdot\varepsilon^*) f_{H_2} m_{H_2}^2 A_0(q^2) }{m_{H_2}^2 - (p+q)^2 } + q^\mu \int_{s_h}^{\infty} ds { \rho_\mu(s, q^2) \over s - (p+q)^2 } \,,
\end{align}
this relation  highlights the pseudoscalar meson $H_{1}(q)$ pole, positioned just below the kinematic threshold $q^2=(m_{H_{1}}+ m_V)^2$.
Putting the expressions in \eqref{eq:double_disp} and \eqref{eq:HVdisp} into the equation for the two correlation functions in \eqref{eq:PScAVc} and multiplying both sides by the denominator of the pole term and taking the limit $q^2\to m_{H_{1}}^2$, we obtain the desired relation:
\begin{eqnarray}
 g_{H_{1}H_{2}V}\,\,&=&\,\, \frac{2 m_V}{c_V f_{H_{1}}} \lim_{q^2\to m_{H_{1}}^2}
\bigg[\bigg( 1 - \frac{q^2}{m^2_{H_{1}}} \bigg) A_0(q^2)\bigg] \,.
\label{eq:HHVlimit}
\end{eqnarray}
this equation establishes a connection between the asymptotic behavior of the $H_2\to V$ transition form factor $A_0$ and the $H_1H_2V$ coupling.

\begin{figure}[tb]
\begin{center}
\includegraphics[width=0.35\columnwidth]{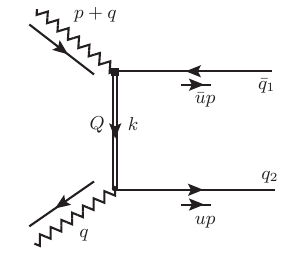} \\
\vspace*{0.1cm}
\caption{Diagrammatical representation of the leading-order (LO) contribution. }
\label{fig:twist-2-LO}
\end{center}
\end{figure}

\section{The light-cone sum rules at twist-two}  \label{sec-4}

As we mentioned before, the OPE remains applicable for the correlation function specified in \eqref{eq:corr} in the vicinity of the light-cone $x^2 \sim 0$,  provided that both external momenta squared $(p+q)^2$ and $q^2$ are considerably below the heavy quark threshold $m_Q^2$.
Ensuring the validity of power counting within the OPE requires satisfying:
\begin{eqnarray}
m_Q^2-q^2 \sim m_Q^2-(p+q)^2\sim {\cal O}(m_Q\tau) \,,n \cdot p \sim {\cal O}(m_Q) \,,
\end{eqnarray}
where $\tau \gg \Lambda_{\text{QCD}}$ represents a parameter independent of $m_Q$. Consequently, the heavy quark propagating in the correlation function is highly virtual and can be expanded near the light-cone.

\subsection{Hard-collinear factorization for the correlation function at leading power}

At leading power of $\delta_V\sim m_V/m_Q$, within the naive dimensional regularization (NDR) scheme for $\gamma_5$, the correlation function in perturbative QCD can be generally expressed as the following factorization form
\begin{align}
\Pi_{\rm LP}^{\rm OPE}((p+q)^2, q^2) & = q^\mu \, \left( \mathcal{T}_V^{(0)} + \mathcal{T}_V^{(1)} \right) \otimes \big\langle O^V_{\mu} \big\rangle \,,
\end{align}
where $\mathcal{T}_V^{(0)}$ and $\mathcal{T}_V^{(1)}$ are the leading order (LO) and next-to-leading order (NLO) hard coefficient function. The matrix element of the vector light-cone operator is
\begin{align}
\big\langle O^V_{\mu} \big\rangle & = \big\langle V(p,\varepsilon^\ast) \big\vert \bar{q}^\prime (x) \gamma_\mu q(0) \big\vert 0 \big\rangle \,,
\end{align}
in the leading power approximation, the light-cone operator can be expanded by using the collinear quark field
\begin{align}
O^V_{\mu} & = \bar{q}^\prime (x) \gamma_\mu q(0) \overset{\rm LP}{=} n_\mu \bar{\chi}(x) \frac{\bar{n}\llap{/}}2 \chi(0) \,,
\end{align}
$\chi(x) = W^\dagger(x)\xi(x)$ stands for the gauge invariant building block, $\xi(x)$ is the collinear quark field in position space, $W(x)$ is the collinear Wilson line
\begin{align}
W(x) = \mathbf{P}\exp\left[ig\int_{-\infty}^0ds \, \bar{n}\cdot A_c(x+s\bar{n})\right] \,.
\end{align}
The collinear operator in position space can be converted into the corresponding momentum label formalism
\begin{align}
O^V_{\mu} & = n_\mu \int d\omega_1 d\omega_2 \, e^{\frac i2 \omega_1 (n\cdot x) } \mathcal{J}_V(\vec{\omega})
 \equiv n_\mu \mathcal{O}_{V} \,,
\end{align}
where the operator $\mathcal{J}_{V}(\vec{\omega})$ \cite{Hardmeier:2003ig} is defined as
\begin{align}
\mathcal{J}_V(\vec{\omega}) \equiv \bar{\chi}_{n,\,\omega_1} \frac{\bar{n}\llap{/}}2\chi_{n,\,\omega_2} \,,
\end{align}
$\chi_{n,\,\omega_1(\omega_2)}$ is the collinear field in the momentum lable space. Then, the matrix element can be obtained as
\begin{align}
\big\langle \mathcal{O}_V \big\rangle
& = \big\langle V(p,\varepsilon^\ast) \big\vert \int d\omega_1 d\omega_2 \, e^{\frac i2 \omega_1 (n\cdot x)} \mathcal{J}_V(\vec{\omega}) \big\vert 0 \big\rangle \nn
& = \int_0^1 du \, e^{\frac i2 u (p\cdot x)} \big\langle \mathcal{J}_{V}(\vec{\omega}) \big\rangle \,,
\end{align}
where
\begin{align}
\big\langle \mathcal{J}_{V}(\vec{\omega}) \big\rangle & = \frac{\bar{n} \cdot \varepsilon^\ast }{2} m_V f_V^{\|} \phi_{2;V}^{\|}(u) \,,
\end{align}
$\phi_{2;V}^{\|}(u)$ is the twist-2 light-cone distribution amplitude of vector meson.
Therefore, the factorization form of the correlation function can be expressed as
\begin{align}
\Pi_{\rm LP}^{\rm OPE}((p+q)^2, q^2) & = (n\cdot q) \int_0^1 du \left( \mathcal{T}_V^{(0)} + \mathcal{T}_V^{(1)} \right) \big\langle \mathcal{J}_{V}(\vec{\omega}) \big\rangle \,.
\end{align}

The hard coefficient functions can be extracted from the parton level correlation function
\begin{align} \label{eq:parton_CRF}
\Pi_{q\bar{q}}((p+q)^2,q^2) & = i \int d^4 x \, e^{i q \cdot x} \big\langle q(u p) \, \bar{q}(\bar{u} p) \big| T\big\{ j_5^\dagger(x),\, j_5(0) \big\} \big| 0 \big\rangle \,,
\end{align}
expanding the correlator to NLO, we have
\begin{align}
\Pi_{q\bar{q}} & = \Pi_{q\bar{q}}^{(0)} + \Pi_{q\bar{q}}^{(1)} + O(\alpha_s^2) \nonumber \\
& = (n\cdot q) \bigg[ H_V^{(0)} \otimes \big\langle \mathcal{O}_V \big\rangle_{q\bar{q}}^{(0)} + \Big( H_V^{(1)} \otimes \big\langle \mathcal{O}_V \big\rangle_{q\bar{q}}^{(0)} + H_{V}^{(0)} \otimes \big\langle \mathcal{O}_V \big\rangle_{q\bar{q}}^{(1)} \Big) + O(\alpha_s^2) \bigg] \,,
\end{align}
then, the NLO renormalized hard-scattering kernel $H_V^{(1)}$ is given by the matching condition
\begin{align}\label{eq:HCF_matching}
H_{V}^{(1)} \otimes \big\langle \mathcal{O}_V \big\rangle_{q\bar{q}}^{(0)} = \mathcal{T}_V^{(1)} \otimes \big\langle \mathcal{O}_V \big\rangle_{q\bar{q}}^{(0)} - H_{V}^{(0)} \otimes \big\langle \mathcal{O}_V \big\rangle_{q\bar{q}}^{(1)} \,,
\end{align}
where the second term is for the subtraction of the infrared divergence. $\big\langle \mathcal{O}_V \big\rangle^{(1)}$ is the UV renormalized  matrix element
\begin{align}\label{eq:HCF_OPR}
\big\langle \mathcal{O}_V \big\rangle_{q\bar{q}}^{(1)} = \Big[ M_V^{(1)R} + Z_V^{(1)} \Big] \big\langle \mathcal{O}_V \big\rangle_{q\bar{q}}^{(0)} \,,
\end{align}
where $M_V^{(1)R}$ is the bare matrix element computed from the infrared regularization scheme $R$ and $Z_V^{(1)}$ represent the renormalization factor for the subtraction of the UV divergence.
Inserting this expression into Eq.(\ref{eq:HCF_matching}), we obtain the renormalized hard coefficient function
\begin{align}\label{eq:HCF_matching_2}
 H_{V}^{(1)} = \mathcal{T}_V^{(1)} - \Big[ M_V^{(1)R} + Z_V^{(1)} \Big] H_{V}^{(0)}  \,.
\end{align}

\subsubsection{The hard-collinear factorization at LO}

At leading order, the leading power contribution ($\delta_V^{1}$) only comes from the twist-2 LCDA, the correlation function can be expressed as the following factorization form
\begin{align}
&\Pi^{\rm LO}_{\rm LP}((p+q)^2, q^2) = (n\cdot q) \int_0^1 du \, \mathcal{T}_V^{(0)}(r_{1},r_{2},u) \, \big\langle \mathcal{J}_{V}(\vec{\omega}) \big\rangle^{(0)} \,,
\end{align}
where the LO hard coefficient function can be calculated from the parton level correlation function \eqref{eq:parton_CRF}, which corresponds to the diagram in Fig.\ref{fig:twist-2-LO}, the matching condition gives
\begin{align} \label{eq:hard_CF_LO}
  H_{V}^{(0)}(r_{1},r_{2},u) = \mathcal{T}_V^{(0)}(r_{1},r_{2},u) = - \frac{1}{u r_{1} + \bar{u}r_{2} - 1} \,
\end{align}
and $r_{1} = (p+q)^2/m_Q^2$, $r_{2} = q^2/m^2_Q$, $r_3=u r_{1} + \bar{u}r_{2}$.


\subsubsection{The hard-collinear factorization at NLO}

\begin{figure}[tb]
\begin{center}
\includegraphics[width=1.0 \columnwidth]{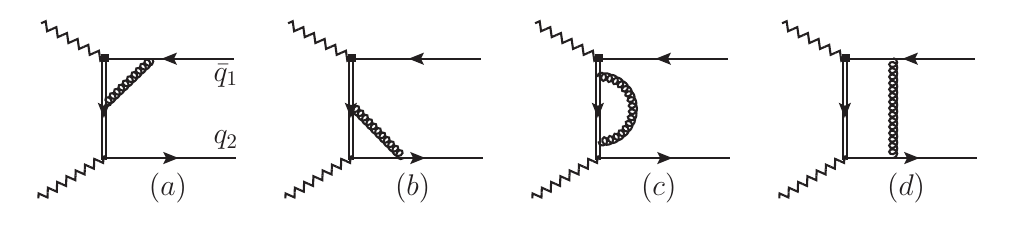} \\
\vspace*{0.1cm}
\caption{NLO QCD corrections of leading-twist contributions.}
\label{fig:twist-2-NLO}
\end{center}
\end{figure}

At next-to-leading order, the leading-power contribution comes from the twist-2 LCDAs of vector mesons, the correlation function can be expressed as the following factorization form
\begin{align}
\Pi_{\rm LP}^{\rm NLO}((p+q)^2, q^2) & = (n\cdot q) \int_0^1 du \, \mathcal{T}_V^{(1)}(r_{1},r_{2},u) \big\langle \mathcal{J}_{V}(\vec{\omega}) \big\rangle^{(0)} \,,
\end{align}
where the NLO hard function is
\begin{align}
\mathcal{T}_V^{(1)}(r_{1},r_{2},u) = \frac{\alpha_s C_F}{4\pi} \Big[ \mathcal{T}_V^{(1),\,a} + \mathcal{T}_V^{(1),\,b} + \mathcal{T}_V^{(1),\,c} + \mathcal{T}_V^{(1),\,d} \Big]\,,
\end{align}
the individual terms stand for the contributions from the expansion of the parton level correlator to $O(\alpha_s)$, that is the four diagrams in Fig.\ref{fig:twist-2-NLO}, by using the integral by parts(IBP) technique, they are expressed as the linear combination of the master integrals
\begin{align}
\mathcal{T}_{V}^{(1),\,a}(r_{1}, r_{2}, u) & = \frac{2}{\left(r_{3}-1\right) m_b^2} \bigg[ \left(\frac{\varepsilon \left(r_{3}-1\right) + 1}{r_{3}-r_{1}} - 1\right) I_1 + \frac{\varepsilon \left(r_{1}-1\right) + 1}{r_{1}-r_{3}} I_2 \nn
& \quad + \left(r_{1}-1\right) m_b^2 I_3\bigg] \,, \label{eq:HCF_Ta_MIexp} \\
\mathcal{T}_{V}^{(1),\,b}(r_{1}, r_{2}, u) & = \frac{2}{\left(r_{3}-1\right) m_b^2} \bigg[ \left(\frac{\varepsilon \left(r_{3}-1\right) + 1}{r_{3}-r_{2}} - 1\right) I_1 + \frac{\varepsilon \left(r_{2}-1\right) + 1}{r_{2}-r_{3}} I_4 \nn
& \quad + m_b^2 \left(r_{2}-1\right) I_5 \bigg] \, \label{eq:HCF_Tb_MIexp} \\
\mathcal{T}_{V}^{(1),\,c}(r_{1}, r_{2}, u) & =  \left(\frac{r_{3}^2-6 r_{3}+1}{\left(r_{3}-1\right)^2 r_{3}} - \frac{\varepsilon }{r_{3}}\right) I_1 + \frac{(r_{3}+1)(\varepsilon-1)}{\left(r_{3}-1\right)^2 r_{3} m_b^2} I_6 \,, \label{eq:HCF_Tc_MIexp}\\
\mathcal{T}_{V}^{(1),\,d}(r_{1}, r_{2}, u) & = m_b^2 \bigg\{ \left(\frac{\varepsilon \left(r_{3}-1\right)}{r_{3}-r_{2}}+\frac{r_{1}-1}{r_{2}-r_{3}}-1\right) I_3 + \left( \frac{\varepsilon \left(r_{3}-1\right)}{r_{3}-r_{1}}+\frac{1-r_{2}}{r_{3}-r_{1}} - 1\right) I_5   \nn
& \quad + \left[\varepsilon \left(\frac{r_{3}-1}{r_{1}-r_{3}} + \frac{r_{3}-1}{r_{2}-r_{3}} +2\right)+\frac{r_{1}-1}{r_{3}-r_{2}}+\frac{r_{2} - 1}{r_{3}-r_{1}} \right] I_{7} \bigg\} \,, \label{eq:HCF_Td_MIexp}
\end{align}
where the master integrals $I_1,\ldots,I_7$ are calculated in the dimensional regularization and are given in App.\ref{app:MI}.
After inserting these master integrals into Eq.\eqref{eq:HCF_Ta_MIexp} to \eqref{eq:HCF_Td_MIexp}, we get the following explicit  form for contributions of each diagram to the hard kerel
\begin{align}
\mathcal{T}_{V}^{(1),\,a}(r_{1}, r_{2}, u)
& = 2 \frac{1-r_{1}}{(r_{3}-1)(r_{3}-r_{1})} \bigg\{ \left[ \frac{r_{1}-r_{3}}{r_{1}-1} - \ln \frac{1-r_{3}}{1-r_{1}} \right]
\bigg( \frac{1}{\tilde \varepsilon} + \ln \left(\frac{\mu^2}{m_b^2}\right) \bigg) \nonumber \\
&\quad + \mathrm{Li}_2\left(r_{3}\right) - \mathrm{Li}_2\left(r_{1}\right) + \ln^2\left(1-r_{3}\right) - \ln^2\left(1-r_{1}\right) \nonumber \\
&\quad + \frac{\left(r_{3}-1\right) \left(r_{3}-r_{1}-1\right) \ln \left(1-r_{3}\right)}{r_{3} \left(r_{1}-1\right)} + \frac{\ln \left(1-r_{1}\right)}{r_{1}} + \frac{r_{1}-r_{3}}{r_{1}-1} \bigg\} \,, \\
\mathcal{T}_{V}^{(1),\,b}(r_{1}, r_{2}, u)
& =  2 \frac{1-r_{2}}{(r_{3}-1)(r_{3}-r_{2})} \bigg\{ \left[\frac{r_{2}-r_{3}}{r_{2}-1} - \ln \frac{1-r_{3}}{1-r_{2}} \right]
\bigg( \frac{1}{\tilde\varepsilon} + \ln \left(\frac{\mu ^2}{m_b^2}\right) \bigg) \nn
& \quad + \mathrm{Li}_2\left(r_{3}\right) - \mathrm{Li}_2\left(r_{2}\right) + \ln ^2\left(1-r_{3}\right) - \ln ^2\left(1-r_{2}\right)  \nn
& \quad + \frac{\left(r_{3}-1\right)
\left(r_{3}-r_{2}-1\right) \ln \left(1-r_{3}\right)}{r_{3} \left(r_{2}-1\right)}+\frac{\ln \left(1-r_{2}\right)}{r_{2}} + \frac{r_{2}-r_{3}}{r_{2}-1}
\bigg\} \,,
\end{align}
\begin{align}
\mathcal{T}_{V}^{(1),\,c}(r_{1}, r_{2}, u)
& = - \frac{1}{r_{3}-1} \bigg\{ \frac{r_{3}-7}{r_{3}-1} \bigg( \frac{1}{\Tilde{\varepsilon}} + \ln\left(\frac{\mu ^2}{m_b^2}\right) \bigg)  + \frac{r_{3}^2-10 r_{3}+1}{r_{3}(r_{3}-1)} \nn
&\quad - \frac{\left(r_{3}^2 - 6 r_{3} + 1\right) \ln \left(1-r_{3}\right)}{r_{3}^2} \bigg\} \,, \\
\mathcal{T}_{V}^{(1),\,d}(r_{1}, r_{2}, u)
& =  2 \bigg\{ \bigg[ \frac{\left(r_{3}-1\right) \ln \left(1-r_{3}\right)}{\left(r_{3}-r_{1}\right) \left(r_{2}-r_{3}\right)}  + \frac{\left(r_{2}-1\right) \ln \left(1-r_{2}\right)}{\left(r_{3}-r_{2}\right) \left(r_{2}-r_{1}\right)} \nonumber \\
& \quad + \frac{\left(r_{1}-1\right) \ln \left(1-r_{1}\right)}{\left(r_{3}-r_{1}\right) \left(r_{1}-r_{2}\right)} \bigg] \bigg( \frac{1}{\Tilde{\varepsilon}} + \ln\left(\frac{\mu ^2}{m_b^2}\right) \bigg) \nonumber \\
& \quad + \frac{\left(r_{3}-1\right)}{\left(r_{3}-r_{1}\right) \left(r_{3}-r_{2}\right)} \Big[ {\rm Li}_2\left( r_{3} \right) + \ln \left(1-r_{3}\right) + \ln^2\left(1-r_{3}\right) \Big]  \nonumber \\
& \quad - \frac{\left(r_{1}-1\right)}{\left(r_{3}-r_{1}\right) \left(r_{1}-r_{2}\right)} \Big[ {\rm Li}_2\left( r_{1} \right) + \ln \left(1-r_{1}\right) + \ln^2\left(1-r_{1}\right) \Big] \nonumber \\
& \quad - \frac{\left(r_{2}-1\right)}{\left(r_{3}-r_{2}\right) \left(r_{2}-r_{1}\right)} \Big[ {\rm Li}_2\left( r_{2} \right) + \ln \left(1-r_{2}\right) + \ln^2\left(1-r_{2}\right) \Big] \bigg\} \,,
\end{align}
then putting these results into the matching equation (\ref{eq:HCF_matching}), and subtracting the UV and IR divergence in the $\overline{\rm MS}$ scheme, we obtain the NLO renormalized hard function
\begin{align} \label{eq:hard_CF_NLO}
H_V^{(1)}(r_{1},r_{2},u)
& = \frac{1}{r_{3}-1} \bigg\{ \bigg[2 \left(2-\frac{1-r_{1}}{r_{3}-r_{1}}\ln\frac{1-r_{3}}{1-r_{1}}
-\frac{1-r_{2}}{r_{3}-r_{2}}\ln\frac{1-r_{3}}{1-r_{2}}\right) \nn
&\hspace{0.5cm} - 2(r_{3}-1)\bigg(\frac{(r_{3}-1)\ln(1-r_{3})}{(r_{3}-r_{1})(r_{3}-r_{2})} + \frac{(1-r_{1})\ln(1-r_{1})}{(r_{3}-r_{1})(r_{1}-r_{2})} \nn
&\hspace{0.5cm} + \frac{(1-r_{2})\ln(1-r_{2})}{(r_{3}-r_{2})(r_{2}-r_{1})}\bigg) - \frac{r_{3} - 7}{r_{3}-1} \bigg] \ln\frac{\mu^2}{m^2_Q} \nn
&\hspace{0.5cm} + \frac{2\left[r_{3}^2-r_{3}(r_{1}+r_{2})+2r_{1} r_{2}-r_{1}-r_{2}+1\right]}{(r_{3}-r_{1})(r_{3}-r_{2})}\Big[{\rm Li}_2(r_{3})+\ln^2(1-r_{3})\Big]\nn
&\hspace{0.5cm} - \frac{2(r_{1}-1)(r_{3}-r_{1}+r_{2}-1)}{(r_{3}-r_{1})(r_{1}-r_{2})}\Big[{\rm Li}_2(r_{1})+\ln^2(1-r_{1})\Big] \nn
&\hspace{0.5cm} - \frac{2(r_{2}-1)(r_{3}+r_{1}-r_{2}-1)}{(r_{3}-r_{2})(r_{2}-r_{1})}\Big[{\rm Li}_2(r_{2})+\ln^2(1-r_{2})\Big] \nn
&\hspace{0.5cm} - \bigg[\frac{r^2_1+r_{3}(2-3r_{1}-3r_{2}) + 3r_{1} r_{2} + 1}{(r_{3}-r_{1})(r_{3}-r_{2})} \nn
&\hspace{0.5cm} + \frac{r_{3}(2r_{1} r_{2}-r_{1}-r_{2})-r_{1} r_{2}}{r_{3}^2(r_{3}-r_{1})(r_{3}-r_{2})} \bigg]\ln(1-r_{3})\nn
&\hspace{0.5cm} - \frac{2(r_{1}-1)(r_{3} r_{1}-r_{2})}{r_{1}(r_{3}-r_{1})(r_{1}-r_{2})} \ln(1-r_{1})\nn
&\hspace{0.5cm} - \frac{2(r_{2}-1)(r_{3} r_{2}-r_{1})}{r_{2}(r_{3}-r_{2})(r_{2}-r_{1})} \ln(1-r_{2}) + \frac{3r^2_1+6r_{3}-1}{r_{3}(r_{3}-1)} \bigg\} \,.
\end{align}

\subsection{The factorization scale independence for the correlation function}
In this part, we show the factorization scale independence at leading power and the twist-two NLO level.
The derivative of the twist-two NLO correlation function on the scale $\mu$ is
\begin{align} \label{eq:mu-independence-tw2}
\frac{d}{d\ln \mu} & \Pi_{\rm LP}^{\rm OPE} (r_{1},r_{2},\mu) = - m_V f^{\|}_V (q\cdot \varepsilon^*)\, \nn
& \times  \int^1_0 du \, \bigg\{ \bigg( \frac{d}{d\ln \mu} \phi_{2;V}^{\|}(u,\mu) \bigg) \left[ \mathcal{T}_V^{(0)}(r_{1},r_{2},u,\mu)  + {\cal T}_V^{(1)}(r_{1},r_{2},u,\mu) \right] \nn
& \hspace{1.6cm} + \,\phi_{2;V}^{\|}(u,\mu)\frac{d}{d\ln \mu} \left[ \mathcal{T}_V^{(0)}(r_{1},r_{2},u,\mu)
 + {\cal T}_V^{(1)}(r_{1},r_{2},u,\mu) \right] \bigg\} \,,
\end{align}
where $\Pi_{\rm LP}^{\rm OPE} \equiv \Pi_{\rm LP}^{\rm LO} + \Pi_{\rm LP}^{\rm NLO}$, we need to use the evolution equations of the twist-2 LCDA and the heavy quark mass
\begin{align}
 &\frac{d}{d \ln \mu} \phi_{2;V}^{\|}(u,\mu) = 2\int^1_0 d u'\, V(u,u') \phi_{2;V}^{\|}(u',\mu) \,, \\
 &\frac{d}{d \ln \mu}m_Q(\mu)=-6\frac{\alpha_s C_F}{4\pi}m_Q(\mu) \,,
\end{align}
where the evolution kernel is given by \cite{Lepage:1980fj,Efremov:1979qk}
\begin{align}
V(u,u') =  \frac{\alpha_s C_F}{4\pi} \bigg[ \frac{2{\bar u}}{\bar u'}\left(1+\frac{1}{u-u'}\right)\theta(u-u')\bigg|_{+} + \frac{2u}{u'} \left(1+\frac{1}{u'-u}\right)\theta(u'-u)\bigg|_{+}\bigg] \,,
\end{align}
the plus distribution function is defined as
\begin{align}
    V(u,u')\bigg\vert_+ = V(u,u') - \delta(u-u') \int_0^1 dt V(t,u') \,.
\end{align}
Inserting these evolution equations into Eq.(\ref{eq:mu-independence-tw2}), then we obtain the first term
\begin{align}
& \int^1_0 du \, \bigg( \frac{d}{d\ln \mu} \phi_{2;V}^{\|}(u,\mu) \bigg) \left[ \mathcal{T}_V^{(0)}(r_{1},r_{2},u,\mu)  + {\cal T}_V^{(1)}(r_{1},r_{2},u,\mu) \right] \nn
& =  4\cdot \frac{\alpha_s C_F}{4\pi} \int^1_0 du \,\frac{\phi_{2;V}^{\|}(u,\mu)}{r_{3}-1} \bigg\{ \left[\frac{1-r_{1}}{r_{3}-r_{1}}\ln\frac{1-r_{3}}{1-r_{1}}+\frac{1-r_{2}}{r_{3}-r_{2}}\ln\frac{1-r_{3}}{1-r_{2}}+2\right] \nn
&\hspace{3cm} + \left[ -\frac1{2}+\frac{(r_{3}-1)^2\ln(1-r_{3})}{(r_{3}-r_{1})(r_{3}-r_{2})} - \frac{(r_{3}-1)(r_{1}-1)\ln{(1-r_{1})}}{(r_{3}-r_{1})(r_{1}-r_{2})} \right.\nn
&\hspace{3cm} \left. + \frac{(r_{3}-1)(r_{2}-1)\ln{(1-r_{2})}}{(r_{3}-r_{2})(r_{1}-r_{2})}\right] \bigg\} \,,
\end{align}
and the second term reads
\begin{align}
& \int^1_0 d u\,\phi_{2;V}^{\|}(u,\mu)\frac{d}{d\ln \mu} \left[ \mathcal{T}_V^{(0)}(r_{1},r_{2},u,\mu)  + {\cal T}_V^{(1)}(r_{1},r_{2},u,\mu) \right] \nn
& = 4\cdot\frac{\alpha_s C_F}{4\pi} \int^1_0 d u\, \frac{\phi_{2;V}^{\|}(u,\mu)}{r_{3}-1} \bigg\{\left[ - \frac{1-r_{1}}{r_{3}-r_{1}}\ln\frac{1-r_{3}}{1-r_{1}} -\frac{1-r_{2}}{r_{3}-r_{2}}\ln\frac{1-r_{3}}{1-r_{2}} - 2 \right] \nn
&\hspace{3cm} + \bigg[ - \frac{(r_{3}-1)^2\ln(1-r_{3})}{(r_{3}-r_{1})(r_{3}-r_{2})}+\frac{(r_{3}-1)(r_{1}-1)\ln(1-r_{1})}{(r_{3}-r_{1})(r_{1}-r_{2})} \nn
&\hspace{3cm} - \frac{(r_{3}-1)(r_{2}-1)\ln(1-r_{2})}{(r_{3}-r_{2})(r_{1}-r_{2})} \bigg] + \frac{1}{2} \bigg\} \,,
\end{align}
Inputting these results into Eq.(\ref{eq:mu-independence-tw2}), we obtain
\begin{align}
\frac{d}{d\ln \mu} \Pi_{\rm LP}^{\rm OPE}(r_{1},r_{2},\mu) = O(\alpha_s^2) \,,
\end{align}
this equation displays the factorization scale independence of the leading power correlator up to $O(\alpha_s^2)$.

\subsection{Double spectral function for LO and NLO at twist-two}
Within the framework of perturbative QCD and light-cone OPE, we have obtained the correlation function at large space-like region of the variables $(p+q)^2$ and $q^2$ at LO and NLO for twist-two contributions of vector meson DAs, by utilizing the dispersion relation, which can be expressed as
\begin{eqnarray} \label{eq:dispersion_OPE}
\Pi_{\rm LP}^{\rm OPE}((p+q)^2,q^2) = ( q \cdot\varepsilon^*) \iint ds_{1} ds_{2}\frac{ \rho_{\rm LP}(r_{1},r_{2}) }{[s_{1}-(p+q)^2][s_{2}-q^2]} \,,
\end{eqnarray}
where
\begin{align}
\rho_{\rm LP}(r_{1},r_{2}) & = \frac{1}{\pi^2} f^{\|}_V m_V {\rm Im}_{s_{1}}{\rm Im}_{s_{2}}\int^1_0 du \, \phi_{2;V}^{\|}(u) \Big[ H_V^{(0)}(r_{1},r_{2},u) + H_V^{(1)}(r_{1},r_{2},u) \Big] \,,
\end{align}
then by using the quark-hadron duality and Borel transformation, matching the Eq.(\ref{eq:dispersion_OPE}) with Eq.(\ref{eq:double_disp}), we obtain the sum rules for the $HHV$ coupling constants
\begin{align} \label{eq:LCSR_LP_exp1}
f_{H_1} f_{H_2} g_{H_1H_2V}^{\rm LP} & = \frac{m_Q^4}{m_{H_1}^2 m_{H_2}^2}  e^{\frac{m_{H_1}^2 + m_{H_2}^2}{2 M^2}} \int^{+\infty}_{-\infty}dr_{1} \int^{2\hat{s}_0-r_{2}}_{-\infty}dr_{2} \, e^{-\frac{r_{1}+r_{2}}{2\hat{M}^2}} \rho_{\rm LP}(r_{1},r_{2}) \,,
\end{align}
where the duality subtraction region is $s_1+s_2<2s_0$, $s_0$ is the relevant threshold. For the calculation of the double spectral density, we use the following variable replacement
\begin{eqnarray}
r\equiv \frac{r_{2}-1}{r_{1}-1},\quad \sigma\equiv r_{1}+r_{2}-2,\quad d r_{1} \, d r_{2}= \frac{\sigma}{(r+1)^2} \,d r \, d\sigma \,,
\end{eqnarray}
this gives
\begin{eqnarray}
    r_{1}=\frac{\sigma}{r+1}+1,\quad r_{2}=\frac{r\sigma}{r+1}+1 \,,
\end{eqnarray}
then we obtain
\begin{align} \label{eq:LCSR_LP_exp2}
f_{H_1} f_{H_2} g_{H_1H_2V}^{\rm LP}
& = \frac{m^4_Q}{m_{H_1}^2 m_{H_2}^2} e^{\frac{m_{H_1}^2 + m_{H_2}^2}{2 M^2}} \int^{2\Hat{s}_0-2}_{-\infty}d \sigma \int^{+\infty}_{-\infty}d r \, e^{-\frac{\sigma+2}{2\Hat{M}^2}} \frac{\sigma}{(r+1)^2} \rho_{\rm LP}(r,\sigma) \,,
\end{align}
where $\Hat{s}_0={s_0}/{m^2_Q}$, $\Hat{M}^2=M^2/m^2_Q$, $\rho_{\rm LP} \equiv \rho_{\rm LP}^{\rm LO} + \rho_{\rm LP}^{\rm NLO}$. As discussed in Ref.\cite{Khodjamirian:1999hb}, our results demonstrate that the double spectral density $\rho_{\rm LP}(r,\sigma)$ can be accurately extracted using the asymptotic vector meson distribution amplitude $\phi_{2;V}^{\|}(u,\mu)=6u\bar{u}$, while neglecting the minor contributions from higher Gegenbauer moments. By applying the spectral representations detailed in Appendix B of Ref.\cite{Li:2020rcg}, the leading power spectral density $\rho_{\rm LP}(r,\sigma)$ is derived as follows:
\begin{align}
& \rho^{\rm LO,\,as}_{\rm LP}(r,\sigma) = f^{\|}_V m_V \frac{3r(r+1)}{\sigma} \delta^{(2)}(r-1) \theta(\sigma) \,, \label{eq:double_SD_LO_asymp} \\
& \rho^{\rm NLO,\,as}_{\rm LP}(r,\sigma) = -\frac{1}{2} \frac{\alpha_s C_F}{4\pi}f^{\|}_V m_V \bigg\{\Big[f_1(r,\sigma)+f_2(r,\sigma)+f_3(r,\sigma)\left(\ln^2 (r)-\pi^2\right)\Big]\delta^{(2)}(r-1)\nn
& \hspace{2.5cm}+\Big[f_2(r,\sigma)+2f_3(r,\sigma)\ln(r)\Big]\Big[\frac{d^3}{dr^3}\ln|1-r|\Big]+\Delta f(r,\sigma)\bigg\}\theta(\sigma)\theta(r) \,, \label{eq:double_SD_NLO_asymp}
\end{align}
where
\begin{align}
f_1(r,\sigma) & = \frac{r+1}{\sigma^2} \bigg\{-\frac{\sigma}{r+\sigma+1}\Big[6+18\sigma+r(72+66r+54\sigma+\pi^2(r+\sigma+1))\Big] \nn
& \hspace{0.5cm} + \frac{12r(r+1)\sigma}{r+\sigma+1}\ln(r)-12\Big[(r+1)^2+2r\sigma\Big]\ln\frac{r+\sigma+1}{r+1} \nn
& \hspace{0.5cm} + \frac{12r\sigma}{(r+\sigma+1)(r+r\sigma+1)}\ln\frac{\sigma}{1+r} \bigg[ 3(r+1)^2 + 4(r+1)^2\sigma + 5r\sigma^2  \nn
& \hspace{0.5cm} - (r+\sigma+1)(r+r\sigma+1)\bigg(3\ln\frac{r+\sigma+1}{r+1} + \ln\frac{r+r\sigma+1}{r+1} \bigg) \bigg] \nn
& \hspace{0.5cm} - 6r\sigma \left[ 3{\rm Li}_2\bigg( -\frac{\sigma}{r+1} \bigg) + 2{\rm Li}_2 \bigg(-\frac{r\sigma}{r+1} \bigg) \right] - 36 r \sigma\ln\frac{\mu^2} {m^2_Q} \bigg\}\nn
f_2(r,\sigma) & = -\frac{12(r+1)r}{\sigma}\Big[\ln\left(\frac{r\sigma(r+\sigma+1)}{(r+1)^2}\right) + \frac{r+1}{r\sigma+r+1}\Big],\nn
f_3(r,\sigma) & = \frac{12(r+1)r}{\sigma},\nn
\Delta f(r,\sigma) & = -12 \bigg( 4 + 3\ln\frac{\mu^2}{m^2_Q} \bigg)m^4_Q \frac{d}{dm^2_Q} \left[ \frac{1}{m^2_Q} \frac{(r+1)r}{\sigma}\delta^{(2)}(r-1) \right] \,,
\end{align}
where the asymptotic form of the twist-2 LCDA, $\phi_{2;V}^{\|} = 6u\bar{u}$, is employed in our calculations. This choice, although simplifying the analysis, requires modifications to the double spectral function to incorporate the higher moments of $\phi_{2;V}^{\|}$ at leading order. Consequently, we suggest the following adjustments to the double spectral function given in \eqref{eq:double_SD_LO_asymp} as
\begin{align}
\rho^{\rm LO}_{\rm LP}(r,\sigma) & = f^{\|}_V m_V \frac{3r(r+1)}{\sigma} \bigg\{ \delta^{(2)}(r-1) + \frac{2}{6!} \Big[ 3 (r+1)(r-1)^3 a_1  \nn
& \quad + 6 (r-1)^2 \left( r^2+3 r+1\right) a_2 + 10 (r+1) (r-1) (r^2+5r+1) a_3  \nn
& \quad + 15 \left(r^4+10r^3+20r^2+10r+1\right) a_4 \Big] \delta^{(6)} (r-1) \bigg\} \theta(\sigma) \,, \label{eq:double_SD_LO}
\end{align}
here we truncate the Gegenbauer expansion of $\phi_{2;V}^{\|}(u)$ to the fifth order.

\subsection{The light-cone sum rules for the $D_{(s)}D_{(s)}V$ and $B_{(s)}B_{(s)}V$ couplings at LP}
With the double spectral density in hand, we can continue to write the final expressions of the NLO LCSRs for the $D_{(s)}D_{(s)}V$ and $B_{(s)}B_{(s)}V$ couplings.
Inserting the double spectral density into the sum rules in Eq.(\ref{eq:LCSR_LP_exp2}) and complete the integral on the variale $r$, we obtain the final expressions for the LCSRs
\begin{align} \label{eq:LCSR_LP_exp3}
f_{H_1} f_{H_2} g_{H_1H_2V}^{\rm LP}
& = \frac{1}{m_{H_1}^2 m_{H_2}^2} e^{\frac{m_{H_1}^2 + m_{H_2}^2}{2 M^2}} \Big[ \mathcal{F}_{\rm LP}^{\rm LO}(M^2,s_0) + \mathcal{F}_{\rm LP}^{\rm NLO,\,as}(M^2,s_0) \Big] \,,
\end{align}
where
\begin{align}
& \mathcal{F}_{\rm LP}^{\rm LO}(M^2,s_0) = f^{\|}_V m_V m_Q^4 \Hat{M}^2\Big( e^{-\frac{1 }{ \hat M^2}}-e^{-\frac{\hat s_0}{\hat M^2}} \Big) \phi^{\|}_{2;V}\Big(\frac12\Big) \,, \\
& \mathcal{F}_{\rm LP}^{\rm NLO,\,as}(M^2,s_0) = - \frac{\alpha_s C_F}{4\pi} f^{\|}_V m_V m_Q^4  \bigg[ \int^{2\Hat{s}_0-2}_0d\sigma\,e^{-\frac{\sigma+2}{2\Hat{M}^2}}g(\sigma) + \Delta g(M^2,m^2_Q) \bigg] \,,
\end{align}
with $\mathcal{F}_{\rm LP}^{\rm LO}$ including the higher moments contributions of $\phi^{\|}_{2;V}$ and
\begin{align}
& g(\sigma) =  \frac{3}{4}\Big[ - 2{\rm Li}_2(-1-\sigma) - 2{\rm Li}_2(-\sigma) - 5 {\rm Li}_2\left(-\frac{\sigma }{2}\right) \Big] - 6\ln\left( \frac{\sigma}{2} \right)\ln\Big(\frac{\sigma+2}{2}\Big)  \nn
& \hspace{1.1cm} - \frac{3}{2} \ln(2+\sigma)\ln(\sigma+1) + \frac{3[\sigma(\sigma+2)(\sigma+8)+8]}{(\sigma+2)^3}\ln(\sigma) + \frac{3}{2}\ln(\sigma) \nn
& \hspace{1.1cm} + \frac{3(\sigma+1)(\sigma+6)}{(\sigma+2)^3}\ln(\sigma+1) - \frac{33}{8}\ln(\sigma+2) - \frac{3(3\sigma+20)}{8(\sigma+2)} \nn
& \hspace{1.1cm} - \frac{12\sigma(\sigma+1)}{(\sigma+2)^3}\ln2 - \frac{3}{8}\ln2 - \frac{5\pi^2}{8} - \frac{9}{2} \ln\bigg(\frac{\mu^2}{m^2_Q}\bigg) \,, \\
& \Delta g(M^2,m^2_Q) = 3 \bigg[ 4+3\ln \bigg(\frac{\mu^2}{m^2_Q}\bigg) \bigg]  e^{-\frac{1}{\hat M^2}} \,.
\end{align}

\subsection{The factorization scale independence for the strong couplings}
Checking factorization scale independence is as straightforward as examining the terms in ${\cal F}^{\rm NLO,\,as}_{\rm LP}$ that relate to the scale $\mu$, denoted as ${\cal F}^{\rm NLO,\,as}_{\rm LP,\,scale}$
\begin{align}
{\cal F}^{\rm NLO,\,as}_{\rm LP,\,scale}(M^2,s_0) & = \frac{\alpha_s C_F}{4\pi} m_Q^4 \bigg\{ \frac{9}{2} \int^{2\Hat{s}_0-2}_0 d\sigma\, e^{-\frac{\sigma+2}{2\Hat{M}^2}} - 9 e^{-\frac{1}{\Hat{M}^2}} \bigg\} \ln\bigg(\frac{\mu^2}{m^2_Q}\bigg) \,,
\end{align}
where
\begin{eqnarray}
 &&\int^{2\Hat{s}_0-2}_0d\sigma\,e^{-\frac{\sigma+2}{2\Hat{M}^2}} = 2\Hat{M}^2 \Big( e^{-\frac{1}{\Hat{M}^2}}-e^{-\frac{\Hat{s}_0}{\Hat{M}^2}} \Big),
\end{eqnarray}
then
\begin{align}
\frac{d}{d\ln\mu} {\cal F}^{\rm NLO,\,as}_{\rm LP,\,scale}(M^2,s_0) & =\frac{\alpha_s C_F}{4\pi} m^4_Q \bigg[ 18 \Hat{M}^2 \Big( e^{-\frac{1}{\Hat{M}^2}}-e^{-\frac{\Hat{s}_0}{\Hat{M}^2}} \Big) - 18 e^{-\frac{1}{\Hat{M}^2}} \bigg] + O(\alpha_s^2) \,.
\end{align}
For the next-to-leading order (NLO) matching, only the leading-order terms with contributions from the asymptotic expansion of the twist-2 DA are required.
\begin{eqnarray}
{\cal F}^{\rm LO,\,as}_{\rm LP}(M^2,s_0)
&& = \frac{3}{2} m_Q^4 \Hat{M}^2\Big( e^{-\frac{1 }{ \hat M^2}}-e^{-\frac{\hat s_0}{\hat M^2}} \Big)
\end{eqnarray}
then
\begin{align}
\frac{d}{d\ln\mu}{\cal F}^{\rm LO}_{\rm LP}(M^2,s_0) & = \frac{3}{2}\Hat{M}^2  m^4_Q \Big(e^{-\frac{1}{\Hat{M}^2}}-e^{-\frac{\Hat{s}_0}{\Hat{M}^2}}\Big) \cdot 2\Big(-\frac{3}{2}\frac{\alpha_s C_F}{\pi}\Big) \nn
&\quad + \frac{3}{2}\Hat{M}^2  m^4_Q \,e^{-\frac{2}{\Hat{M}^2}}\cdot \left(-\frac{1}{M^2}\right)\cdot 2m_Q^2\left(-\frac{3}{2}\frac{\alpha_s C_F}{\pi}\right)  + O(\alpha_s^2)\nn
& = \frac{\alpha_s C_F}{4\pi} m^4_Q \bigg[ - 18 \Hat{M}^2 \Big( e^{-\frac{1}{\Hat{M}^2}} - e^{-\frac{\Hat{s}_0}{\Hat{M}^2}} \Big) + 18 e^{-\frac{\Hat{s}_0}{\Hat{M}^2}}\bigg]  + O(\alpha_s^2)  \,,
\label{eq:LO-mu-dep}
\end{align}
Summing these terms together, we obtain:
\begin{align}
\frac{d}{d\ln\mu} f_{H_1} f_{H_2} g_{H_1H_2V}
& = \frac{1}{m_{H_1}^2 m_{H_2}^2} e^{\frac{m_{H_1}^2 + m_{H_2}^2}{2 M^2}} \frac{d}{d\ln\mu} \bigg[ {\cal F}^{\rm LO,\,as}_{\rm LP} + {\cal F}^{\rm NLO,\,as}_{\rm LP} \bigg] = O(\alpha_s^2) \,,
\end{align}
this equation shows the scale independence of the $H_1H_2V$ couplings at LP up to $O(\alpha_s^2)$.

\section{The light-cone sum rules for the higher-twist corrections} \label{sec-5}
In this section, we develop the light-cone sum rules for the subleading power corrections up to $\delta_V^4$. These corrections involve contributions from the two-particle twist-2 to 4 and the three-particle twist-3 and 4 LCDAs of vector mesons, which can be expressed in a compact form,
\begin{align} \label{eq:HTLP_1}
\Pi_{\rm NLP}^{\rm LO}(r_{1}, r_{2}) & = m_Q (1+\hat{m}_{q_1}+\hat{m}_{q_2}) \left(q\cdot \varepsilon^\ast\right)   \nonumber \\
& \times\sum_{i=2}^4 \sum_{j=1}^4 \bigg\{\int_0^1 du \, \delta^i_V \, (-1)^j H_V^{(0)\,j}(r_{1},r_{2},u) \mathcal{A}_{\mathrm{2p},\,ij}(u)   \nonumber \\
& \quad + \int_0^1 dv \int_0^1 d\alpha_2 \int_0^{1-\alpha_2} d\alpha_g \, \delta^i_V \, (-1)^j H_V^{(0)\,j}(r_{1},r_{2},\alpha)\mathcal{A}_{\mathrm{3p},\,ij}(v,\underline{\alpha}) \bigg\} \,,
\end{align}
where $H_V^{(0)}$ denotes the leading-order hard kernel, defined as in\eqref{eq:hard_CF_LO}. The part involving soft functions is denoted using the following notations:
\begin{align} \label{eq:HTLP_2}
& \mathcal{A}_{2\mathrm{p},\,22}(u) =  f_V^{\perp} \psi_{3;V}^{\|}(u) \,, \quad
\mathcal{A}_{2\mathrm{p},\,31}(u) =  - f^{\|}_V \Delta\phi_{2;V}^{\|}(u) \,, \\
& \mathcal{A}_{2\mathrm{p},\,32}(u) = - f^{\|}_V \Big[ u {\bar u}\phi_{2;V}^{\|}(u) + \frac{1}{4} \phi_{4;V}^{\|}(u) + 2 \widehat{\widehat{\mathbb{C}}}(u) \Big],\\
& \mathcal{A}_{2\mathrm{p},\,33}(u) = \frac{1}{2} f^{\|}_V \Big[ \phi_{4;V}^{\|}(u) + 8 \widehat{\widehat{\mathbb{C}}}(u) \Big] \,, \quad
 \mathcal{A}_{2\mathrm{p},\,43}(u) =  2 f_V^{\perp} u \bar{u} \psi_{3;V}^{\|}(u) \,, \\
& \mathcal{A}_{3\mathrm{p},\,33}(v,\underline{\alpha}) = \frac{ 2 f_V^{\|}}{\bar{\alpha }} \Big[ \left(v-\bar{v}\right) \widehat{\Psi}^{\|}_{4;V}(\underline{\alpha})  + \widehat{\widetilde{\Psi }}^{\|}_{4;V}(\underline{\alpha}) + 2  \left(v-\bar{v}\right) \widehat{\Phi }^{\|}_{4;V}(\underline{\alpha}) + 2 \widehat{\widetilde{\Phi }}^{\|}_{4;V}(\underline{\alpha}) \Big]\,, \\
& \mathcal{A}_{3\mathrm{p},\,34}(v, \underline{\alpha}) = - \frac{ 2 f_V^{\|}}{\bar{\alpha }} (r_{2}-1) \Big[ \left(v-\bar{v}\right) \widehat{\Psi }^{\|}_{4;V}(\underline{\alpha}) + \widehat{\widetilde{\Psi }}^{\|}_{4;V}(\underline{\alpha}) + 2\left( v-\bar{v}\right)\widehat{\Phi }^{\|}_{4;V}(\underline{\alpha}) + 2 \widehat{\widetilde{\Phi }}^{\|}_{4;V}(\underline{\alpha}) \Big] \,,
\end{align}
and we also introduce the following notations:
\begin{align}
& \mathbb{C}(u) = \phi_{2;V}^{\|}(u) - 2 \phi_{3;V}^{\perp}(u) +  \psi_{4;V}^{\|}(u) \,, \quad \Delta\phi_{2;V}^{\|}(u)=\frac{2 u \phi_{2;V}^{\|}(u) }{r_{1}-r_{2}} \,,
\end{align}
The DAs with a hat are defined as follows:
\begin{align}
& \widehat{\widehat{\mathbb{C}}}(u) \equiv \int_0^u dv \int_0^{v} dw \, \mathbb{C}( w ) \,, \\
& \widehat{\Phi}_{\rm 3p}(\underline{\alpha}) \equiv \int_0^{\alpha_1} d\alpha_1^{\prime}\int_0^{1-\alpha_1^{\prime}} d\alpha_3^\prime \;\delta(\alpha_3^\prime - \alpha_3) \Phi_{\rm 3p}( \alpha_1^{\prime}, 1-\alpha_1^{\prime}-\alpha_3,  \alpha_3 ) \,.
\label{3p-redefine}
\end{align}
Our subsequent task involves deriving the LCSRs for subleading power contributions.
\begin{align} \label{eq:LCSR_NLP_exp1}
f_{H_1} f_{H_2} g_{H_1H_2V}^{\rm NLP} & = \frac{m_Q^4}{m_{H_1}^2 m_{H_2}^2}  e^{\frac{m_{H_1}^2 + m_{H_2}^2}{2 M^2}} \int^{+\infty}_{-\infty}dr_{1} \int^{2\hat{s}_0-r_{2}}_{-\infty}dr_{2} \, e^{-\frac{r_{1}+r_{2}}{2\hat{M}^2}} \rho_{\rm NLP}(r_{1},r_{2}) \,,
\end{align}
where the double spectral density corresponding to these subleading power contributions is formulated as:
\begin{eqnarray}
\rho_{\rm NLP}(r_{1},\,r_{2}) ={1 \over \pi^2} \, {\rm Im}_{s_{1}} \, {\rm Im}_{s_{2}} \Pi_{\rm NLP}^{\rm LO}(r_{1},\,r_{2}) \,,
\end{eqnarray}
here $r_{1} = s_{1}/m_Q^2,\, r_{2}= s_{2}/m_Q^2$. Extending the derivation of the dispersion representation within the framework of the leading power factorization formula \eqref{eq:dispersion_OPE}, we consider the general expression for double spectral densities governing invariant amplitudes in the two-particle case. This expression is provided as:
\begin{align}
\rho_{\rm 2p}^j(r_{1},\,r_{2}) & \equiv {1 \over \pi^2} \, {\rm Im}_{s_{1}} \, {\rm Im}_{s_{2}}  \,
\int_0^1 d u \, { \big( m_Q^2 \big)^j \phi_{2p}(u)   \over [ u \, s_{1} + \bar u \, s_{2} - m_Q^2 + i \, 0]^{j}}  \nn
& = {\big( m_Q^2 \big)^j  \over \Gamma(j)} \, {d^{j - 1} \over (d \, m_Q^2)^{j -1}}  \,
\sum_k \, {(-1)^{k+1} \, c^{(\phi_{2p})}_k\over  \Gamma(k+1) } \, (s_{2} - m_Q^2)^k  \, \delta^{(k)}(s_{1}-s_{2}) \,\nn
& = {\big( m_Q^2 \big)^j  \over \Gamma(j)} \, {d^{j - 1} \over (d \, m_Q^2)^{j -1}} \frac{1}{m_Q^2}  \,
\sum_k \, {(-1)^{k+1} \, c^{(\phi_{2p})}_k\over  \Gamma(k+1) } \, (r_{2} - 1)^k  \, \delta^{(k)}(r_{1}-r_{2}) \,\nn
& = \big( m_Q^2 \big)^j \sum_k c^{(\phi_{2p})}_k\rho_{j k}(r_{1},r_{2}) \,,
\label{eq:rho-2p}
\end{align}
where $\displaystyle\delta^{(k)}(r_{1}-r_{2})\equiv \frac{d^k}{dr_{1}^k}[\delta(r_{1}-r_{2})]$, the function $\phi_{2p}(u)$  is given by a Taylor expansion around $u=0$,
\begin{eqnarray}
\phi_{2p}(u) = \sum_k \, c^{(\phi_{2p})}_k \, u^k \,.
\label{eq:expansion}
\end{eqnarray}
Similar to the two-particle case, the three-particle scenario adheres to a comparable framework,
\begin{eqnarray}
&&\{\rho_{\rm 3p}^j(r_{1},r_{2}),\,\widehat{\rho}_{\rm 3p}^j(r_{1},r_{2})\}~\nn
&&= {1 \over \pi^2} \, {\rm Im}_{s_{1}} \, {\rm Im}_{s_{2}}  \,
\int_0^1 d v \, \int {\cal D} \alpha \,\,
\frac{v^\ell\,\big( m_Q^2 \big)^j \{\Phi_{3p}(\underline{\alpha}),\widehat\Phi_{3p}(\underline{\alpha})/\bar{\alpha} \}} {\big[  \alpha \, s_{1} + \bar{\alpha} \, s_{2} - m_Q^2 + i \, 0 \big]^j}\nn
 && = {\big( m_Q^2 \big)^j  \over \Gamma(j)} \, {d^{j - 1} \over (d \, m_Q^2)^{j -1}}  {1 \over \pi^2} \, {\rm Im}_{s_{1}} \, {\rm Im}_{s_{2}}  \,
\int_0^1 d \alpha \,
\frac{\{\overline{\Phi}_{3p}(\alpha,\ell),\overline{\widehat\Phi}_{3p}(\alpha,\ell) \}}{\alpha \, s_{1} + \bar{\alpha} \, s_{2} - m_Q^2 + i \, 0 }
 \nn
&& = {\big( m_Q^2 \big)^j  \over \Gamma(j)} \, {d^{j - 1} \over (d \, m_Q^2)^{j -1}}  \frac{1}{m_Q^2}  \,
\sum_k \, {(-1)^{k+1} \, \{c^{{\Phi}_{3p}}_{\ell, k},c^{\widehat{\Phi}_{3p}}_{\ell, k}\}\over  \Gamma(k+1) } \, (r_{2} - 1)^k  \, \delta^{(k)}(r_{1}-r_{2}) \,\nn
&& = \big( m_Q^2 \big)^j \sum_k \{c^{{\Phi}_{3p}}_{\ell, k} \rho_{jk,3p}(r_{1},r_{2}),\,c^{\widehat{\Phi}_{3p}}_{\ell, k} \widehat{\rho}_{jk,3p}(r_{1},r_{2})\},
\label{eq:3p-rho}
\end{eqnarray}
where the coefficients, denoted as$c^{{\Phi}{3p}}_{\ell, k}$ and $c^{\widehat{\Phi}{3p}}_{\ell, k}$ stem from the series expansion of two ``effective" DAs, namely ${\overline{\Phi}_{3p}(\alpha,\ell),\overline{\widehat\Phi}_{3p}(\alpha,\ell) }$,
\begin{align}
\{\overline{\Phi}_{3p}(\alpha,\ell),\overline{\widehat\Phi}_{3p}(\alpha,\ell) \} & = \int_0^{\alpha} \, d \alpha_1 \, \int_{\alpha-\alpha_1}^{1-\alpha_1} \, d \alpha_3 \,
{ (\alpha - \alpha_1)^\ell \over\alpha_3^{\ell+1} } \, \{\Phi_{3p}(\underline{\alpha}),\widehat{\Phi}_{3p}(\underline{\alpha})/\bar{\alpha} \} \nonumber \\
& = \sum_k \,\{ c^{{\Phi}_{3p}}_{\ell, k},c^{{\widehat{\Phi}}_{3p}}_{\ell, k}\} \, \alpha^k \,, \quad \text{with}~ k\geq 2 \,.
\end{align}

To address the terms proportional to $q^2$ in Eq.\eqref{eq:HTLP_1}, we introduce supplementary functions $\widetilde{\widehat{\rho}}_{jk,3p}(r_{1},r_{2})$ as outlined in Eq.\eqref{eq:3p-rho}:
\begin{equation}
    \widetilde{\widehat{\rho}}_{jk,3p}(r_{1},r_{2}) = r_{2} \, \widehat{\rho}_{jk,3p}(r_{1},r_{2}).
\end{equation}
Replacing $\alpha$ with $u$, we perform a variable substitution,
\begin{eqnarray}
\{\overline{\Phi}_{3p}(\alpha,\ell),\, \overline{\widehat\Phi}_{3p}(\alpha,\ell) \} \leftrightarrow
\{\overline{\Phi}_{3p}(u,\ell),\, \overline{\widehat\Phi}_{3p}(u,\ell) \} \,,
\end{eqnarray}
simplifies the integration of contributions from three-particle processes with those from the previously analyzed two-particle contributions. Substituting each twist and multiplicity component in the sum (\ref{eq:HTLP_1}) with its corresponding double dispersion form, we obtain the LO double spectral density of the correlation function,
\begin{eqnarray}
\rho_{\rm NLP}^{\rm LO}(r_{1},r_{2})
= && m_Q \big( 1 + {\Hat{m}_{q_1}}+{\Hat{m}_{q_2}} \big) \Bigg\{
\delta_V^2 \rho^{(\psi^{\|}_{3;V})}_1
+ \delta_V^3 \rho^{({\Delta\phi}^{\|}_{2;V})}_1  \nn
&& + \, \delta_V^3 \bigg( \rho^{(\phi_{2;V}^{\|})}_1+\rho^{(\phi^{\|}_{4;V})}_1 + \rho_1^{(\mathbb{C})} + \rho^{(\phi^{\|}_{4;V})}_2 + \rho_2^{(\mathbb{C})}\bigg)+ \, \delta_V^4 \rho^{(\psi_{3;V}^{\|})}_2 \nn
&&  +\,\delta_V^3\Bigg[\rho^{({\overline{\widehat{\Psi}}}^{\|}_{4;V})}_1+\rho^{({\overline{\widehat{\widetilde{\Psi}}}}^{\|}_{4;V})}_1+\rho^{({\overline{\widehat{\Phi}}}^{\|}_{4;V})}_1+\rho^{({\overline{\widehat{\widetilde{\Phi}}}}^{\|}_{4;V})}_1\Bigg]  \nn
&&  +\,\delta_V^3\Bigg[\rho^{({\overline{\widehat{\Psi}}}^{\|}_{4;V})}_2+\rho^{({\overline{\widehat{\widetilde{\Psi}}}}^{\|}_{4;V})}_2+\rho^{({\overline{\widehat{\Phi}}}^{\|}_{4;V})}_2+\rho^{({\overline{\widehat{\widetilde{\Phi}}}}^{\|}_{4;V})}_2\Bigg]\Bigg \}(s_1,s_2)\,.
\label{eq:rhoLO}
\end{eqnarray}
This expression takes an expansion form (\ref{eq:expansion}) for each term, where the coefficients $c_k^{(\phi)}$ are easily determined from the polynomial form of the DAs, as explicitly outlined in Appendix \ref{app:lcda}.
The resulting expression (\ref{eq:rhoLO}) contributes to a new and comprehensive understanding. We illustrate, for instance, the contributions to $\rho^{\rm (LO)}$ from the twist-2 and twist-3 DAs, emphasizing their asymptotic behavior,
\begin{align}
&\rho_1^{(\psi^{\|}_{3;V})} (r_{1},r_{2})= 3 m^4_Q f^{\perp}_V\frac{d}{d m^2_Q} \frac{1}{m_Q^2}
\Big[ (r_{2} - 1) \delta^{(1)}(r_{1}-r_{2}) + \frac{1}{2}
(r_{2} - 1)^2 \delta^{(2)}(r_{1}-r_{2}) \Big] \, \theta(r_{1}-1)
\,,\nn
&\rho_1^{(\phi^{\|}_{2;V})} (r_{1},r_{2})=\rho_1^{(\phi^{\|}_{4;V})} (r_{1},r_{2}) = -3 m^4_Q f^{\|}_V\frac{d}{d m^2_Q} \frac{1}{m_Q^2}
\Bigg[ \frac{1}{2} (r_{2} - 1)^2\delta^{(2)}(r_{1}-r_{2})\nn
&\hspace{2.5cm} +\frac{2}{3}
(r_{2} - 1)^3\delta^{(3)}(r_{1}-r_{2})+\frac{1}{4}
(r_{2} - 1)^4\delta^{(4)}(r_{1}-r_{2})\Bigg ] \, \theta(r_{1}-1),\nn
&\rho_2^{(\phi^{\|}_{4;V})} (r_{1},r_{2}) = 4 m^6_Q f^{\|}_V \frac{d^2}{d (m_Q^2)^2}
\frac{1}{m_Q^2} \Bigg[  \frac{1}{2}
(r_{2} - 1)^2\delta^{(2)}(r_{1}-r_{2}) +\frac{2}{3}
(r_{2} - 1)^3\delta^{(3)}(r_{1}-r_{2})\nn
&\hspace{4cm}+\frac{1}{4}
(r_{2} - 1)^4\delta^{(4)}(r_{1}-r_{2})\Bigg ] \, \theta(r_{1}-1),\nn
&\rho_2^{(\psi^{\|}_{3;V})} (r_{1},r_{2})=\frac{f^{\perp}_V}{f^{\|}_V}\rho_2^{(\phi^{\|}_{4;V})} (r_{1},r_{2}) \,.
\label{eq:rhophiV}
\end{align}
After incorporating  the obtained NLP double spectral density into Eq.\eqref{eq:LCSR_NLP_exp1} and carrying out the necessary the dual integrals, we arrive at the NLP LCSRs for the concentrate couplings,
\begin{align} \label{eq:LCSR_NLP_exp2}
f_{H_1} f_{H_2} g_{H_1H_2V}^{\rm NLP}
& = \frac{1}{m_{H_1}^2 m_{H_2}^2} e^{\frac{m_{H_1}^2 + m_{H_2}^2}{2 M^2}} \mathcal{F}_{\rm NLP}^{\rm LO}(M^2,s_0) \,,
\end{align}
where
\begin{align}
\mathcal F_{\rm NLP}^{\rm LO}(M^2,s_0) =\mathcal F_{2p}^{\rm LO}(M^2,s_0)+\mathcal F_{3p}^{\rm LO}(M^2,s_0)
+\widetilde{\mathcal F}_{3p}^{\rm LO}(M^2,s_0) \,,
\label{eq:LONLP}
\end{align}
this produces the following simplified expression:
\begin{align}
& \mathcal F_j^{(\phi)}(M^2,s_0) = \frac{1}{(j-1)!} \bigg[(-1)^{j} \big(M^2\big)^{2-j} \,
e^{-\frac{m^2_Q}{M^2}} + \, \delta_{j1} \, M^2\, e^{- \frac{s_0}{M^2}} \bigg] \,\phi(u) \,\bigg|_{u=\frac{1}{2}} \,,\\
& \widetilde{\mathcal F}_{j}^{(\phi)}(M^2,s_0)
 =  -\frac{1}{(j-1)!}\, M^4 \frac{d^{j-1}}{d{m^2_Q}^{j-1}} \bigg\{
 e^{-\frac{m_Q^2}{M^2}} \bigg[ \bigg(1+\frac{m_Q^2}{M^2}\bigg) \phi(u) + u\phi^\prime(u)\bigg]  \nn
&\hspace{2.6cm} - e^{-\frac{s_0}{M^2}} \bigg[ \left(1+ \frac{s_0}{M^2}\right) \phi(u) + \bigg( 1+\frac{s_0-m_Q^2}{M^2} \bigg) u\phi^\prime(u) \bigg]
\bigg\} \bigg|_{u=\frac{1}{2}} \,.\label{eq:Fellphi}
\end{align}
%
Applying these formulas, we isolate the unique contributions from individual DAs to the NLP LCSRs,
\begin{align}
&\mathcal F_{2p}^{\rm (LO)}(M^2,s_0)
= m^5_Q (1+{\Hat{m}_{q_1}}+{\Hat{m}_{q_2}}) e^{-\frac{m^2_Q}{M^2}}\bigg\{
\delta^2_V f^{\perp}_{V}{\psi}^{\|}_{3;V}(u)
  \nn
&\hspace{2.8cm} + \, \delta^3_V f^{\|}_V \Delta {\cal F}_{\phi^{\|}_{2;V}} - \, \delta_V^3 f_V^{\parallel} \Big[  u \bar{u} {\phi}^{\|}_{2;V}(u)+{1 \over 4} {\phi}^{\|}_{4;V}(u)+2\hat{\hat{\mathbb{C}}}(u) \Big] \nn
&\hspace{2.8cm} - \, \frac{1}{4}\frac{m^2_Q}{M^2} \delta^3_V f^{\|}_V \Big[\phi^{\|}_{4;V}(u)+8\hat{\hat{\mathbb{C}}}(u) \Big] -\delta^4_V f^{\perp}_V \frac{m^2_Q}{M^2}u{\bar u}\psi^{\|}_{3;V}\bigg\}\bigg|_{u=\frac{1}{2}},\label{eq:2pLO}\\
& \mathcal F_{3p}^{\rm (LO)}(M^2,s_0) = m^5_Q (1+{\Hat{m}_{q_1}}+{\Hat{m}_{q_2}})\delta_V^3 f_V^{\|}  e^{-\frac{m^2_Q}{M^2}} \frac{m^2_Q}{M^2} \bigg( \frac{m_Q^2}{3M^2}-1 \bigg) \overline{\chi}_{\Psi\Phi}(u)\bigg|_{u=\frac{1}{2}} \,,\\
&\widetilde{\mathcal F}_{3p}^{\rm (LO)}(M^2,s_0) = \frac{m^5_Q (1+{\Hat{m}_{q_1}}+{\Hat{m}_{q_2}}) }{3M^2} \delta_V^3 f_V^\perp e^{-\frac{m^2_Q}{M^2}} \nn
&\hspace{2.8cm} \times \bigg[ \bigg(\frac{m^2_Q}{M^2}-2\bigg)\,\overline{\chi}_{\Psi\Phi}(u) + u\,\frac{\partial}{\partial u}\overline{\chi}_{\Psi\Phi}(u)\bigg]\bigg|_{u=\frac{1}{2}} \,,
\end{align}
where
\begin{align}
\overline{\chi}_{\Psi\Phi}(u) & =\overline{\widehat{\Psi}}^{\|}_{4;V}(u,\ell)\Big|_{\ell=1} -\overline{\widehat{\Psi}}^{\|}_{4;V}(u,\ell)\Big|^{\ell=0}_{\ell=1}+\overline{\widehat{\widetilde{\Psi}}}^{\|}_{4;V}(u,\ell)\Big|_{\ell=0}\nn
&\hspace{0.4cm}+2\bigg[ \overline{\widehat{\Phi}}^{\|}_{4;V}(u,\ell)\Big|_{\ell=1} -\overline{\widehat{\Phi}}^{\|}_{4;V}(u,\ell)\Big|^{\ell=0}_{\ell=1}+\overline{\widehat{\widetilde{\Phi}}}^{\|}_{4;V}(u,\ell)\Big|_{\ell=0} \bigg] \,, \\
\Delta {\cal F}_{\phi^{\|}_{2;V}} & = - \frac{1}{80} \Big[25 + 21 a^{\|}_1{(\mu)} - 15 a^{\|}_3{(\mu)} + 25 a^{\|}_4{(\mu)} \Big] \,.
\end{align}

Finally, summing the the leading and subleading power LCSRs in Eq.\eqref{eq:LCSR_LP_exp3} and \eqref{eq:LCSR_NLP_exp2} together we obtain the final expression for the $H_1H_2V$ couplings
\begin{align} \label{eq:LCSR_final_expression}
f_{H_1} f_{H_2} g_{H_1H_2V}
& = \frac{1}{m_{H_1}^2 m_{H_2}^2} e^{\frac{m_{H_1}^2 + m_{H_2}^2}{2 M^2}} \Big[ \mathcal{F}_{\rm LP}^{\rm LO}(M^2,s_0) + \mathcal{F}_{\rm LP}^{\rm NLO} (M^2,s_0) +  \mathcal{F}_{\rm NLP}^{\rm LO}(M^2,s_0) \Big] \,.
\end{align}

\section{Numerical analysis \label{sec-6}}

This section addresses the determination of various input parameters and the extraction of the strong couplings $g_{D_{(s)}D_{(s)}V}$ and $g_{B_{(s)}B_{(s)}V}$ from the established LCSRs in Eq.(\ref{eq:LCSR_final_expression}). The mass of have-light mesons and light vector mesons are  referenced from the PDG \cite{ParticleDataGroup:2022pth}. Careful consideration is given to the decay constants of pseudoscalar heavy-light mesons, employing three distinct approaches. The first approach utilizes Lattice QCD (LQCD) values for the decay constants of charmed and bottom mesons, as detailed in Table\ref{tab:decay-constant}. This includes $N_f=2+1+1$ results from \cite{FlavourLatticeAveragingGroupFLAG:2021npn} and heavy pseudoscalar meson constant ratios from \cite{Lubicz:2017asp}.
The second approach employs two-point QCD sum rules to determine the decay constants $f_H$, ($H=D_{(s)},B_{(s)}$), as described in \cite{Gelhausen:2013wia}. This method meticulously incorporates ${\cal O}(\alpha^2_s)$ contributions to the perturbative component and calculates ${\cal O}(\alpha_s)$ corrections to the quark-condensate term, thereby enhancing the accuracy and reliability of the resulting decay constants. The third approach derives from experimental data reported in the PDG 2022 compilation \cite{ParticleDataGroup:2022pth}. This approach uses averages of inclusive and exclusive determinations for the CKM matrix elements, specifically $|V_{cd}|=0.221\pm 0.004$, $|V_{cs}|=0.976\pm 0.006$ and  $|V_{ub}|=(3.82\pm 0.20)\times 10^{-3}$.  Additionally, experimental averages include of $|V_{cd}f_{D^+}|=45.82\pm 1.10$ MeV, $|V_{cs}f_{D_s^+}|=243.5\pm 2.7$ MeV  and  $|V_{ub}f_{B^+}|=0.77\pm 0.07$ MeV. These values are used to calculate the decay constants by taking their respective ratios, and the results are summarized in Table \ref{tab:decay-constant}.

\begin{table}[h!]
\setlength\tabcolsep{3pt}
\renewcommand{\arraystretch}{1.5}
\centering
\begin{tabular}{|c|c|c|c|c|}
\hline
\multirow{2}{*}{Methods}&\multicolumn{2}{|c|}{ charmed meson}
&
\multicolumn{2}{|c|}{bottom meson }\\[1mm]
\cline{2-5}
&$f_D~[\mbox{MeV}]$ & $f_{D_s}~[\mbox{MeV}]$
&$f_B~[\mbox{MeV}]$ & $f_{B_s}~[\mbox{MeV}]$
\\
\hline
two-point QCDSRs
&$201^{+12}_{-13}$ & $238^{+13}_{-23}$
&$207^{+17}_{-09}$& $242^{+17}_{-12}$\\[1mm]
\hline
LQCD & $212.0\pm 0.7$ & $249.9\pm 0.5$
& $190.0\pm 1.3$& $230.3\pm 1.3$ \\[1mm]
\hline
Experiment
&$207.3\pm 6.2$ & $249.5\pm 3.16$
&$201.6\pm 21.2$& $--$\\[1mm]
\hline
\end{tabular}
\caption{Decay constants for charmed and bottom mesons using three methods.}
\label{tab:decay-constant}
\end{table}

\begin{table}[tb]
\centering
\renewcommand{\arraystretch}{1.2}
\resizebox{\columnwidth}{!}{
\begin{tabular}{|c|c|c|c|}
\hline
parameter & input value & [Ref.]& rescaled values
\\[1mm]\hline
\multirow{2}{*}{$\alpha_s(m_Z)$} &
\multirow{2}{*}{$0.1179 \pm 0.0010$} &
\multirow{6}{*}{\cite{ParticleDataGroup:2022pth}} &
$\alpha_s(1.5\,\mbox{GeV})=0.3487^{+0.0102}_{-0.0097}$
\\
~& ~& ~&
$\alpha_s(3.0\,\mbox{GeV})=0.2527^{+0.0050}_{-0.0048}$
\\[1mm]
$\overline{m}_c(\overline{m}_c)$    &  1.280 $\pm$ 0.025 \mbox{GeV} &
~ &
$\overline{m}_c(1.5\,\mbox{GeV})=1.205 \pm 0.034$ \mbox{GeV}
\\[1mm]
$\overline{m}_b(\overline{m}_b)$   &    4.18  $\pm$ 0.03 \,\mbox{GeV} &
&$\overline{m}_b(3.0\,\mbox{GeV})=4.473  \pm 0.04$  \mbox{GeV}
\\[1mm]
\multirow{2}{*}{($\overline{m}_u+\overline{m}_d$)(2 \,\mbox{GeV}) }
&
\multirow{2}{*}{$6.78 \pm 0.08 $ \mbox{MeV}}& \multirow{2}{*}{\cite{FlavourLatticeAveragingGroupFLAG:2021npn,ParticleDataGroup:2022pth}} & ($\overline{m}_u+\overline{m}_d$)(1.5 \,\mbox{MeV}) =
7.305 $\pm$ 0.09 \mbox{MeV}\\
~ & ~& ~&
($\overline{m}_u+\overline{m}_d$)(3.0 \,\mbox{GeV}) =
6.331 $\pm$ 0.07 \mbox{MeV}
\\[1mm]
\multirow{2}{*}{($\overline{m}_s$)(2 \,\mbox{GeV}) }
&
\multirow{2}{*}{$93.1 \pm 0.6 $ \mbox{MeV}}& \multirow{2}{*}{\cite{FlavourLatticeAveragingGroupFLAG:2021npn,ParticleDataGroup:2022pth}} & ($\overline{m}_s$)(1.5 \,\mbox{MeV}) =
100.305 $\pm$ 0.65 \mbox{MeV}\\
~ & ~& ~&
($\overline{m}_s$)(3.0 \,\mbox{GeV}) =
86.936 $\pm$ 0.56 \mbox{MeV}
\\[1mm]
\hline
\end{tabular}}
\caption{QCD parameters used in the LCSRs and two-point QCDSRs.}
\label{tab:QCD}
\end{table}

A detailed discussion on the determination of the QCD parameters can be found in our prior study \cite{Jin:2024zyy}. In this paper, we directly incorporate the table from the cited work \cite{Jin:2024zyy}, as presented in Table \ref{tab:QCD}.
As discussed in the previous section, it is essential to determine the values of the distribution amplitudes (DAs) or their derivatives at the midpoint $u=1/2$. This determination relies on a comprehensive set of coefficients in the conformal expansion, which are influenced by the symmetry properties dictated by the renormalization group (RG) equations governing their scale dependence \cite{Agaev:2010aq}. Moreover, considering the next-to-leading order (NLO) for asymptotic twist-2 two-particle DAs necessitates the renormalization of Gegenbauer coefficients to NLO. This renormalization process is complex and multiplicative, as detailed in  Eq.\eqref{eq:evoaNLO} in Appendix \ref{app:lcda}.

In addition to providing the decay constants $f^{\perp,\parallel}_V(\mu)$ and twist-2 parameters $a^{\perp,\parallel}_n(\mu)$, we present the Gegenbauer moments for the twist-3 and twist-4 DAs of vector mesons in our previous paper \cite{Jin:2024zyy}.  These DAs are constructed up to next-to-leading order (NLO) in the conformal expansion as discussed in Ref.\cite{Ball:1998ff,Ball:1998sk}. The normalization and non-asymptotic coefficients at $\mu=1\,{\rm GeV}$, serving as input parameters, are computed using QCD sum rules (refer to  Ref.\cite{Ball:1998ff,Ball:1998sk,Ball:2007zt} and references therein).

We subsequently specify the renormalization and  factorization scale for the quark-gluon coupling and quark mass, the Borel parameter, and the quark-hadron duality threshold relevant to the LCSRs in Eq.\eqref{eq:LCSR_final_expression}. These parameters are pivotal for the accurate determination of sum rule behavior and the extraction of strong couplings.
Our parameter choices ensure strong convergence and stability of the sum rules, reducing theoretical uncertainties in the results. Owing to the finite mass of heavy quarks, these parameters vary between couplings involving charmed and bottom mesons.

\begin{table}[h!]
\setlength\tabcolsep{3pt}
\centering
\begin{tabular}{|c|c|c|c|c|c|}
\hline
Parameter & default value (interval) & [Ref.]
&Parameter & default value (interval) & [Ref.] \\[1mm]
\hline
\multicolumn{3}{|c|}{ charmed meson sum rules}
&
\multicolumn{3}{|c|}{bottom meson sum rules}\\
\hline
$\mu$ (GeV) & 1.5~(1.0\,-\,3.0)&
\multirow{3}{*}{\cite{Khodjamirian:2009ys}}
&$\mu$  (GeV) & {3.0~(2.5\,-\,4.5)}&
\multirow{3}{*}{\cite{Khodjamirian:2011ub}} \\[1mm]
$M^2$ (GeV$^2$)  & 4.5~(3.5\,-\,5.5) &
& $M^2$  (GeV$^2$)  & {16.0~(12.0\,-\,20.0)} &\\[1mm]
$ s_0$  (GeV$^2$) & 7.0~(6.5\,-\,7.5) &
&$ s_0$  (GeV$^2$) &  {37.5\,(35.0\,-\,40.0)} &\\[1mm]
\hline
\end{tabular}
\caption{
The renormalization scale $\mu$, Borel parameter $M^2$, along with duality threshold $s_0$, are employed in the LCSR for both charmed and bottom particles.}
\label{tab:scales}
\end{table}

The Borel parameter $M^2$ and the effective threshold $s_0$ in LCSRs are determined based on established criteria that prioritize the minimization of subleading power contributions in the light-cone OPE and the suppression of excited state contributions. These criteria focus on ensuring convergence and stability within the sum rule framework. Our adherence to these guidelines ensures the reliability and precision of our results. Table \ref{tab:scales} lists these parameters,  consistent with with those used in previous LCSRs analyses of heavy-to-light decay form factors and couplings \cite{Khodjamirian:2020mlb,Khodjamirian:2009ys,Khodjamirian:2011ub}.

In addition, we select the renormalization scale $\mu$ equal to the factorization scale used in the OPE, which is approximately by $\mu\sim\sqrt{m^2_H-m^2_Q}\sim \sqrt{2m_Q \bar{\Lambda}}$ ($\bar{\Lambda}=m_H-m_Q$). To ensure the perturbative expansion of the correlation function in  $\alpha_s$ remains within reasonable limits, we constrain the the NLO twist-2 and higher power terms are constrained to be no more than $30\%$ of LO counterparts. Consequently, we adopt default values of $\mu_c = 1.5\, {\rm GeV}$  for charm and $\mu_b = 3\, {\rm GeV}$ for bottom, exploring variations within the intervals  $(1.0\sim3.0){\rm GeV}$  for charm and $(2.5\sim4.5){\rm GeV}$  for bottom to quantify the uncertainties. This approach, motivated by theoretical consistency, aligns with previous studies in the field \cite{Khodjamirian:1997ub,Bagan:1997bp,Khodjamirian:2000ds,Ball:2004ye,Duplancic:2008ix,Khodjamirian:2009ys,Khodjamirian:2011ub}.  By harmonizing these scales, we ensure coherence in our calculations and facilitate comparisons with prior research.

\begin{figure}[htb]
\centering
\includegraphics[width=0.45 \columnwidth]{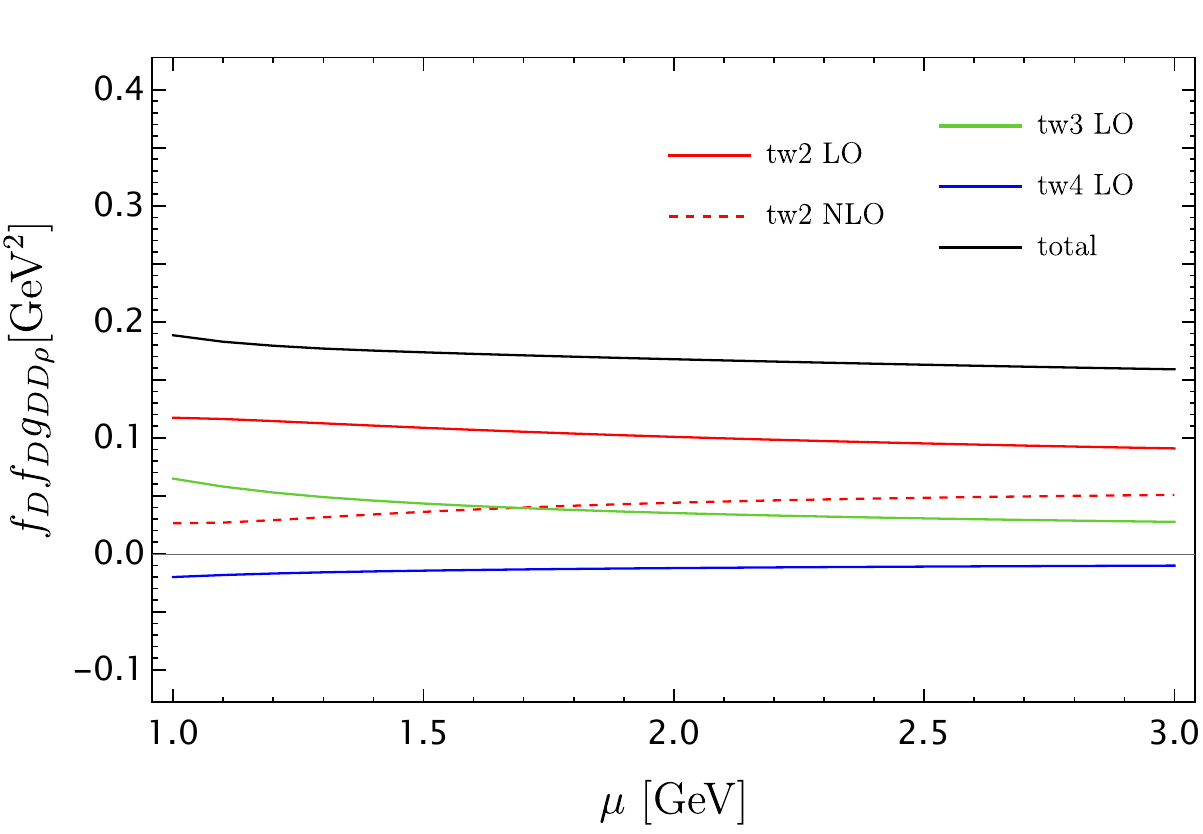} ~~
\includegraphics[width=0.45 \columnwidth]{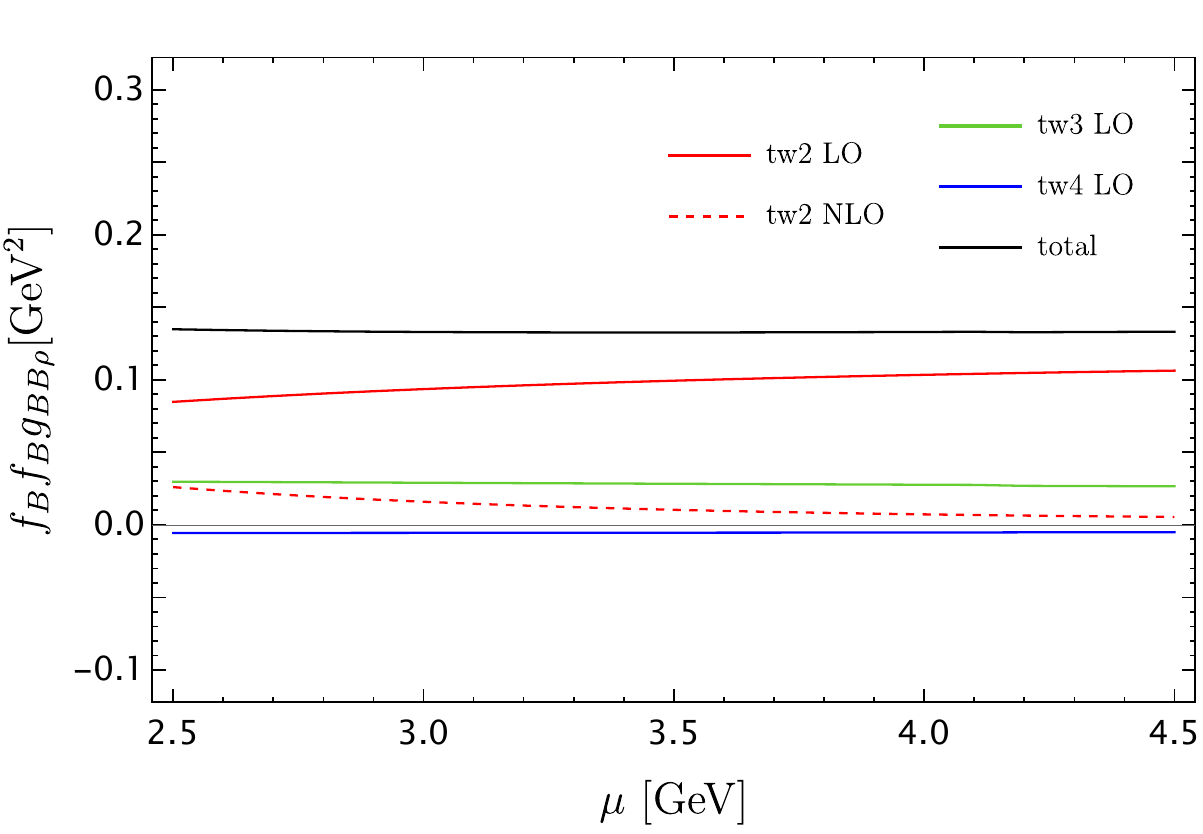} ~~
\includegraphics[width=0.45 \columnwidth]{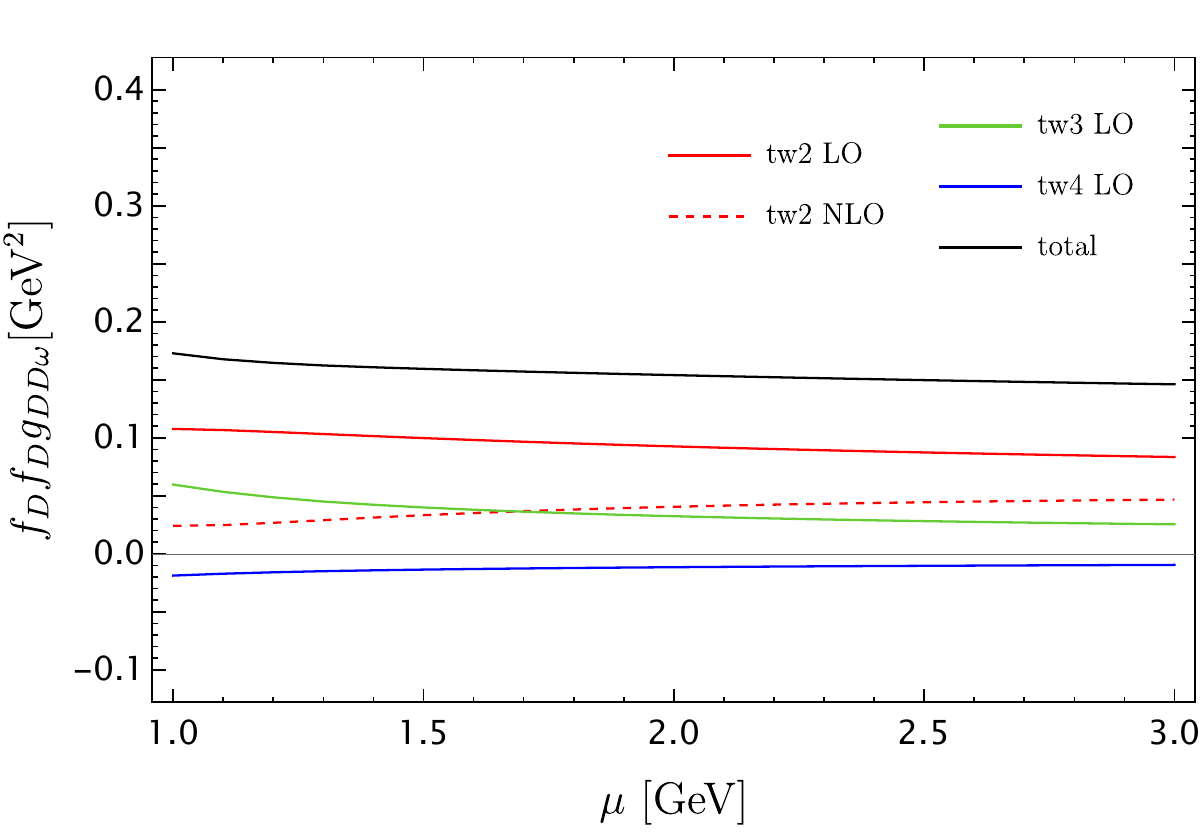} ~~
\includegraphics[width=0.45 \columnwidth]{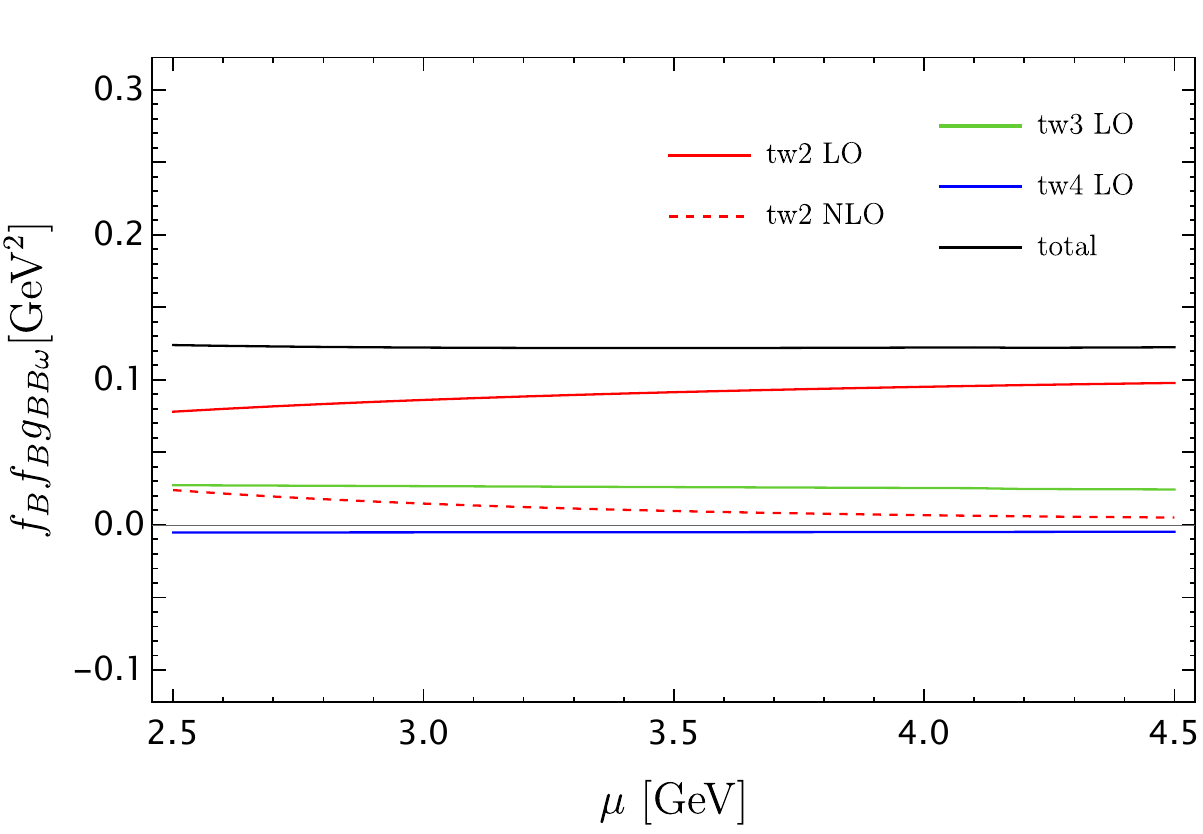} ~~
\includegraphics[width=0.45 \columnwidth]{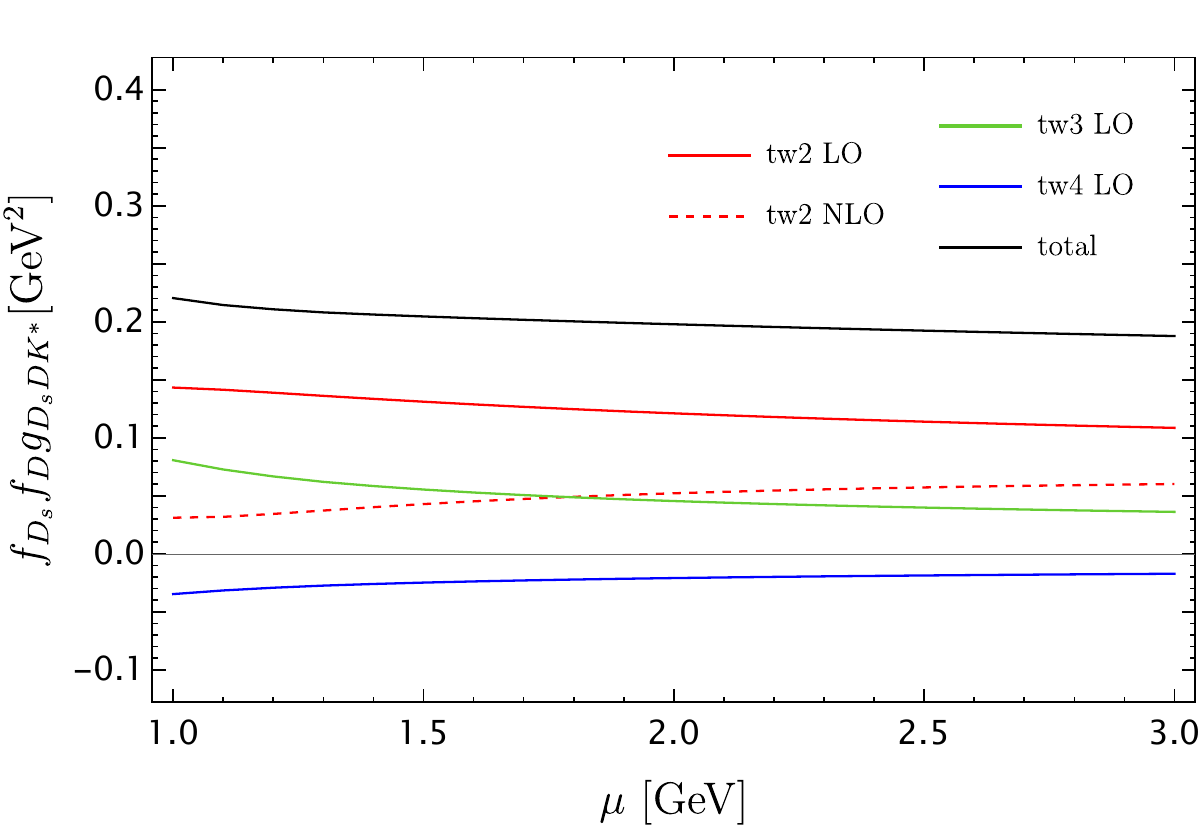} ~~
\includegraphics[width=0.45 \columnwidth]{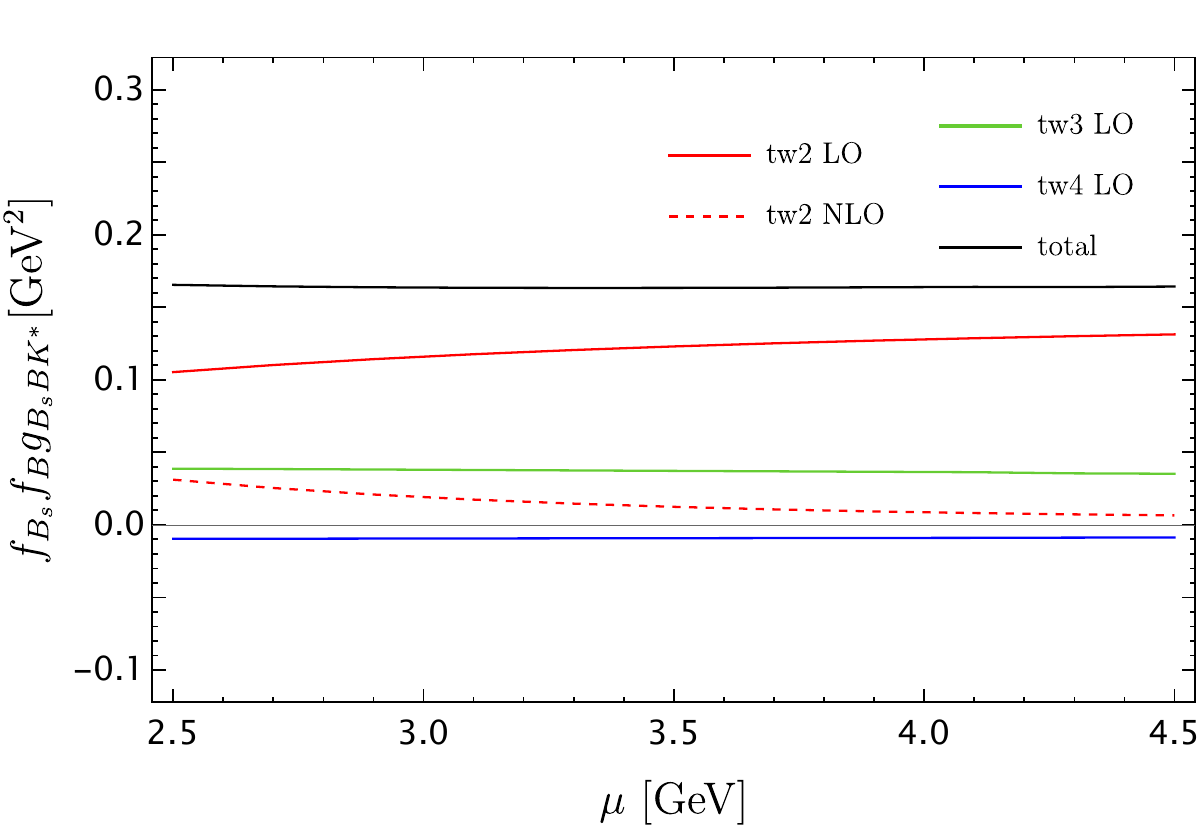} ~~
\includegraphics[width=0.45 \columnwidth]{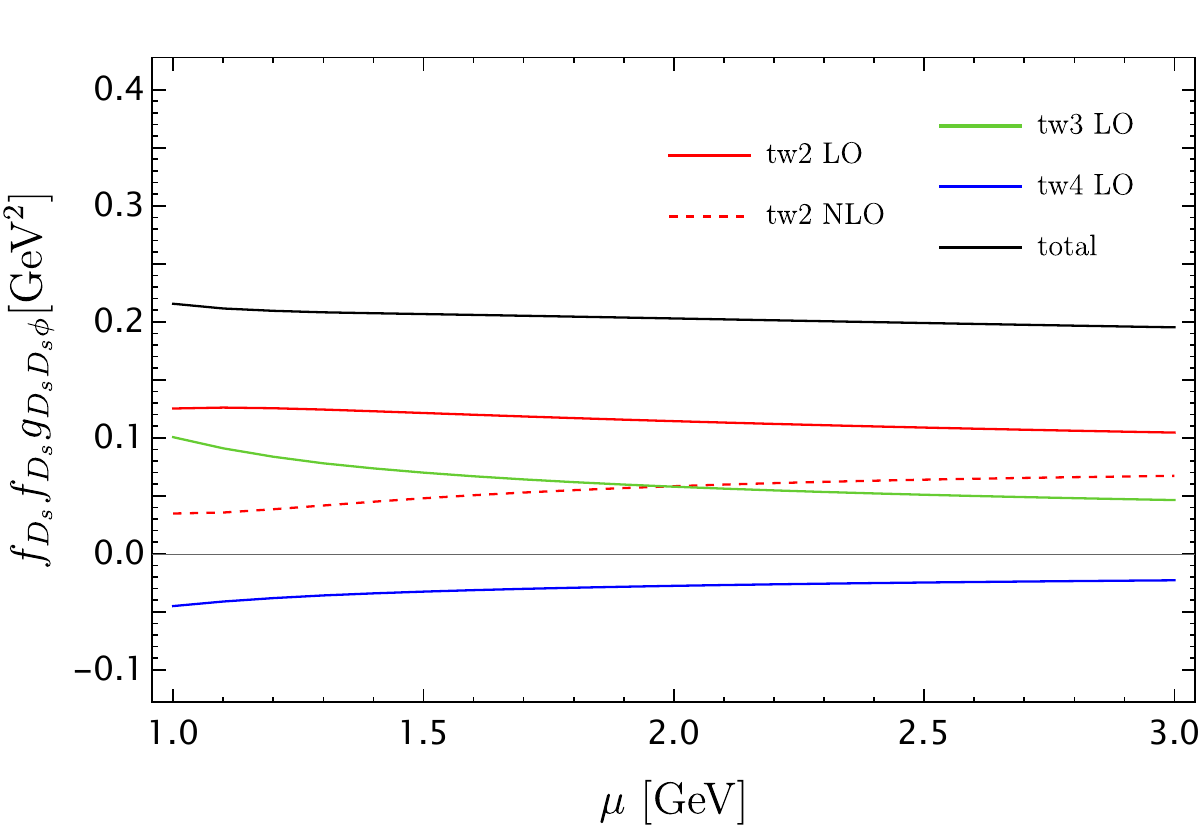} ~~
\includegraphics[width=0.45 \columnwidth]{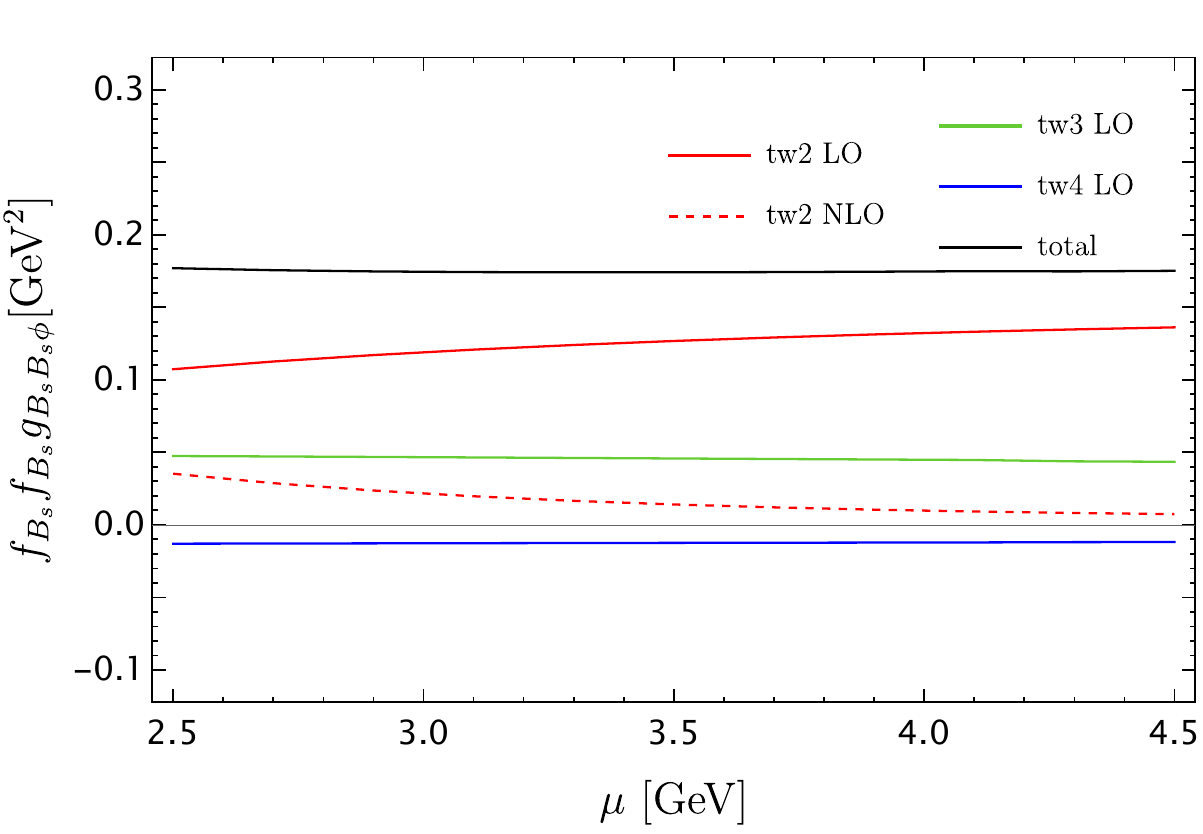}
\caption{Scale dependence of the products
$ f_{D_{(s)}} f_{D_{(s)}} g_{D_{(s)}D_{(s)}V}$ and $ f_{B_{(s)}}f_{B_{(s)}} g_{B_{(s)}B_{(s)}V}$ calculated from LCSR.
Displayed are the total values and separate  contributions from  twist-2 to twist-4
 of the vector meson($\rho,K^*,\phi$) DAs, considering central values for all other input parameters.}
\label{fig:mudep}
\end{figure}

\begin{figure}[htb]
\centering
\includegraphics[width=0.45 \columnwidth]{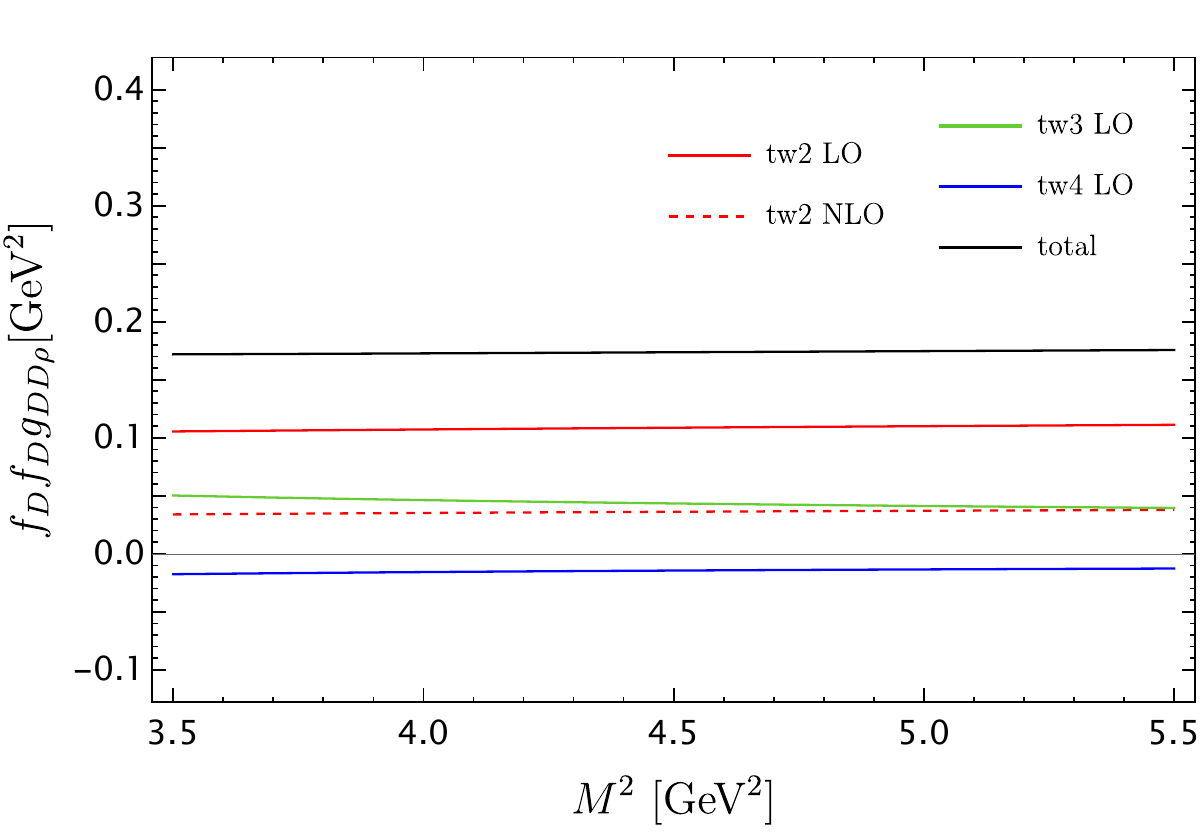} ~~
\includegraphics[width=0.45 \columnwidth]{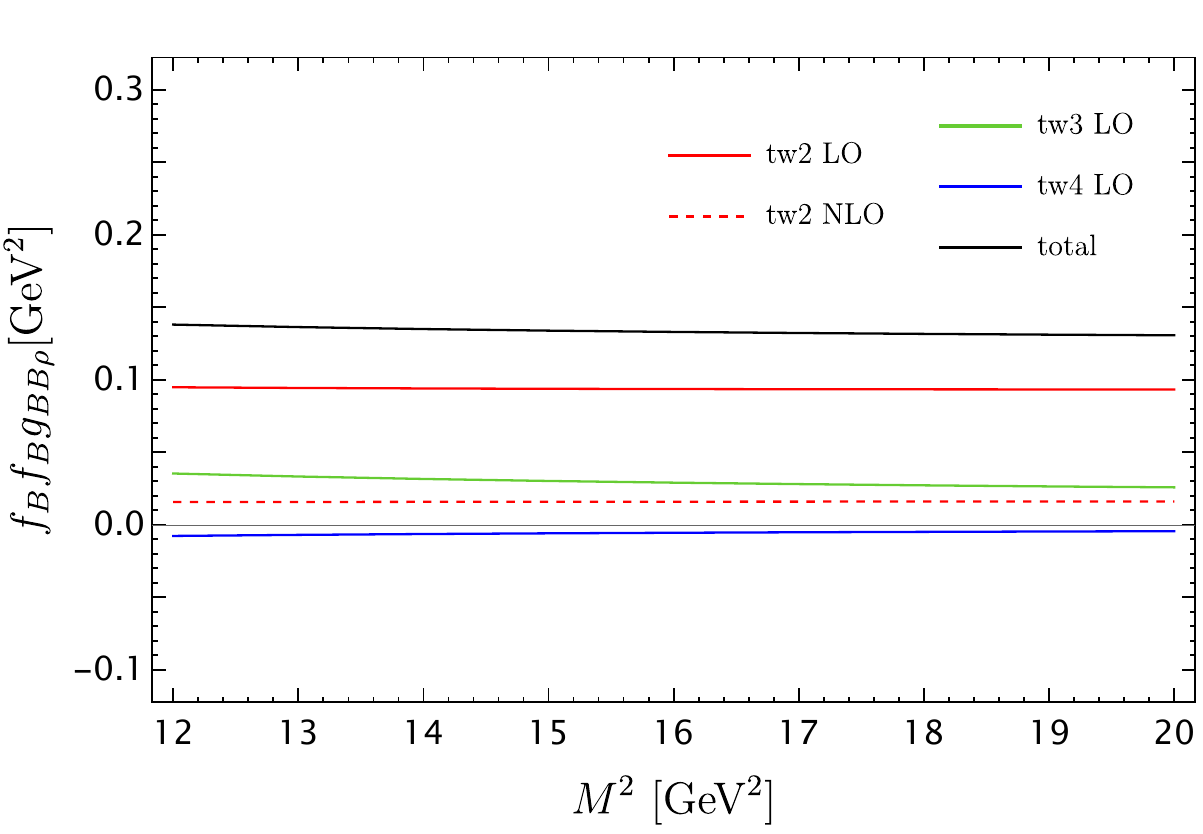} ~~
\includegraphics[width=0.45 \columnwidth]{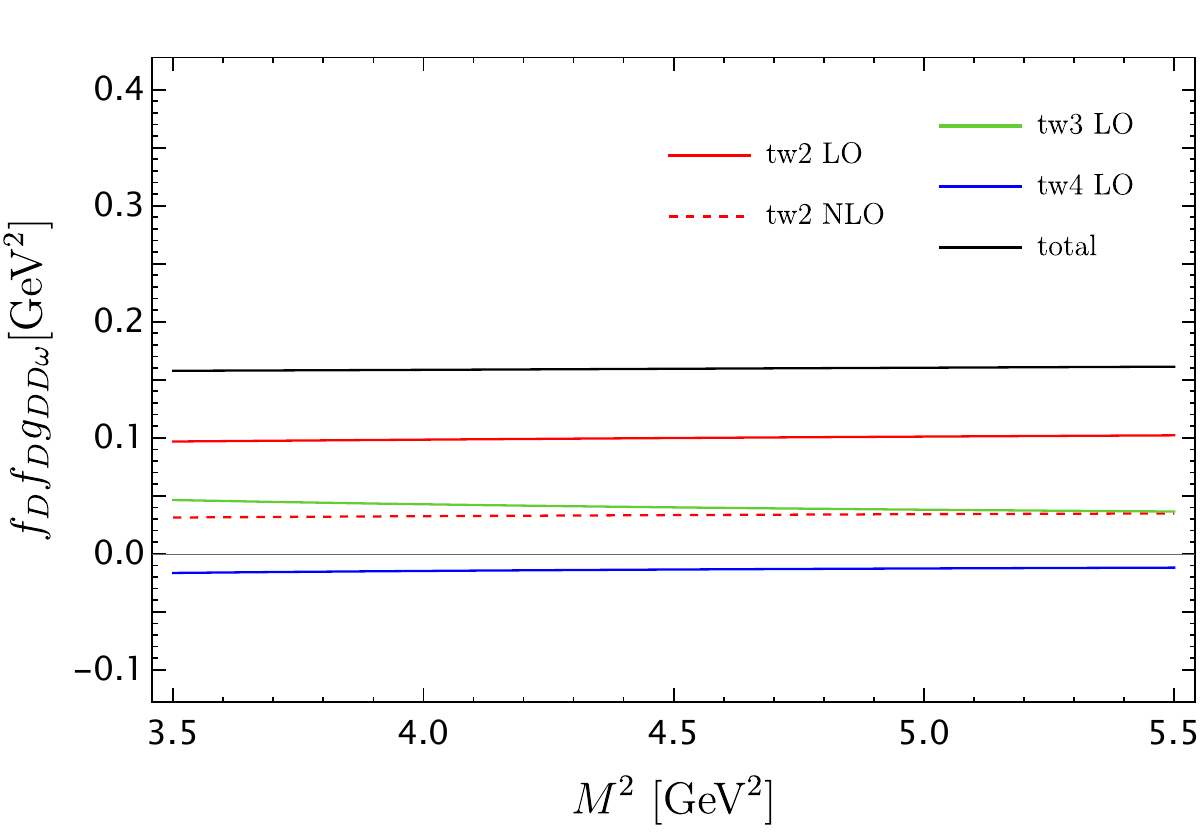} ~~
\includegraphics[width=0.45 \columnwidth]{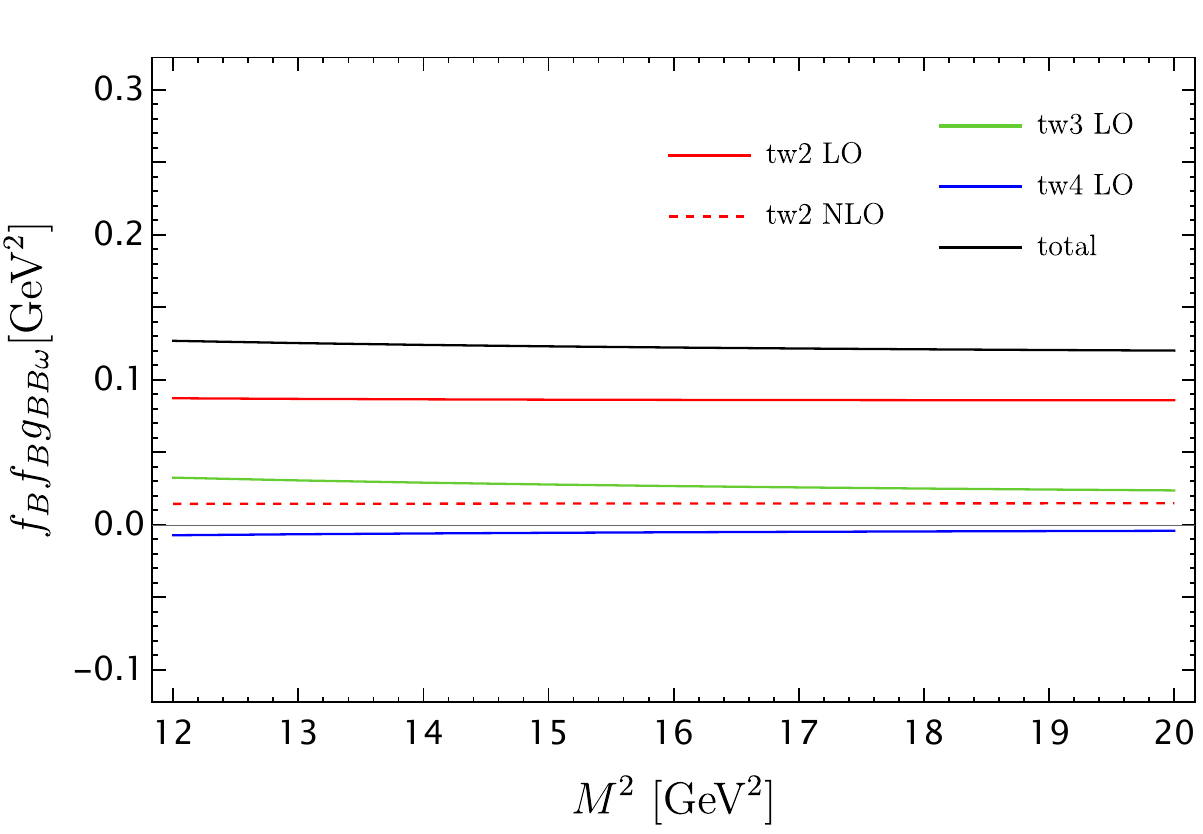} ~~
\includegraphics[width=0.45 \columnwidth]{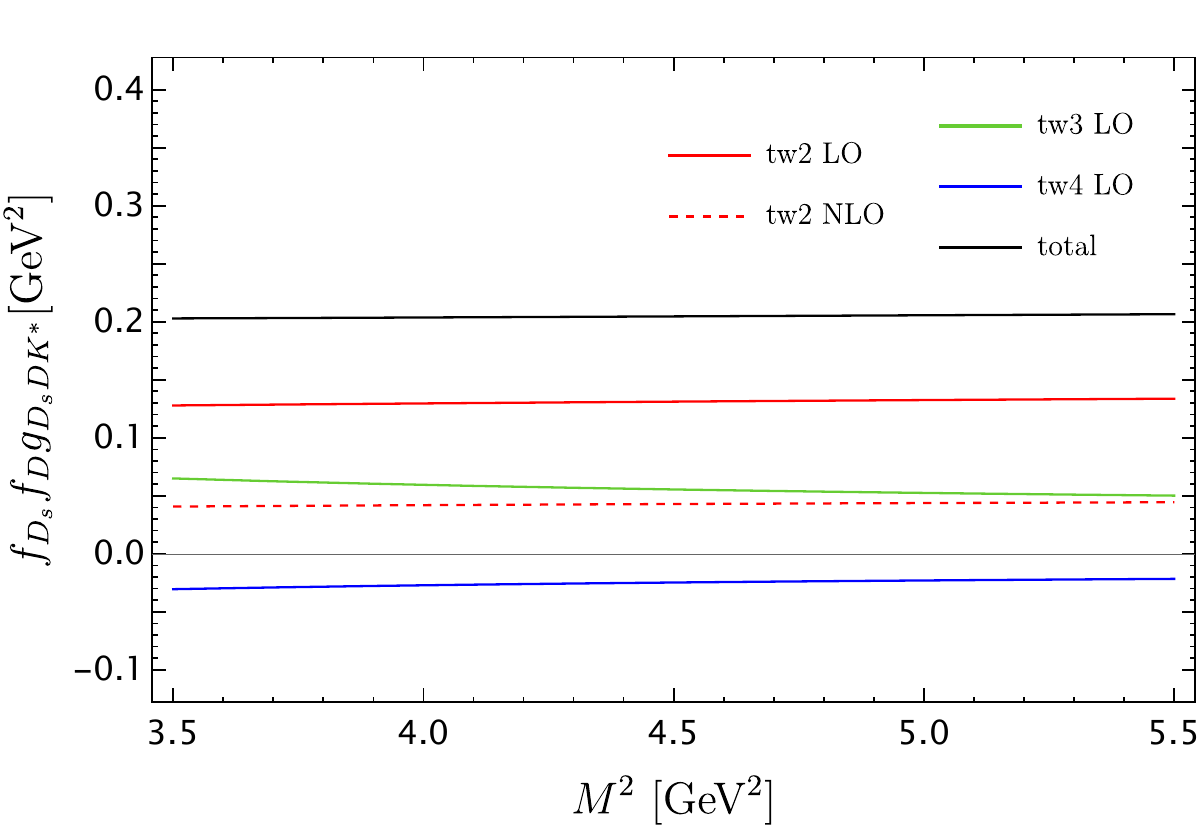} ~~
\includegraphics[width=0.45 \columnwidth]{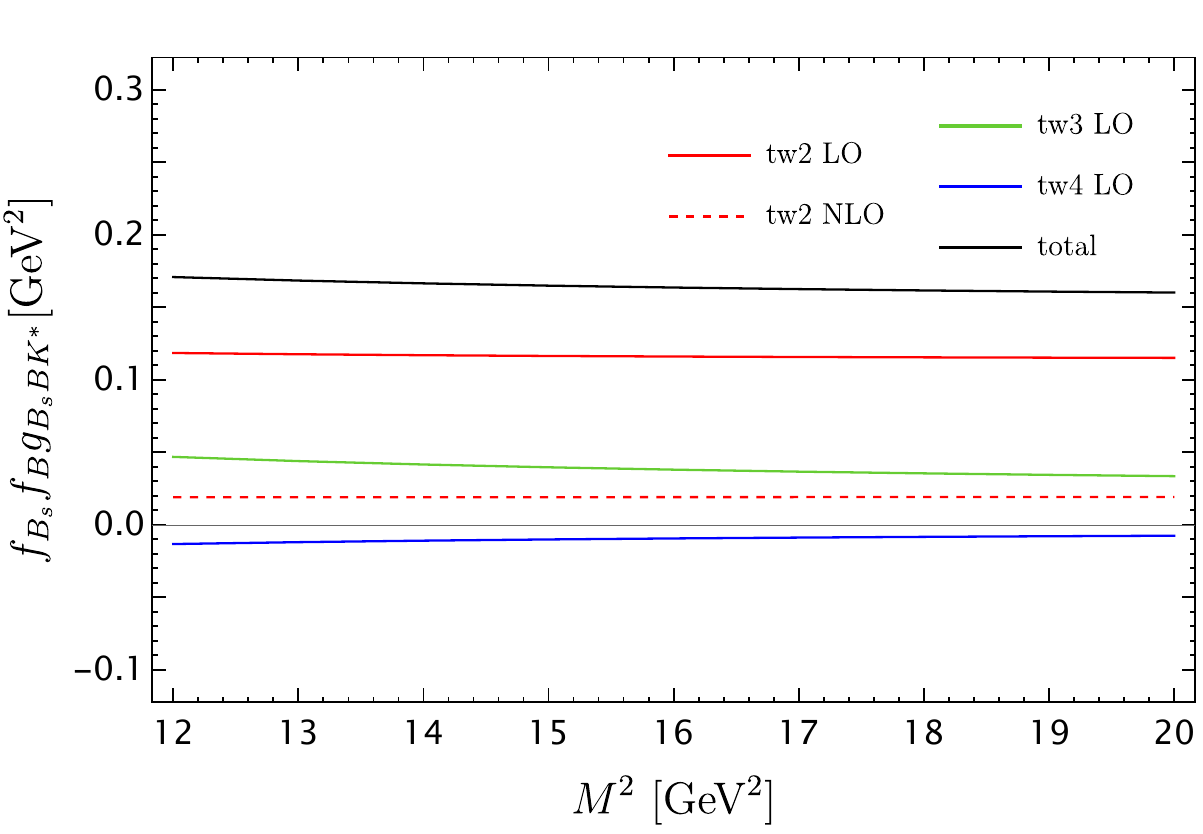} ~~
\includegraphics[width=0.45 \columnwidth]{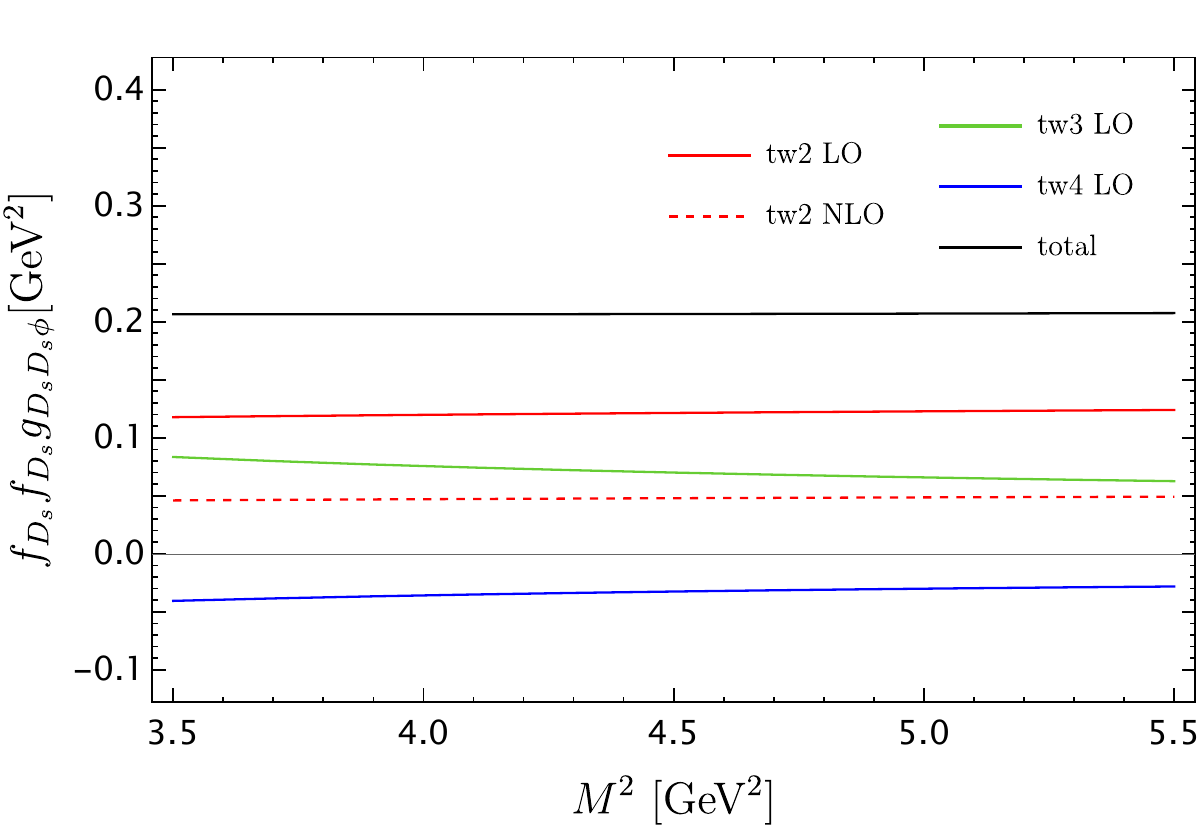} ~~
\includegraphics[width=0.45 \columnwidth]{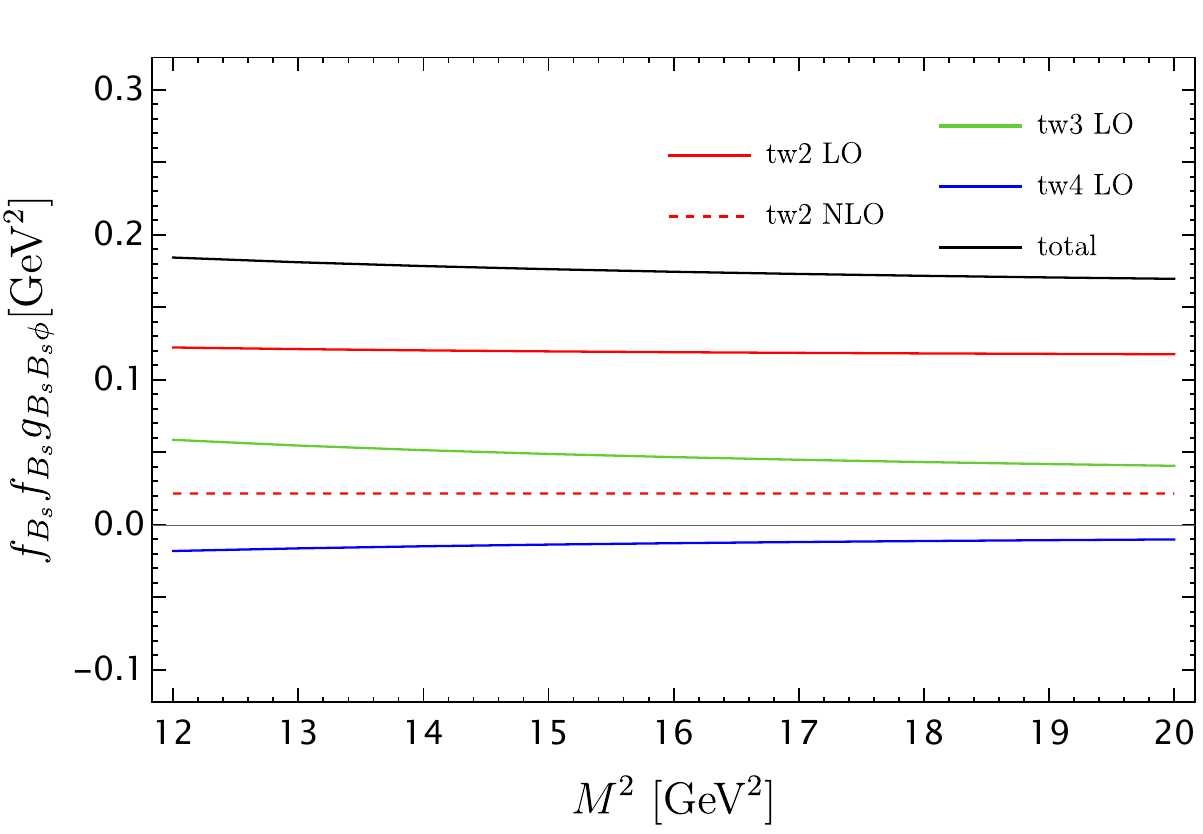}
\caption{Borel parameter dependence of the products
$ f_{D_{(s)}} f_{D_{(s)}} g_{D_{(s)}D_{(s)}V}$ and $ f_{B_{(s)}}f_{B_{(s)}} g_{B_{(s)}B_{(s)}V}$ calculated from LCSR. Displayed are the total values and separate  contributions from  twist-2 to twist-4
 of the vector meson ($\rho,K^*,\phi$) DAs, considering central values for all the other input parameters.}
\label{fig:M2dep}
\end{figure}

\begin{figure}[htb]
\centering
\includegraphics[width=0.45 \columnwidth]{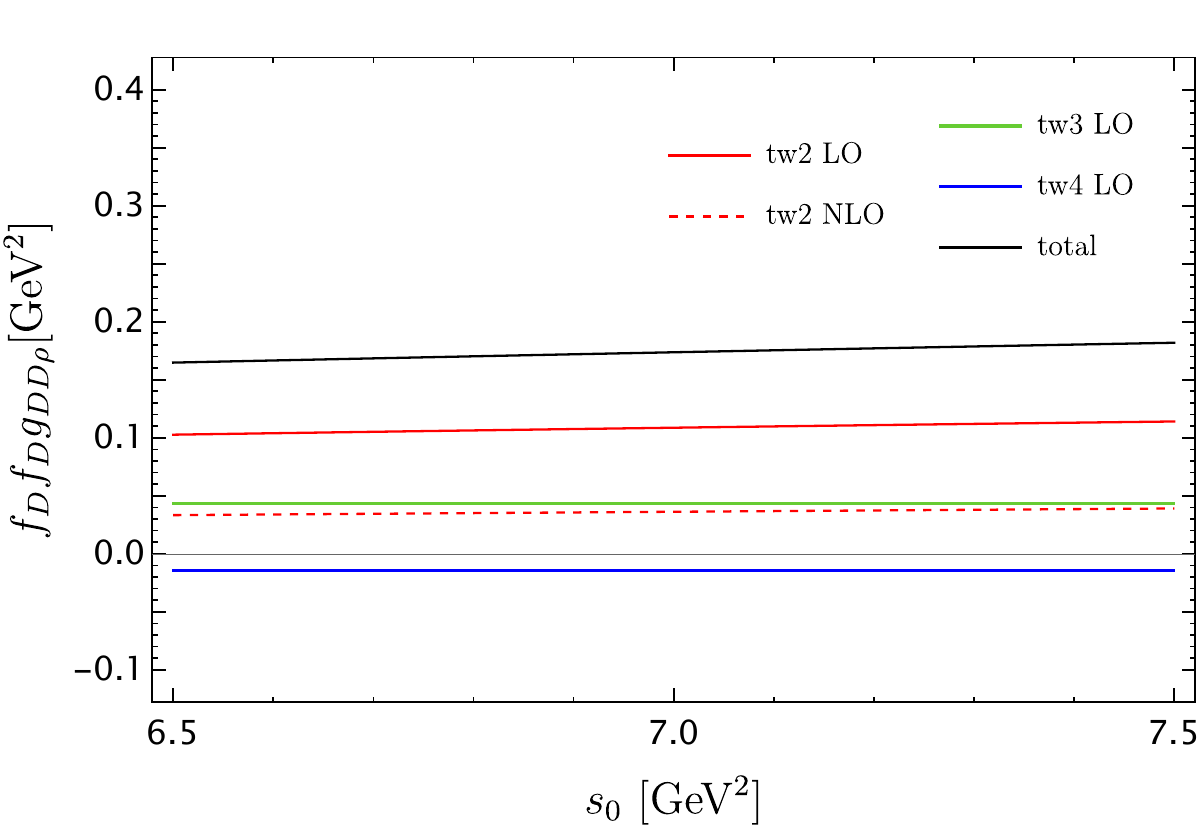} ~~
\includegraphics[width=0.45 \columnwidth]{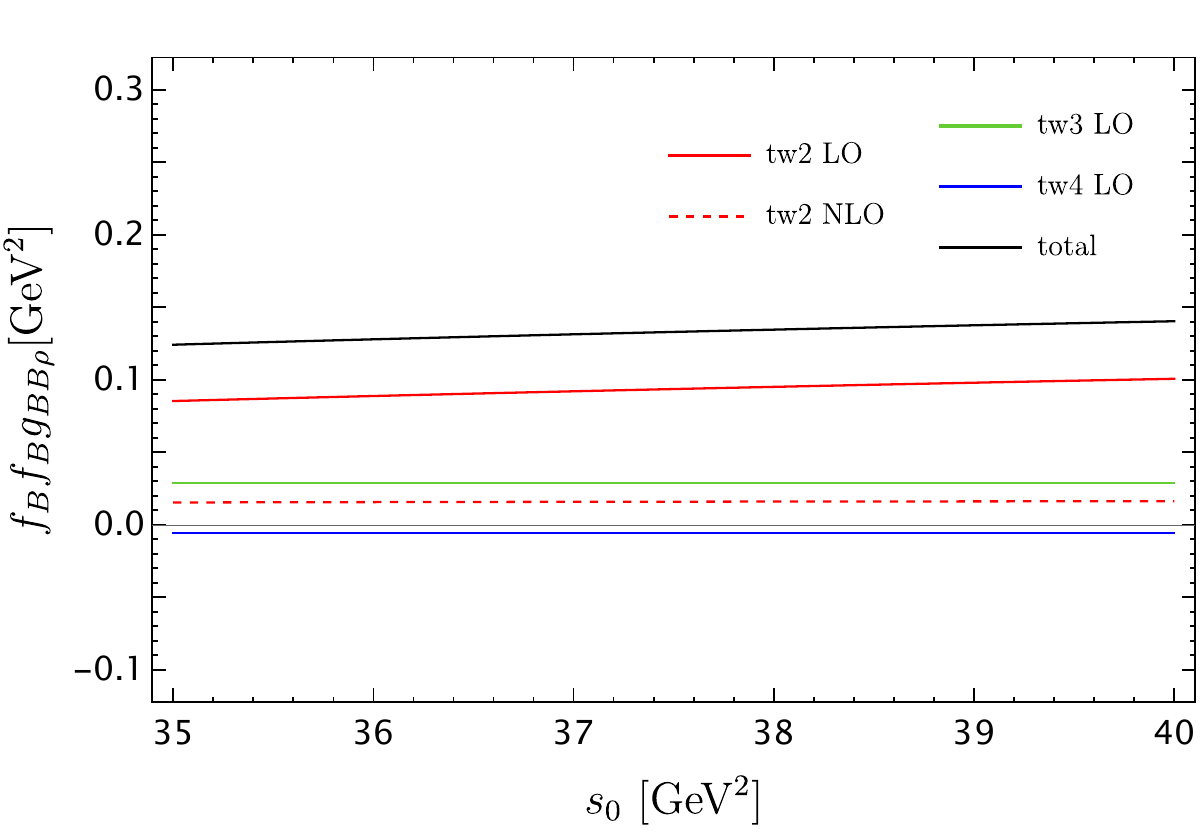} ~~
\includegraphics[width=0.45 \columnwidth]{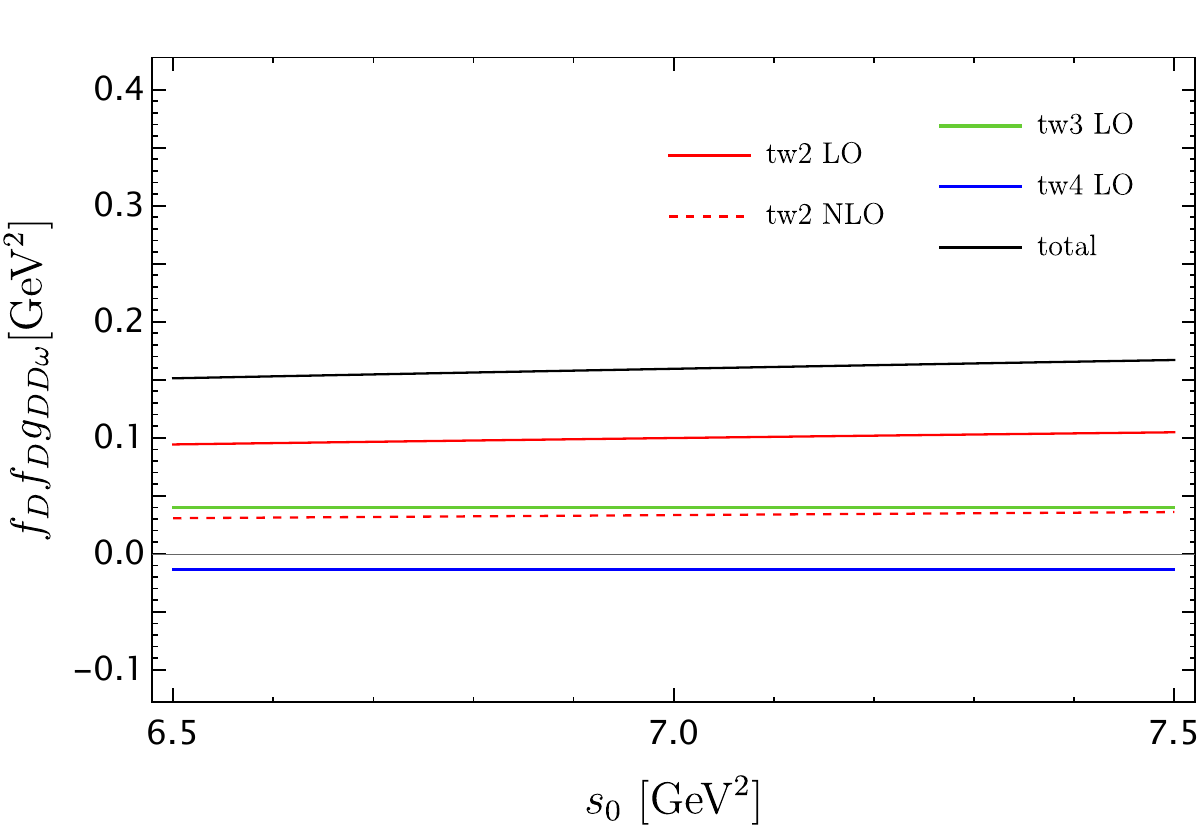} ~~
\includegraphics[width=0.45 \columnwidth]{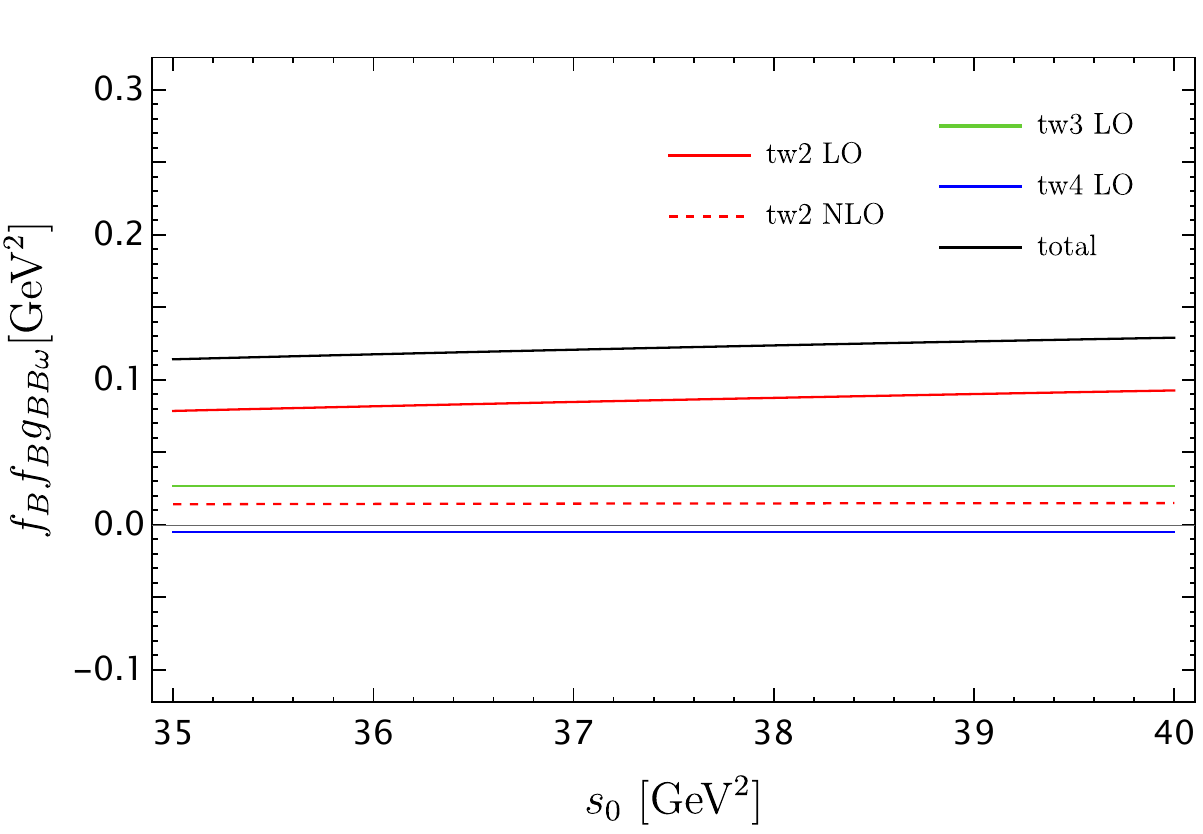} ~~
\includegraphics[width=0.45 \columnwidth]{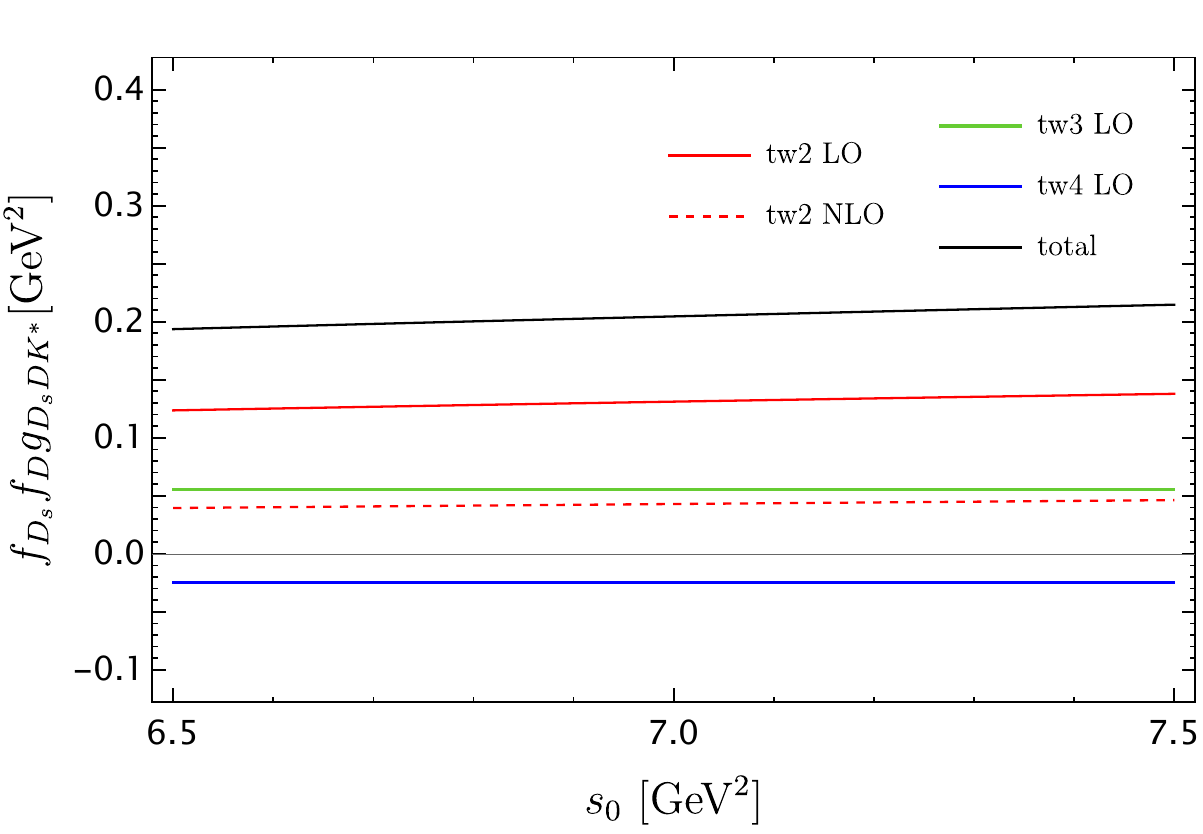} ~~
\includegraphics[width=0.45 \columnwidth]{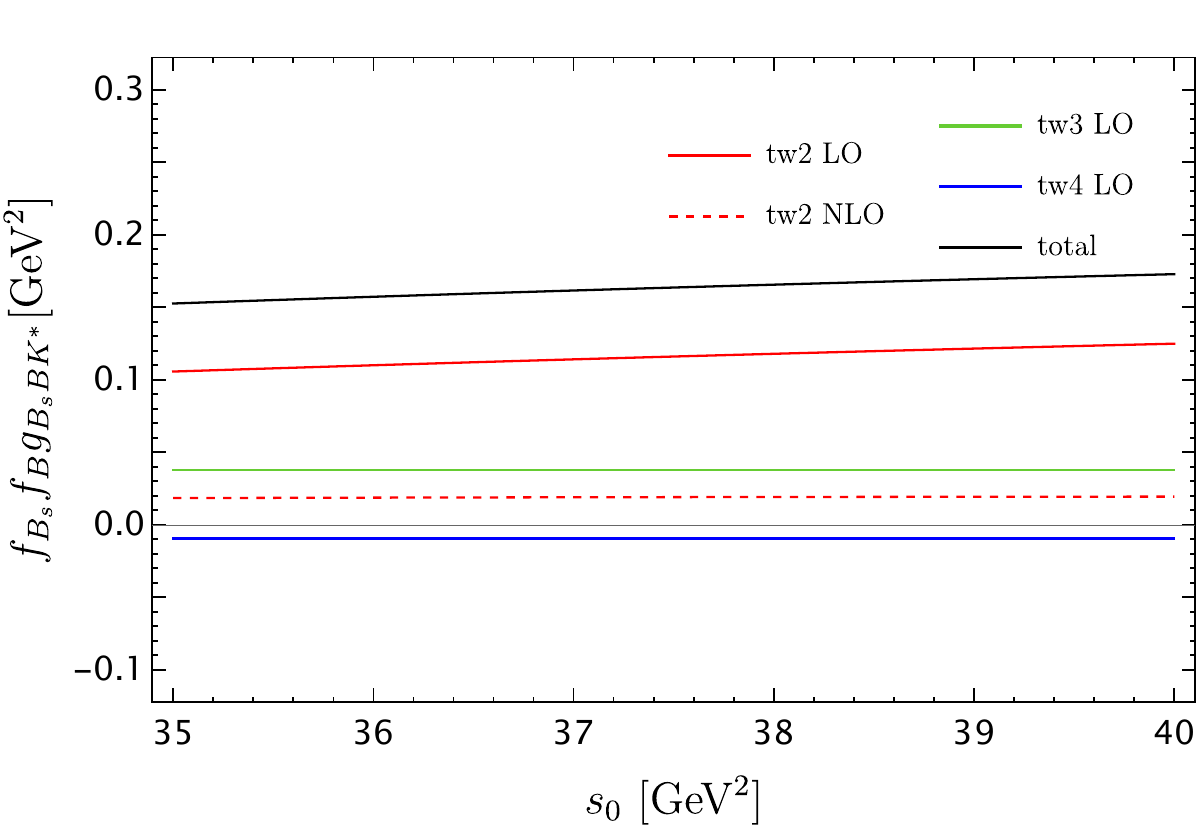} ~~
\includegraphics[width=0.45 \columnwidth]{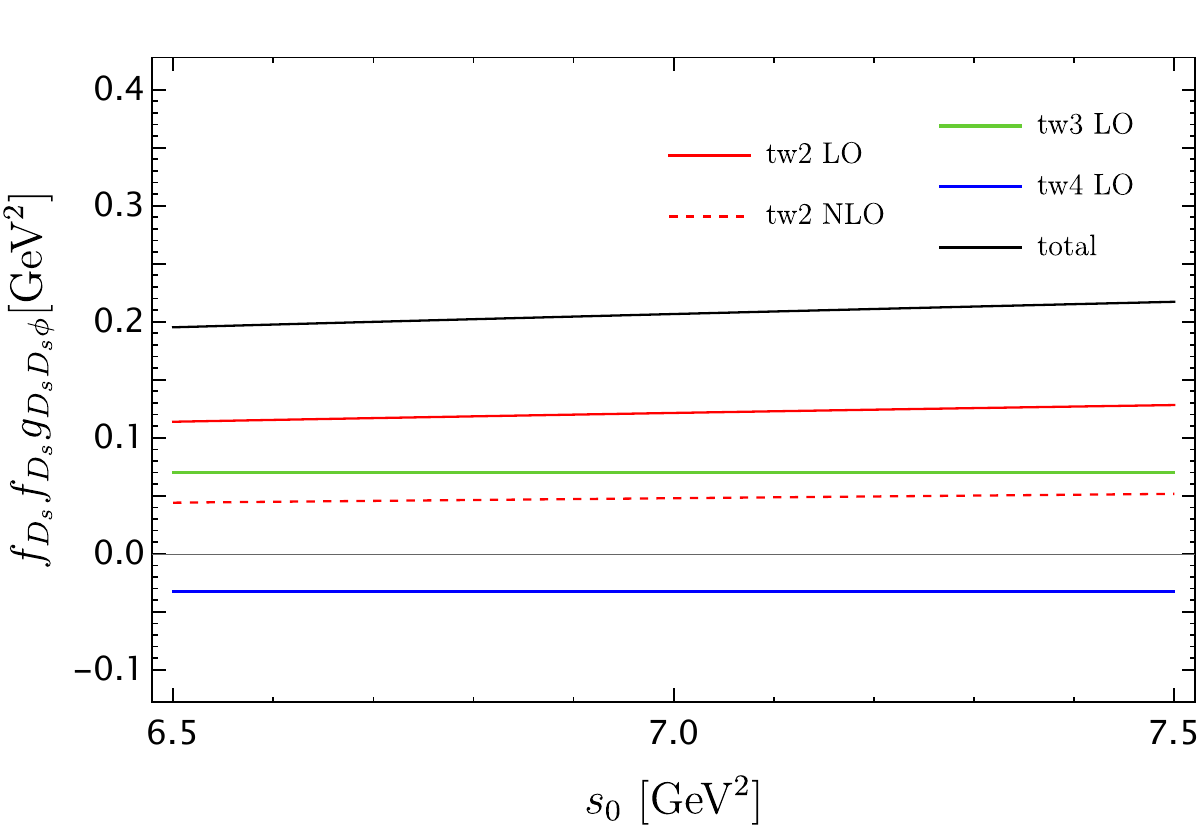} ~~
\includegraphics[width=0.45 \columnwidth]{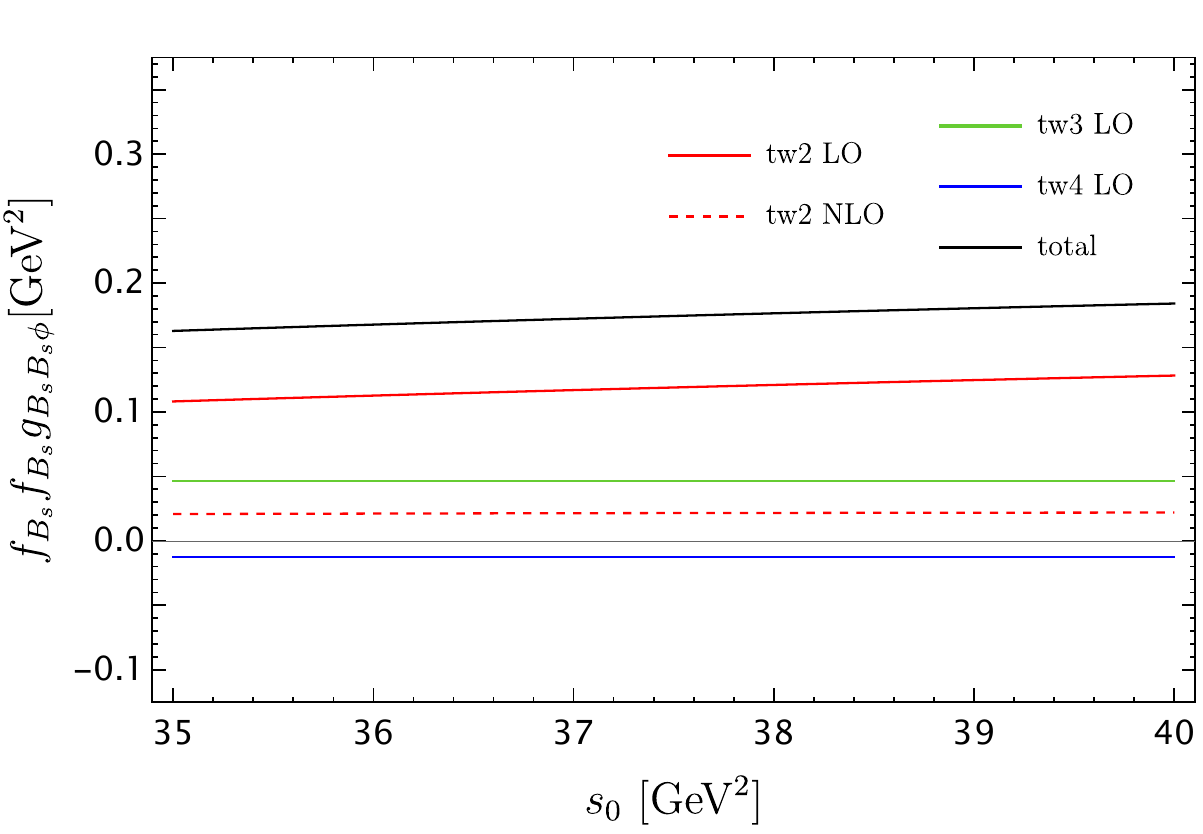}
\caption{Threshold dependence of the products
$ f_{D_{(s)}} f_{D_{(s)}} g_{D_{(s)}D_{(s)}V}$ and $ f_{B_{(s)}}f_{B_{(s)}} g_{B_{(s)}B_{(s)}V}$ calculated from LCSR. Displayed are the total values and separate  contributions from  twist-2 to twist-4
 of the vector meson ($\rho,K^*,\phi$) DAs, considering central values for all the other input parameters.}
\label{fig:s0dep}
\end{figure}
Moreover, it is informative to explore how LCSRs for $D_{(s)}D_{(s)}V$ and $B_{(s)}B_{(s)}V$ couplings, alongside decay constant products, respond to variations in  the scale $\mu$, the Borel parameter $M^2$ and the  effective threshold $s_0$. Firstly, Figure \ref{fig:mudep}  underscores the scale-dependence of individual contributions. Notably, however, we observe contrasting trends for the charm and bottom scenarios.  Specifically, the results on the left diminish as $\mu$ increases, while those on the right escalate with
$\mu$. This phenomenon can be attributed to Eq.\eqref{eq:LO-mu-dep}, where $\Hat{M}^2\equiv M^2/m^2_Q$. For charm, $\Hat{M}^2>1$, resulting in a negative slope for the LO curve; conversely, for bottom, $\Hat{M}^2<1$, leading to a positive slope for the LO curve. Additionally, we also observe a pronounced cancellation between the contributions of twist-2 NLO and LO. For contributions from other twists, particularly twist-3, there is indeed a noticeable dependence on $\mu$. However, since we have not computed the contributions from twist-3 NLO, this dependency cannot be offset. (We hope that our future work can address this shortfall.) Secondly, as illustrated in Figure \ref{fig:M2dep} and \ref{fig:s0dep}, the behavior concerning the Borel parameter $M^2$ and the effective threshold $s_0$ exhibits a nearly uniform pattern within the specified range, indicative of negligible sensitivity to variations in $M^2$ and $s_0$. This observation reinforces the resilience of the extracted quantities to alterations in the Borel parameter and effective threshold, thereby amplifying the trustworthiness of our analyses.

\begin{table}[h!]
\centering
\setlength\tabcolsep{5pt}
\def\arraystretch{1.4}
\begin{tabular}{|c||c|c|c|c|c|c||}
\hline
 Power  & $\delta^1_{V}$ & $\delta^1_{V}\alpha_s$ & $\delta^2_{V}$ & $\delta^3_{V}$ & $\delta^4_{V}$ &  total\\
\hline
$g_{D D\rho}$
& 2.76
& 0.80
& 0.65
& $-0.34$
& $-0.022$
& 3.86
\\
$g_{B B \rho}$
& 2.69
& 0.44
& 0.65
& $-0.86$
& $-0.006$
& 3.68
\\
\hline
$g_{D_s D_s\phi}$
& 2.52
& 0.77
& 0.65
& $-0.59$
& $-0.037$
& 3.31
\\
$g_{B_s B_s\phi} $
& 2.39
& 0.41
& 0.66
& $-0.16$
& $-0.011$
& 3.29
\\
\hline
\end{tabular}
\caption{Power corrections for the strong couplings $g_{HH\rho}$ and $g_{HH\phi}$ using the decay constats from LQCD. The expansion extends to the power of $\delta_{V}$ ranging from 1 to 4, where $\delta_{V}=\delta_{V}^{(Q)}=m_{V}/m_Q$, for the coupling with $D_s$ meson $\delta_{\rho}^{(c)}=0.64$ and $\delta_{\phi}^{(c)}=0.85$ and for the coupling with $B_s$ meson $\delta_{\rho}^{(b)}=0.17$ and $\delta_{\phi}^{(b)}=0.23$. }
\label{tab:power}
\end{table}

In order to assess the efficacy of power corrections, we present an examination of the individual contributions within the power correction series, employing the $\rho$ and
$\phi$ meson as representative cases. From the data provided in Table \ref{tab:power},  it becomes evident that for the  $D$ meson, despite the notable magnitude of $\delta_{V}$, the contributions generally exhibit a decreasing trend in power. Specifically, contributions at  $\delta^1_{V}$ and $\delta^2_{V}$, compared to $\delta^0_{V}$, diminished by an order of magnitude, although $\delta^1_{V}$ and $\delta^2_{V}$ demonstrate similar magnitudes. Similarly, $\delta^3_{V}$ and $\delta^4_{V}$, relative to $\delta^2_{V}$ and $\delta^3_{V}$, are also suppressed by an order of magnitude, with $\delta^2_{V}$ and $\delta^3_{V}$ displaying comparable magnitudes. This observed trend can be attributed to the fact that in Eq.\eqref{eq:2pLO}, higher-power contributions entail a decrease in the ratio $m^2_c/M^2$. In contrast to the scenario observed for the $D$ meson, while the magnitude of $\delta_{\phi}$ for the $B$ channel is considerably smaller, the ratio $m_b^2/M^2$ assumes a compensatory role. This results in a power expansion behavior similar to that of the $B_s B_s\phi$ coupling for the  $D_s D_s\phi$ counterpart.
The observation of a coincidental similarity in the $\delta^2_V$ term, exhibiting a value of 0.65, within both the $D D\rho$ and $D_s D_s\phi$ contexts invites further investigation. This coincidence arises from a delicate balance between the parameters characterizing the $\phi$ and $\rho$ mesons. While the parameters associated with the $\phi$ meson surpass those of the $\rho$ meson, the ratio necessitates consideration of the decay constants $f_{Ds}$ and $f_D$, wherein their proximity renders the ratio nearly equivalent.

\begin{table}[h!]
\centering
\setlength\tabcolsep{3.5pt}
\def\arraystretch{1.1}
\begin{tabular}{|c|c|c|c|c|c|c|}
\hline
LCSRs results  & tw~2 LO  & tw~2 NLO & tw~3 LO & tw~4 LO
 & total \\
\hline
$f_{D}\,f_{D}\,g_{D D\rho}$
& $0.108^{+0.017}_{-0.023}$ & $0.036^{+0.015}_{-0.011}$
& $0.043^{+0.023}_{-0.016}$ & $-0.018^{+0.007}_{-0.010}$
& $0.173^{+0.022}_{-0.022}$
\\
\hline
$f_{D}\,f_{D}\,g_{D D \omega}$
& $0.071^{+0.011}_{-0.015}$ & $0.023^{+0.010}_{-0.007}$
& $0.028^{+0.015}_{-0.011}$ & $-0.012^{+0.004}_{-0.006}$
& $0.113^{+0.014}_{-0.015}$
\\
\hline
$f_{D_s}\,f_{D}\,g_{D_s D K^{*}}$
& $0.131^{+0.024}_{-0.031}$ & $0.043^{+0.018}_{-0.013}$
&$0.059^{+0.030}_{-0.022}$& $-0.033^{+0.013}_{-0.019}$
& $0.206^{+0.028}_{-0.028}$
\\
\hline
$f_{D_s}\,f_{D_s}\,g_{D_s D_s\phi}$
& $0.121^{+0.021}_{-0.027}$  & $0.048^{+0.020}_{-0.014}$
& $0.070^{+0.034}_{-0.025}$ & $-0.041^{+0.015}_{-0.023}$
 & $0.207^{+0.024}_{-0.026}$
\\
\hline
\hline
$f_{B}\,f_{B}\,g_{B B \rho}$
& $0.094^{+0.017}_{-0.014}$ & $0.016^{+0.010}_{-0.010}$
& $0.029^{+0.006}_{-0.004}$ & $-0.009^{+0.002}_{-0.004}$
& $0.133^{+0.012}_{-0.012}$
\\
\hline
$f_{ B}\,f_{B}\,g_{B B \omega}$
& $0.061^{+0.011}_{-0.010}$ & $0.010^{+0.007}_{-0.007}$
& $0.019^{+0.004}_{-0.003}$ & $-0.006^{+0.001}_{-0.003}$
& $0.086^{+0.008}_{-0.009}$
\\
\hline
$f_{ B_s}\,f_{B}\,g_{ B_s B K^{*}}$
& $0.115^{+0.022}_{-0.019}$  & $0.019^{+0.012}_{-0.013}$
& $0.039^{+0.010}_{-0.006}$ & $-0.016^{+0.004}_{-0.008}$
& $0.163^{+0.016}_{-0.016}$
\\
\hline
$f_{ B_s}\,f_{ B_s}\,g_{ B_s B_s\phi}$
& $0.119^{+0.023}_{-0.020}$  & $0.022^{+0.014}_{-0.014}$
& $0.047^{+0.012}_{-0.007}$ & $-0.020^{+0.005}_{-0.010}$
& $0.174^{+0.019}_{-0.018}$
\\
\hline
\end{tabular}
\caption{Summary of the contribution from individual twist and NLO effects on the products of the strong couplings $g_{H_1H_2 V}$ and decay constants presented with both central values and associated data errors.}
\label{tab:products}
\end{table}

The numerical impacts of perturbative QCD and distinct higher-twist corrections on the products of strong couplings and decay constants, as computed in Eq.\eqref{eq:LCSR_final_expression}, are now investigated. The outcomes of individual twists and NLO effects are summarized in Table \ref{tab:products}, alongside pertinent data uncertainties. Notably, the NLO corrections of twist-2 magnitude demonstrate numerical smaller than the LO contributions of twist-3, constituting approximately $30\%$ of the leading twist for both charm and bottom scenarios.

\begin{table}[h!]
\centering
\setlength\tabcolsep{5pt}
\def\arraystretch{1.2}
\begin{tabular}{|c|c||c|c|c|c|c||}
\hline
\multirow{3}{*}{Charm}&decay constants & $g_{D D\rho}$ & $g_{D D\omega}$ & $g_{D_s D K^*}$ &   $g_{D_s D_s \phi}$\\
\hline
&  two-point QCDSRs
& $4.30^{+0.82}_{-0.72}$
& $2.80^{+0.54}_{-0.48}$
& $4.30^{+0.80}_{-0.67}$
& $3.65^{+0.93}_{-0.59}$
\\
& LQCD
& $3.86\pm 0.49$
& $2.52\pm 0.33$
& $3.88^{+0.52}_{-0.53}$
& $3.31^{+0.39}_{-0.41}$
\\
& Experiment
& $4.21\pm 0.58$
& $2.74\pm 0.39$
& $3.98\pm 0.55$
& $3.32^{+0.40}_{-0.42}$
\\
\hline
\hline
\multirow{3}{*}{bottom}&decay constants & $g_{B B\rho}$ &$g_{B B\omega}$ & $g_{B_s B K^*}$  &$g_{B_s B_s \phi}$\\
\hline
&  two-point QCDSRs
& $3.10^{+0.40}_{-0.54}$
& $2.02^{+0.27}_{-0.36}$
& $3.26^{+0.39}_{-0.46}$
& $2.98^{+0.45}_{-0.49}$
\\
& LQCD
& $3.68^{+0.34}_{-0.35}$
& $2.39^{+0.24}_{-0.24}$
& $3.73^{+0.37}_{-0.37}$
& $3.29^{+0.36}_{-0.34}$\\
& Experiment
& $3.27^{+0.87}_{-0.67}$
& $2.13^{+0.57}_{-0.44}$
& $--$
& $--$\\
\hline
\end{tabular}
\caption{LCSR  results for the strong couplings $g_{H_1H_2V}$ for the three methods of dividing out the decay constants.}
\label{tab:res}
\end{table}

Our comprehensive analysis involves normalizing couplings using heavy meson decay constants via three distinct methods: the two-point QCDSRs, LQCD and experiment results, as summarized in Table \ref{tab:decay-constant}. The primary findings are highlighted in Table \ref{tab:res} highlights the main findings, while additional details on parameters for each set of decay constants are presented in Tables \ref{tab:paraerr1} and \ref{tab:paraerr2}.
Within the  $D$ meson sector, predominant sources of uncertainty are attributed to the Gegenbauer moment
 $a^{\|}_2$ and  the effective threshold  $s_0$ from LQCD. Moreover, when employing two-point sum rules and experimental data for decay constant computations, an additional uncertainty emerges from the decay constant itself, which exhibits comparable magnitude. Conversely, in the $B$ meson sector, uncertainties pertaining to the Borel parameter display an increase relative to those observed in the $D$ meson sector.

\begin{table}[h!]
\centering
\setlength\tabcolsep{4pt}
\def\arraystretch{1.3}
\begin{tabular}{|c|c||c|c|c|c|c|c|c|c|c|c|}
\hline
 &{c.v.} & $\Delta f_{H_1}$ & $\Delta f_{H_2}$ & $\Delta \mu$ &  $\Delta m_Q$ &  $\Delta s_0$ & $\Delta M^2$ &  $\Delta f^{\perp}_V$ &
 $\Delta f^{\|}_V$ & $\Delta a^{\|}_2$    & $\Delta_{\rm tot}$\\
\hline
\multirow{3}{*}{$g_{D D\rho}$}
& $4.30$
&  \multicolumn{2}{|c|}{$^{+0.62}_{-0.47}$}
& $\pm 0.36$
& $^{+0.09}_{-0.10}$
& $^{+0.20}_{-0.22}$
&$^{+0.05}_{-0.04}$
& $\pm{0.04}$
& $\pm{0.05}$
& $\pm 0.32$
&$^{+0.82}_{-0.72}$
\\
\cline{2-12}
&$3.86$
&  \multicolumn{2}{|c|}{$\pm 0.03$}
&$\pm 0.33$
& $\pm 0.08$
& $^{+0.18}_{-0.20}$
& $\pm 0.04$
& $\pm 0.03$
& $\pm 0.04$
& $\pm 0.29$
&$\pm 0.49$
\\
\cline{2-12}
&$4.21$
&  \multicolumn{2}{|c|}{$^{+0.24}_{-0.22}$}
&$\pm 0.35$
& $\pm 0.09$
& $^{+0.20}_{-0.22}$
& $^{+0.05}_{-0.04}$
& $\pm 0.04$
& $\pm 0.05$
& $\pm 0.32$
&$\pm 0.58$
\\
\hline
\multirow{3}{*}{$g_{D D\omega}$}
& $2.80$
&  \multicolumn{2}{|c|}{$^{+0.40}_{-0.31}$}
& $\pm 0.23$
& $\pm 0.06$
& $^{+0.13}_{-0.14}$
&$\pm 0.03$
& $\pm{0.04}$
& $^{+0.09}_{-0.11}$
& $\pm 0.21$
&$^{+0.54}_{-0.48}$
\\
\cline{2-12}
&$2.52$
&  \multicolumn{2}{|c|}{$^{+0.16}_{-0.14}$}
& $\pm 0.23$
& $\pm 0.06$
& $^{+0.12}_{-0.13}$
& $\pm 0.03$
& $\pm 0.04$
& $^{+0.09}_{-0.10}$
& $\pm 0.19$
&$\pm 0.33$
\\
\cline{2-12}
&$2.74$
&  \multicolumn{2}{|c|}{$\pm 0.02$}
& $\pm 0.21$
& $\pm 0.06$
& $^{+0.13}_{-0.14}$
& $\pm 0.03$
& $\pm 0.04$
& $^{+0.09}_{-0.10}$
& $\pm 0.21$
&$\pm 0.39$
\\
\hline
\multirow{3}{*}{$g_{D_s D K^*}$}
&$4.30$
&$^{+0.30}_{-0.24}$
&$^{+0.46}_{-0.22}$
& $\pm 0.38$
& $\pm 0.09$
& $^{+0.21}_{-0.23}$
&$\pm 0.03$
& $\pm 0.04$
& $\pm{0.08}$
& $\pm 0.35$
&$^{+0.80}_{-0.67}$
\\
\cline{2-12}
&$3.88$
&$\pm 0.01$
&$\pm 0.01$
&$^{+0.35}_{-0.34}$
& $\pm 0.08$
& $^{+0.19}_{-0.21}$
& $\pm 0.03$
& $\pm 0.04$
& $\pm 0.07$
& $\pm 0.31$
&$^{+0.52}_{-0.53}$
\\
\cline{2-12}
&$3.98$
&$\pm 0.12$
&$\pm 0.05$
&$\pm 0.35$
& $\pm 0.08$
& $^{+0.19}_{-0.21}$
& $^{+0.032}_{-0.027}$
& $\pm 0.04$
& $\pm 0.07$
& $\pm 0.32$
&$\pm 0.55$
\\
\hline
\multirow{3}{*}{$g_{D_s D_s \phi}$}
& $3.65$
&  \multicolumn{2}{|c|}{$^{+0.82}_{-0.37}$}
& $^{+0.16}_{-0.20}$
& $\pm 0.06$
& $^{+0.19}_{-0.20}$
& $\pm 0.01$
& $\pm 0.04$
& $\pm 0.06$
& $\pm 0.34$
& $^{+0.93}_{-0.59}$
\\
\cline{2-12}
&$3.31$
&  \multicolumn{2}{|c|}{$\pm 0.01$}
&$^{+0.14}_{-0.18}$
& $\pm 0.06$
& $^{+0.17}_{-0.18}$
& $\pm 0.01$
& $\pm 0.04$
& $\pm 0.06$
& $\pm 0.31$
&$^{+0.39}_{-0.41}$
\\
\cline{2-12}
&$3.32$
&  \multicolumn{2}{|c|}{$\pm 0.09$}
&$^{+0.14}_{-0.18}$
& $\pm 0.06$
& $^{+0.17}_{-0.18}$
& $\pm 0.01$
& $\pm 0.04$
& $\pm 0.06$
& $\pm 0.31$
&$^{+0.40}_{-0.42}$
\\
\hline
\end{tabular}
\caption{Summary of the individual uncertainties for the strong couplings $g_{H_1H_2 V}$
predicted from the LCSR (\ref{eq:LCSR_final_expression}) .
The negligible uncertainties arising from variations in the remaining input parameters have been considered but are not explicitly enumerated here, but are already taken into account in the determinations
of the total errors. The c.v. (central value) represents the best estimate derived from the LCSR.}
\label{tab:paraerr1}
\end{table}

\begin{table}[h!]
\centering
\setlength\tabcolsep{4.5pt}
\def\arraystretch{1.3}
\begin{tabular}{|c|c||c|c|c|c|c|c|c|c|c|c|}
\hline
 &{ c.v.} & $\Delta f_{H_1}$ & $\Delta f_{H_2}$ & $\Delta \mu$ &  $\Delta m_Q$ &  $\Delta s_0$ & $\Delta M^2$ &  $\Delta f^{\perp}_V$ &
 $\Delta f^{\|}_V$ &$\Delta a^{\|}_{2}$    & $\Delta_{\rm tot}$\\
\hline
\multirow{3}{*}{$g_{B B \rho}$}
&$3.10$
&  \multicolumn{2}{|c|}{$^{+0.29}_{-0.45}$}
&$^{+0.01}_{-0.04}$
& $\pm 0.03$
&$^{+0.17}_{-0.21}$
& $^{+0.12}_{-0.06}$
& $\pm 0.03$
&$\pm 0.04$
& $\pm 0.18$
&$^{+0.40}_{-0.54}$
\\
\cline{2-12}
&$3.68$
&  \multicolumn{2}{|c|}{$\pm 0.05$}
& $^{+0.01}_{-0.05}$
& $\pm 0.03$
& $^{+0.20}_{-0.24}$
& $^{+0.14}_{-0.06}$
&$\pm 0.04$
& $\pm 0.04$
& $\pm 0.22$
&$^{+0.34}_{-0.35}$
\\
\cline{2-12}
&$3.27$
&  \multicolumn{2}{|c|}{$^{+0.81}_{-0.59}$}
& $^{+0.01}_{-0.05}$
& $\pm 0.03$
& $^{+0.18}_{-0.21}$
& $^{+0.12}_{-0.06}$
&$\pm 0.03$
& $\pm 0.04$
& $\pm 0.19$
&$^{+0.87}_{-0.67}$
\\
\hline
\multirow{2}{*}{$g_{B B\omega}$}
&$2.02$
&  \multicolumn{2}{|c|}{$^{+0.19}_{-0.29}$}
&$^{+0.01}_{-0.03}$
& $\pm 0.02$
& $^{+0.11}_{-0.13}$
&$^{+0.08}_{-0.04}$
& $\pm 0.03$
& $\pm 0.07$
&$\pm0.12$
&$^{+0.27}_{-0.36}$
\\
\cline{2-12}
&$2.39$
&  \multicolumn{2}{|c|}{$\pm 0.03$}
& $^{+0.01}_{-0.03}$
& $\pm 0.02$
& $^{+0.13}_{-0.16}$
& $^{+0.09}_{-0.04}$
&$\pm 0.04$
& $^{+0.08}_{-0.09}$
& $\pm 0.14$
&$\pm 0.24$
\\
\cline{2-12}
&$2.13$
&  \multicolumn{2}{|c|}{$^{+0.53}_{-0.39}$}
& $^{+0.01}_{-0.03}$
& $\pm 0.02$
& $^{+0.12}_{-0.14}$
& $^{+0.08}_{-0.04}$
&$\pm 0.03$
& $^{+0.07}_{-0.08}$
& $\pm 0.13$
&$^{+0.57}_{-0.44}$
\\
\hline
\multirow{2}{*}{$g_{B_s B K^*}$}
& $3.26$
& $^{+0.15}_{-0.25}$
& $^{+0.17}_{-0.21}$
&$^{+0.01}_{-0.04}$
& $\pm 0.03$
&$^{+0.18}_{-0.22}$
& $^{+0.15}_{-0.07}$
& $\pm 0.03$
& $\pm 0.06$
& $\pm 0.21$
& $^{+0.39}_{-0.46}$
\\\cline{2-12}
&$3.73$
&$\pm 0.03$
& $\pm 0.02$
& $^{+0.01}_{-0.05}$
& $\pm 0.03$
& $^{+0.21}_{-0.25}$
& $^{+0.17}_{-0.08}$
& $\pm 0.04$
& $\pm 0.07$
& $\pm 0.24$
&$\pm 0.37$
\\
\hline
\multirow{2}{*}{$g_{B_s B_s \phi}$}
& $2.98$
&  \multicolumn{2}{|c|}{$^{+0.32}_{-0.38}$}
& $^{+0.01}_{-0.04}$
& $\pm 0.03$
& $^{+0.17}_{-0.20}$
& $^{+0.17}_{-0.08}$
& $\pm 0.04$
& $\pm 0.05$
& $\pm 0.21$
& $^{+0.45}_{-0.49}$
\\\cline{2-12}
& $3.29$
&  \multicolumn{2}{|c|}{$\pm 0.04$}
& $^{+0.01}_{-0.05}$
& $\pm 0.03$
& $^{+0.18}_{-0.22}$
& $^{+0.18}_{-0.09}$
&$\pm 0.04$
& $\pm 0.06$
& $\pm 0.23$
& $^{+0.36}_{-0.34}$
\\\hline
\end{tabular}
\caption{Summary of the individual theory uncertainties for the strong couplings $g_{H_1H_2 V}$
predicted from the LCSR (\ref{eq:LCSR_final_expression}).
The negligible uncertainties arising from variations in the remaining input parameters have been considered but are not explicitly enumerated here, but are already taken into account in the determinations of the total errors. The c.v. (central value) represents the best estimate derived from the LCSR.}
\label{tab:paraerr2}
\end{table}

The evaluation of strong couplings for  $D_{(s)}D_{(s)}V$ and $B_{(s)}B_{(s)}V$ through LCSRs reveals uncertainty ranges of ($16\%\sim 25\%$, $12\%\sim 13\%$, $12\%\sim 14\%$) and ($13\%\sim 14\%$, $10\%\sim 11\%$, $20\%\sim 27\%$), respectively, when utilizing heavy-meson decay constants obtained from two-point QCDSRs, LQCD and experimental data. These intervals of uncertainty are determined based on variations in input parameters and methodologies employed in the LCSR analysis. Despite inconsistencies in the uncertainties for $g_{D_{(s)}D_{(s)}V}$ and $g_{B_{(s)}B_{(s)}V}$ due to the choice of decay constants, the discrepancy in central values between the $D_{(s)}D_{(s)}V$ and $B_{(s)}B_{(s)}V$ couplings primarily arises from a notable deviation among the lattice-QCD values of $f_{D_{(s)}}$ and $f_{B_{(s)}}$ and the value predicted by the two-point sum rule and experimental data.

\begin{table}[h!]
\centering
\setlength\tabcolsep{5pt}
\def\arraystretch{1.3}
\begin{tabular}{|c|c||c|c|c|c||}
\hline
$\Delta r_{D_{(s)}D_{(s)}V}$ & decay constants & $\Delta r_{D D \omega}$& $\Delta r_{D_s D K^*}$  &  $\Delta r_{D_s D_s \phi}$\\
\hline
charm&  two-point QCDSRs
& $-8\%$
& $1\%$
& $-15\%$
\\
& LQCD
& $-8\%$
& $1\%$
& $-14\%$
\\
& Experiment
& $-8\%$
& $2\%$
& $-19\%$
\\
\hline
\hline
$\Delta r_{B_{(s)}B_{(s)}V}$ & decay constants &  $\Delta r_{B B \omega}$&  $\Delta r_{B_s B K^*}$  &$\Delta r_{B_s B_s \phi}$\\
\hline
bottom&  two-point QCDSRs
& $-8\%$
& $5\%$
& $-4\%$
\\
& LQCD
& $-8\%$
& $1\%$
& $-11\%$
\\
\hline
\end{tabular}
\caption{Summary of the extent of $SU(3)$ breaking effects as determined by LCSR, with $\Delta r_{HHV}$ serving as a measurement parameter.}
\label{tab:SU3}
\end{table}

To analyze the $SU(3)_F$ symmetry breaking effects, we introduce the parameter  $\Delta r_{HHV}$, which quantifies the extent of these breaking effects. The definitions are given by the following expressions:
\begin{eqnarray}
\Delta r_{D_{(s)}D_{(s)}V}= \frac{C_V\cdot  g_{D_{(s)}D_{(s)}V}-g_{D^*D\rho}}{g_{DD\rho}},\quad \Delta r_{B_{(s)}B_{(s)}V}=\frac{C_V\cdot   g_{B_{(s)}B_{(s)}V}-g_{BB\rho}}{g_{BB\rho}},
\end{eqnarray}
where $g_{DD\rho}$ and $g_{BB\rho}$ are defined in Eq.\eqref{eq:coupling-def1} and \eqref{eq:coupling-def2}, respectively. The factor $C_V$ equals $\sqrt{2}$ for the $\omega$ meson, and 1 otherwise.


Table \ref{tab:SU3} provides a detailed account of the relative magnitudes of $SU(3)_F$ symmetry breaking effects on various strong couplings. Both $\Delta r_{D D \omega }$ and $\Delta r_{B B \omega }$ are observed to be consistently $-8\%$. This stability arises from specific reductions in $f^{\parallel}_{\omega}$,$f^{\perp}_{\omega}$, and $m_{\omega}$, which decrease at rates of $7\%$, $1.8\%$, and $0.2\%$, respectively. The remaining parameters are consistent with those of the $\rho$ meson,leading to a cancellation of terms in both the numerator and denominator. It is important to emphasize that the breaking effects observed in $\phi$, as determined  from decay constants derived from experimental data, tends to exceed that calculated via LQCD and two-point QCDSRs when considering their absolute magnitudes. This discrepancy arises from the substantial difference in the decay constant $f_{D}$  and $f_{D_s}$ obtained from the three methods, which causes the negative breaking effects up to $21\%$ from experimental data, larger than $18.4\%$ from the two-point QCDSRs and $17.9\%$ from the LQCD. Consequently, the discrepancy associated with LQCD-calculated decay constants reaches $-19\%$, in contrast to $-15\%$ from  two-point QCDSRs and $-14\%$ from the LQCD. By accounting for additional factors influencing the observed symmetry breaking effects, we present a comprehensive analysis of  $SU(3)$ flavour symmetry deviations. The pattern of $SU(3)$ symmetry breaking in the $B$ sector  is similar to that in the $D$ sector, though with a milder reduction.
\begin{table}[h!]
\centering
\begin{tabular}{|c||c|c|c|c|c|}
\hline
Method  &$g_{D D\rho}$  & $g_{D D\omega}$ & $g_{D_s D K^*}$ &  $g_{D_s D_s \phi}$ &  $g_{B B\rho}$\\
\hline
LCSR\cite{Wang:2007mc}  &$2.62\pm 0.58$ &-- &$3.22\pm 0.64$ &$2.9\pm 0.68$ &\\
\hline
LCSR\cite{Li:2007dv}& $4.61$ & & --&--&$5.34\pm1.15$\\
\hline
QCDSR\cite{Holanda:2006xq}&$--$&$2.9$ & --&--&\\
\hline
QCDSR\cite{Janbazi:2017mpb}&$--$&-- & $3.26\pm0.43$&--&\\
\hline
QCDSR\cite{Khosravi:2013ad}&$--$&-- & --&$4.00\pm 1.09$&\\
\hline
LQCD\cite{Can:2012tx} & $4.84\pm0.34$ &--  & $--$&-- &\\[1mm]
\hline
\hline
\multirow{2}{*}{LCSR (this work)}
& $4.30^{+0.82}_{-0.72}$
& $2.80^{+0.54}_{-0.48}$
& $4.30^{+0.80}_{-0.67}$
& $3.65^{+0.93}_{-0.59}$
& $3.10^{+0.40}_{-0.54}$\\[1mm]
\cline{2-6}
& $3.86\pm 0.49$
& $2.52\pm 0.33$
& $3.88^{+0.52}_{-0.53}$
& $3.31^{+0.39}_{-0.41}$
&$3.68^{+0.34}_{-0.35}$\\[1mm]
\cline{2-6}
& $4.21\pm 0.58$
& $2.74\pm 0.39$
& $3.98\pm 0.55$
& $3.32^{+0.40}_{-0.42}$
&$3.27^{+0.87}_{-0.67}$\\[1mm]
\hline
\end{tabular}
\caption{Numerical values of coupling  $g_{D_{(s)} D_{(s)}V}$ from several methods.}
\label{tab:compar1}
\end{table}

When comparing our numerical outcomes with those of Ref.\cite{Wang:2007mc}, depicted in Table \ref{tab:compar1}, we observed a reduction of the reported values $2.62\pm 0.58$, $3.22\pm 0.64$ and $2.9\pm 0.68$ for $g_{DD\rho}$, $g_{D_s DK^*}$ and $g_{D_s D_s \phi}$, respectively.
The primary reason for this discrepancy is the variation in the decay constants. The references they cited experimental works \cite{CLEO:2007fgd}, providing values of $f_D=0.223\pm 0.017$ and set $f_{D_s}/f_D\simeq 1.1$. Since the coupling constants are computed by dividing by the product of these decay constants, the differences between the two sets of results are considerable.  Comparing our results with those in Ref.\cite{Li:2007dv}, we observe an opposite effect, with an enhancement in the reported value of $5.34 \pm 1.15$, also due to a different selection of the decay constant, specifically $f_B = 140$ MeV. It is important to note that the LCSR studies presented in \cite{Wang:2007mc,Li:2007dv} were restricted to leading-order calculations. The significant enhancement effects attributable to twist-2 NLO calculations were not incorporated. In this work, we extend the analysis by incorporating NLO and subleading power corrections, specifically considering $\delta^3_V$ and $\delta^4_V$, along with higher twist contributions, encompassing twist-3 to twist-4, into Eq.\eqref{eq:LCSR_final_expression}. For the additional results presented in  Table \ref{tab:compar1}, it is apparent that there is a satisfactory concordance with our findings within the error margins. Particularly, the value of $4.84\pm 0.34$ derived from lattice QCD \cite{Can:2012tx}  is  consistent with our result of $4.30^{+0.82}_{-0.72}$ obtained through QCDSR decay constants.

\begin{table}[h!]
\centering
\begin{tabular}{|c|c||c|c|c|}
\hline
Coupling &decay constants & This work& $A_0(q^2)$ \cite{Bharucha:2015bzk} & $A_0(q^2)$ \cite{Gao:2019lta} \\
\hline
\multirow{3}{*}{$g_{B B\rho}$}
& two-point QCDSRs
&  $3.10^{+0.40}_{-0.54}$
& $6.11^{+1.17}_{-1.23}$
& $6.15^{+2.47}_{-2.05}$
\\[1mm]
\cline{2-5}
& LQCD
& $3.68^{+0.34}_{-0.35}$
& $6.65^{+1.25}_{-1.25}$
& $6.70^{+2.79}_{-2.18}$
\\[1mm]
\cline{2-5}
& Experiment
& $3.27^{+0.87}_{-0.67}$
& $6.27^{+1.39}_{-1.32}$
& $6.32^{+2.73}_{-2.14}$
\\
\hline
\multirow{3}{*}{$g_{B B\omega}$}
& two-point QCDSRs
&$2.02^{+0.27}_{-0.36}$
& $4.24^{+1.04}_{-1.07}$
& $4.74^{+2.01}_{-1.53}$
\\[1mm]
\cline{2-5}
& LQCD
&$2.39^{+0.24}_{-0.24}$
& $4.62^{+1.12}_{-1.12}$
& $5.17^{+2.18}_{-1.63}$
\\[1mm]
\cline{2-5}
& Experiment
&$2.13^{+0.57}_{-0.44}$
& $4.36^{+1.17}_{-1.13}$
& $4.87^{+2.13}_{-1.60}$
\\
\hline
\multirow{2}{*}{$g_{B_s B K^*}$}
& two-point QCDSRs
& $3.26^{+0.39}_{-0.46}$
& $6.30^{+1.47}_{-1.49}$
& $7.46^{+3.10}_{-2.22}$
\\[1mm]
\cline{2-5}
& LQCD
& $3.73^{+0.37}_{-0.37}$
& $6.62^{+1.50}_{-1.50}$
& $7.84^{+3.23}_{-2.27}$
\\
\hline
\multirow{2}{*}{$g_{B_s B_s \phi}$}
& two-point QCDSRs
& $2.98^{+0.45}_{-0.49}$
& $7.13^{+1.43}_{-1.46}$
& $--$
\\[1mm]
\cline{2-5}
& LQCD
& $3.29^{+0.36}_{-0.34}$
& $7.49^{+1.45}_{-1.45}$
& $--$
\\
\hline
\end{tabular}
\caption{Coupling constant $g_{B_{(s)}B_{(s)}V}$ extracted from the residue of the transition form factors $A_0$ in the framework of LCSR fit, compared with our obtained values.}
\label{tab:FFcompare}
\end{table}

In addition, we  introduce an alternative technique  to determine the $B_{(s)} B_{(s)} V$ coupling similar to that described in Ref.\cite{Jin:2024zyy}. This method, as extensively described in \cite{Khodjamirian:2020mlb}, relies on the hadronic dispersion relation for the $B\to V$ axial vector form factor $A_0$ as a foundational element. Through the utilization of Eq.\eqref{eq:HHVlimit}, we obtain various numerical values for the coupling $g_{B_{(s)}B_{(s)}V}$ using the $B\to V$ form factor $A_0$ from two distinct LCSRs \cite{Bharucha:2015bzk,Gao:2019lta}, with a summary of these findings provided in Table \ref{tab:FFcompare}. our investigation uncovers a discrepancy between our central estimate and the anticipated outcomes derived from the transition form factors $A_0$. A thorough analysis of the underlying reasons can be found in our previous publication \cite{Jin:2024zyy}. When extending Eq.\eqref{eq:HHVlimit} to include transition form factor $A_0$ from $D$ to $V$, we can derive the coupling for $DDV$ as well. Utilizing the $D$ to $\rho$ form factor detailed in Ref.\cite{Melikhov:2000yu}, our computations yielded $g_{DD\rho}$ couplings of $7.52$ employing heavy meson decay constants from lattice QCD, $7.93$ utilizing decay constants from two-point QCD sum rules, and $7.69$ employing experimental decay constants. These values notably surpass the corresponding results obtained from LCSRs: $4.30^{+0.82}_{-0.72}$, $3.86\pm 0.49$, and $4.21\pm 0.58$, respectively, reflecting a trend akin to $B\to\rho$ transitions.

The investigation of the limit of infinitely heavy-quark mass using the LCSR (\ref{eq:LCSR_final_expression}) for strong coupling yields valuable insights. . This sum rule, based on the correlation function at finite $m_Q$, accurately reflects the leading-power behavior as $m_Q\to \infty$  and facilitates the evaluation of $1/m_Q$ corrections.
For the heavy-to-light form factors obtained from the LCSRs, the heavy-quark mass  expansion has been
investigated in the early papers \cite{Li:2007dv}.
To  continue, we apply to the sum rule (\ref{eq:LCSR_final_expression}) the
 standard scaling relations (valid up to the inverse heavy-quark mass corrections):
\begin{eqnarray}
f_H=\frac{\hat{f}}{\sqrt{m_Q}},~~m_H=m_Q+\overline{\Lambda},~~
M^2= 2m_Q\tau,~~ s_0=m_Q^2+2m_Q\omega_0 \,,
\end{eqnarray}
where $\hat{f}$ and $\overline{\Lambda}$ are, respectively, the static decay constant
and the binding energy of heavy meson in HQET,
and the parameters $\tau$ and $\omega_0$
do not scale with $m_Q$. We obtain at LO
\begin{align}\label{eq:HQET}
\hat{g}_{HHV} & = \frac{e^{\overline{\Lambda}/\tau}}{\hat{f}^2}\bigg\{2\tau \left(1-
e^{-\omega_0/\tau}\right)m_V f^{\|}_V \phi^{\|}_{2;V}(1/2) \nn
&\hspace{1.5 cm} + m^2_V f^{\perp}_V \psi^{\|}_{3;V}(1/2)+m^3_V f^{\|}_V \Delta {\cal F}_{\phi^{\|}_{2;V}}
\bigg\} + \dots \,,
\end{align}
where the ellipsis represents the incorporation of inverse heavy-mass corrections.
Furthermore,  the expression (\ref{eq:LCSR_final_expression}) derived from the LCSR framework allows for the estimation of both the static coupling and its associated inverse mass corrections.  However, the inclusion of NLO terms in this analysis poses a challenge, as it necessitates the resummation of logarithms of the heavy-quark mass. A systematic approach entails deriving the LCSR for the strong coupling directly within the framework of HQET, a task that falls beyond the scope of our current endeavor.

Expanding the rescaled sum rule (\ref{eq:LCSR_final_expression})
in the powers of $1/m_Q$,
and parameterize the LCSR result for the strong coupling  a form with an added inverse heavy-mass correction:
\begin{eqnarray}
g_{HHV}= \hat{g}_{HHV}\left(1+\frac{\delta}{m_H}\right), ~~~(H=D_{(s)},B_{(s)}) \,.
\end{eqnarray}

\begin{table}[h!]
\centering
\setlength\tabcolsep{5pt}
\def\arraystretch{1.4}
\begin{tabular}{|c||c|c|c||}
\hline
Parameter  & $\hat{g}_{HHV}$ & $\delta$ & $\beta$\\
\hline
QCDSRs
& $2.44^{+0.73}_{-0.95}$
& $1.42^{+2.29}_{-1.15}$
& $0.30^{+0.09}_{-0.12}$
\\
LQCD
& $3.58^{+0.59}_{-0,60}$
&$0.15^{+0.57}_{-0.46}$
& $0.44\pm 0.07$
\\
Experiment
& $2.49^{+0.70}_{-0.89}$
&$1.29^{+1.86}_{-0.97}$
& $0.30^{+0.08}_{-0.11}$
\\
\hline
\end{tabular}
\caption{ Summary of computed values for $\hat{g}_{HHV}$, $\delta$, and $\beta$ utilizing three distinct methods for determining decay constants. }
\label{tab:beta}
\end{table}

Finally, we conduct a comparative evaluation between our results and other model-dependent analyses pertaining to chiral and heavy quark limits.  The coupling parameter $\beta$ characterizes the $HH\rho$ interaction, and it is intricately linked to $g_{HH\rho}$ through the expression,
\begin{align}
   \beta=\frac{\sqrt{2}}{2}\frac{\hat{g}_{HH\rho}}{{g_V}} \,,
\end{align}
where $g_V = m_\rho/f_\pi$ ($m_\rho$ represents the $\rho$ meson mass to pion decay constant ratio, as determined utilizing the first and second KSRF relations \cite{Casalbuoni:1996pg,Kawarabayashi:1966kd,Riazuddin:1966sw}.
The results stemming from our computational analyses, delineated in Table \ref{tab:beta}, encompassing $\hat{g}_{HHV}$, $\delta$, and $\beta$ utilizing three distinct methodologies for determining decay constants, are detailed herein. We ascertain $\beta=0.44\pm 0.07$ utilizing decay constants derived from lattice QCD, $\beta=0.30^{+0.09}_{-0.12}$ utilizing decay constants derived from two-point QCDSRs, and $\beta=0.30^{+0.08}_{-0.11}$ utilizing decay constants derived from experimental data. These determinations are predicated on our computation of $\hat{g}_{HH\rho}$ and $\delta$. These values significantly smaller than those predicted by the  vector meson dominance(VMD) model \cite{Isola:2003fh},  where suggests  $\beta=0.9$.  One plausible explanation for this discrepancy could be attributed to the empirical observation, as suggested by the Ref.\cite{Li:2007dv}, that the results for $\beta$ with finite heavy quarks should be smaller than those obtained in the heavy quark limit, in addition to the potential for larger uncertainties inherent in model predictions, thereby affirming the reliability of the LCSRs method.  However, uncertainties persist in quantifying NLO corrections to higher-twists, such as twist-3 NLO and sub-sub-leading power contributions remain unknown factors contributing to the observed deviation. Advancements in computational techniques, including refinements in sum rule methods and modeling,  hold the potential to offer valuable insights into resolving these discrepancies.

\section{Conclusion}  \label{sec-7}
In this paper, we have conducted a comprehensive investigation into the improved calculation of the strong couplings $g_{D_{(s)} D_{(s)}V}$ and $g_{B_{(s)} B_{(s)} V}$, where $V=\rho, \omega, K^*$ and $\phi$, within the framework of LCSRs.

Our investigation has advanced the understanding of the vacuum-to-vector-meson correlation function near the light-cone by establishing a rigorous factorization scheme, achieving leading power in $\delta_V$ and next-to-leading order in $\alpha_s$ through hard-collinear factorization.
The observed cancellation of the factorization scale up to $O(\alpha_s^2)$ for the correlation function, facilitated by the utilization of the evolution equation of twist-two distribution amplitudes, underscores the robustness of our calculation. We analytically derived the double spectral representation of the leading power QCD factorization formula, employing the parton-hadron duality ansatz and the double Borel transformation, then obtained the LCSRs for the couplings. We further confirmed the factorization scale independence for the leading power LCSRs governing the strong coupling constants $g_{HHV}$ up to $O(\alpha_s^2)$, which bolsters the reliability of our methodology.
Our comprehensive analysis, which includes subleading power corrections (up to $\delta_V^4$, $\delta_s^1$, and $(\delta_V\delta_s)^1$) and higher-twist corrections (up to two-particle and three-particle twist-4) of the LCDAs of light vector mesons at leading order, culminates in refined computations for the $D_{(s)}D_{(s)}V$ and $B_{(s)}B_{(s)}V$ couplings, thereby advancing the state-of-the-art in this field.

Through numerical exploration of the derived LCSRs governing the strong couplings $g_{HHV}$, it was evident that the twist-two LO terms played a pivotal role in shaping the contributions to the  $D_{(s)}D_{(s)}V$ and $B_{(s)}B_{(s)}V$ couplings, validating earlier results from LCSR calculations \cite{Wang:2007mc,Li:2007dv}. The predicted NLO QCD corrections to twist-2 contributions for bottom-meson couplings resulted in an approximate $30\%$ enhancement relative to tree-level determinations. Additionally, subleading twist contributions from two-particle and three-particle vector meson DAs are more significant for charm-meson couplings than for bottom-meson couplings. Comparing our LCSR predictions with different QCD techniques \cite{Holanda:2006xq,Janbazi:2017mpb,Khosravi:2013ad,Can:2012tx} generally yielded a satisfactory agreement for the obtained values of the $HHV$ couplings within the theoretical uncertainties.

Have these couplings in hand, we extracted the effective coupling $\beta$ in the HM$\chi$PT Lagrangian, yielding results consistent with those reported in Ref.\cite{Wang:2007mc}, albeit smaller than those predicted by the vector meson dominance (VMD) model \cite{Isola:2003fh}. Furthermore, we fulfilled a detailed exploration to the $SU(3)$ flavour symmetry breaking effects by comparing the $HHV$ (for $V$ is $K^\ast$, $\omega$ and $\phi$) couplings with $g_{D D \rho}$ and $g_{B B \rho}$.

The analysis revealed breaking effects ranging from $-14\%$ to $1\%$ when employing the decay constants of heavy mesons calculated from LQCD, and from $-15\%$ to $+5\%$ when utilizing the decay constants obtained from two-point QCDSRs  and  with a range of $-19\%$ to $+2\%$ observed when employing decay constants obtained from experimental data.  Specifically,  $g_{H H\omega}$ and  $g_{H H\phi}$ showcases negative symmetry breaking effects, while $g_{H H K^*}$ demonstrates positive effects. The most obvious $SU(3)_F$ breaking effects are observed in the coupling $g_{D_s D_s\phi}$ which reach to $-19\%$, this is primarily due to the involvement of the maximum number of strange quarks in this channel.

We introduced a dispersion relation that links the residues of $H \to V$ transition form factors $A_0$ at the $H$ pole with the corresponding $HHV$ strong coupling. The couplings $g_{B_{(s)} B_{(s)} V}$, incorporating systematic uncertainties, were determined from this dispersion relation using form factors obtained from two distinct LCSRs methods \cite{Bharucha:2015bzk,Gao:2019lta}. We also included the $g_{D_{(s)} D_{(s)} \rho}$ coupling using a similar approach with $D \to \rho$ transition form factors\cite{Melikhov:2000yu}. Furthermore, our detailed investigation into $g_{B_{(s)} B_{(s)} V}$ and $g_{D D\rho}$ derived from $B \to V$ and $D \to \rho$ transition form factors disclosed significant deviations from previous studies \cite{Li:2007dv,Wang:2007mc,Holanda:2006xq,Janbazi:2017mpb,Khosravi:2013ad,Can:2012tx}, indicating the importance of further research to resolve these differences. Our study presents common features with the $H^\ast HV$ couplings calculated from LCSRs and derived from the $H \to V$ transition form factors $V$ and $T_1$.

Advancing the LCSRs for the strong coupling $g_{H HV}$ requires several critical steps. First, calculating the NLO QCD corrections to higher-twist vector meson DAs is crucial for addressing discrepancies between direct LCSRs calculations and indirect $H\to V$ transition form factor $A_0$ predictions. Second, updating the non-perturbative parameters in the conformal expansion of vector meson DAs is also essential for improving the accuracy of LCSRs predictions. Third, a twist-2 model for the vector meson, similar to the pion LCDA obtained from lattice QCD data \cite{RQCD:2019osh}, can be constructed to significantly improve the predictions for the relevant couplings \cite{Khodjamirian:2020mlb}. Additionally, integrating highly excited heavy-meson states into the hadronic components of the sum rules, as suggested for the $D^* D\pi$  couplings \cite{Becirevic:2002vp}, offers another promising approach to resolve these challenges.

\newpage

\appendix

\section {The master integrals used in the calculation of NLO hard coefficient function}\label{app:MI}
The master integrals $I_1,\ldots,I_7$ are used to calculate the hard coefficient function at NLO in
$\alpha_s$, which are calculated in the dimensional regularization
\begin{align}
I_1 & = \int \frac{d^{D}l}{(2\pi)^D} \frac{1}{\left(l^{2}-m_b^2\right)(l-p_1-q)^{2}} \nn
& = \frac{i}{16 \pi^2} \bigg[ \ln \left(\frac{\mu^2}{m_b^2}\right) + \frac{1}{\tilde{\varepsilon}} + 2 + \frac{1-r_{3}}{r_{3}} \ln \left(1-r_{3}\right) \bigg] \,, \\
I_2 & = \int \frac{d^{D}l}{(2\pi)^D} \frac{1}{\left(l^{2}-m_b^2\right)(l-p-q)^{2}} \nn
& = \frac{i}{16 \pi^2} \bigg[ \ln \left(\frac{\mu^2}{m_b^2}\right) + \frac{1}{\tilde{\varepsilon}} + 2 + \frac{1-r_{1}}{r_{1}} \ln \left(1-r_{1}\right) \bigg] \,, \\
I_3 & = \int \frac{d^{D}l}{(2\pi)^D} \frac{1}{l^{2}(l-p_2)^{2}\left((l-p-q)^{2}-m_b^2\right)} \nn
& = \frac{i}{32\pi^2 \left(r_{3}-r_{1}\right) m_b^2} \bigg[ 2 \bigg( \ln \left(\frac{\mu ^2}{m_b^2}\right) + \frac{1}{\tilde{\varepsilon}} \bigg) \ln \left(\frac{1-r_{1}}{1-r_{3}}\right)  \nn
& \quad - 2 \mathrm{Li}_2\left(\frac{r_{3}}{r_{3}-1}\right) + 2 \mathrm{Li}_2\left(\frac{r_{1}}{r_{1}-1}\right) + \ln^2\left(1-r_{3}\right)-\ln^2\left(1-r_{1}\right) \bigg] \,, \\
I_4 & = \int \frac{d^{D}l}{(2\pi)^D} \frac{1}{l^{2}\left((l-q)^{2} - m_b^2\right)} \nn
& = \frac{i }{16 \pi ^2 } \bigg[ \ln \left(\frac{\mu^2}{m_b^2}\right) + \frac{1}{\tilde{\varepsilon}} + 2 + \frac{ (1-r_{2})}{r_{2}} \ln \left(1-r_{2}\right) \bigg] \,, \\
I_5 &  = \int \frac{d^{D}l}{(2\pi)^D} \frac{1}{l^{2}(l+p_1)^{2}\left((l-q)^{2}-m_b^2\right)}  \nn
& = \frac{i}{32\pi^2 \left(r_{3}-r_{2}\right) m_b^2} \bigg[ 2 \bigg( \ln \left(\frac{\mu ^2}{m_b^2}\right) + \frac{1}{\tilde{\varepsilon}} \bigg) \ln \left(\frac{1-r_{2}}{1-r_{3}}\right)  \nn
& \quad - 2 \mathrm{Li}_2\left(\frac{r_{3}}{r_{3}-1}\right) + 2 \mathrm{Li}_2\left(\frac{r_{2}}{r_{2}-1}\right) + \ln^2\left(1-r_{3}\right)-\ln^2\left(1-r_{2}\right) \bigg] \,, \\
I_6 & = \int \frac{d^{D}l}{(2\pi)^D} \frac{1}{l^{2}-m_b^2}
= \frac{i m_b^2}{16 \pi ^2} \bigg[ \ln \left(\frac{\mu^2}{m_b^2}\right) + \frac{1}{\tilde{\varepsilon}} + 1\bigg] \,, \\
I_7 & = \int \frac{d^{D}l}{(2\pi)^D} \frac{1}{\left(l^{2} - m_b^2\right) (l-q)^{2} (l-p-q)^{2}} \nn
& = \frac{i}{32\pi^2 \left(r_{1}-r_{2}\right) m_b^2} \bigg[ 2 \bigg( \ln \left(\frac{\mu ^2}{m_b^2}\right) + \frac{1}{\tilde{\varepsilon}} \bigg) \ln \left(\frac{1-r_{2}}{1-r_{1}}\right)  \nn
& \quad - 2 \mathrm{Li}_2\left(\frac{r_{1}}{r_{1}-1}\right) + 2 \mathrm{Li}_2\left(\frac{r_{2}}{r_{2}-1}\right) + \ln^2\left(1-r_{1}\right)-\ln^2\left(1-r_{2}\right) \bigg] \,,
\end{align}
where $p_1 = up$, $p_2=\bar{u}p$ and $\displaystyle\frac{1}{\tilde{\varepsilon}} \equiv \frac{1}{\varepsilon } - \gamma_E +\ln (4\pi) $.

\section {The definitions of the vector meson LCDAs}\label{app:def}
Vector meson LCDAs are defined through expansions based on matrix elements, specifically involving the DAs of quark-antiquark (or quark-antiquark-gluon) within the vector meson. These DAs are characterized by increasing twists.
The widely accepted standard definition is applied for the two-particle LCDAs of vector mesons from twist-$2$ to twist-$5$:
\begin{align} \label{eq:DAvec}
\big\langle V(p,\eta^*)\big| & \bar{q_1}(x) \gamma_{\mu} q_2(0)\big| 0\big\rangle=  f^{\parallel}_{V} m_{V} \int_{0}^{1} d u e^{i u p \cdot x}   \nn
&\hspace{1.8cm} \times\bigg\{ p_{\mu}\frac{\eta^* \cdot x}{p \cdot x}  \left[\phi^{\|}_{2;V}(u) -\phi^{\perp}_{3;V}(u) + \frac{m_{V}^{2} x^{2}}{16} (\phi^{\|}_{4;V}(u)-\phi^{\perp}_{5;V}(u))\right] \nn
&\hspace{1.8cm}+\eta^*_{\mu}\left(\phi^{\perp}_{3;V}(u) +\frac{m_{V}^{2} x^{2}}{16}\phi^{\perp}_{5;V}(u)\right)  \nn
&\hspace{1.8cm} -  \frac{\eta^* \cdot x}{2(p \cdot x)^{2}} x_{\mu}m_{V}^{2} \left(\psi^{\|}_{4;V}(u)-2\phi^{\perp}_{3;V}(u)+\phi^{\|}_{2;V}(u)\right)\bigg\}, \\
\big\langle V(p,\eta^*)\big|&\bar{q_1}(x) \sigma_{\mu \nu} q_2(0)\big| 0\big\rangle
 =-i f_{V}^{\perp} \int_{0}^{1} d u e^{i u p \cdot x} \nn
&\hspace{1.8cm} \times \bigg\{\big(\eta^*_{\mu} p_{\nu}-\eta^*_{\nu} p_{\mu}\big) \big[\phi^{\perp}_{2;V}(u)+\frac{m_{V}^{2} x^{2}}{16} \phi^{\perp}_{4;V}(u)\big]\big. \nn
&\hspace{1.8cm}+ \big(p_{\mu} x_{\nu} -p_{\nu} x_{\mu}\big)\frac{\eta^* \cdot x}{(p \cdot x)^{2}}  m_{V}^{2} \Big(\phi^{\|}_{3;V}(u)-\frac{1}{2}\phi^{\perp}_{2;V}(u)-\frac{1}{2}\psi^{\perp}_{4;V}(u)\Big)\nn
&\hspace{1.8cm}+\big. \frac{\eta^*_{\mu} x_{\nu}-\eta^*_{\nu} x_{\mu}}{2 \,p \cdot x} m_{V}^{2} \Big(\psi^{\perp}_{4;V}(u)-\phi^{\perp}_{2;V}(u)\Big) \bigg\},\,\\
\big\langle V(p,\eta^*)\big|&\bar{q_1}(x) \gamma_{\mu} \gamma_{5} q_2(0)\big| 0\big\rangle  = \frac{1}{4} f^{\parallel}_{V} m_{V} \epsilon_{\mu \nu \rho \sigma} \eta^{*\nu}p^{\rho}x^{\sigma} \int_{0}^{1} d u e^{i u p \cdot x} \Big(\psi^{\perp}_{3,V}(u)
 +\frac{m_{V}^{2} x^{2}}{16}\psi^{\perp}_{5,V}(u) \Big),\\
\big\langle V(p,\eta^*)|&\bar{q_1}(x) q_2(0)| 0\big\rangle  = -\frac{i}{2} f_{V}^{\perp}\big(\eta \cdot x\big) m_V^2 \int_0^1 d u e^{i u p \cdot x} \psi^{\parallel}_{3;V}(u).
\end{align}
The definitions of $f^{\parallel}_V$ and $f^\perp_V$ are given by:
\begin{align}
 &\left\langle V(p,\eta^*)\left|\bar{q}_1(0) \gamma_{\mu} q_2(0)\right| 0\right\rangle= f^{\|}_V m_V \eta^*_\mu,\\ &\left\langle V(p,\eta^*)\left|\bar{q}_1(0) \sigma_{\mu \nu} q_2(0)\right| 0\right\rangle=-if^{\perp}_V\left(\eta^*_\mu p_\nu-\eta^*_\nu p_\mu\right).
\end{align}
We only list the used three-particle chiral-even DAs \cite{Ball:1998ff} which are defined by the following matrix elements
\begin{align}
& \left\langle V(p,\eta^*)\left|\bar{q}_1(x)g_s G_{\mu\nu}(v x) \gamma_\alpha q_2(-x)\right| 0\right\rangle = i f_V^{\|} m_V P_\alpha \left[ P_\nu \eta_{\perp\mu}^\ast - P_\mu \eta_{\perp\nu}^\ast \right]  \Phi_{3;V}^{\|}(v, Px) \nn
& \hspace{5.5cm} + i f_V^{\|} m_V^3 {\eta^\ast \cdot x \over P\cdot x} \left[ P_\mu g_{\alpha\nu}^{\perp} - P_\nu g_{\alpha\mu}^\perp \right] \Phi^{\|}_{4;V}(v, Px) \nn
& \hspace{5.5cm} + i f_V^{\|} m_V^3 {\eta^\ast \cdot x \over (P\cdot x)^2} P_\alpha \left[ P_\mu x_{\nu} - P_\nu x_{\mu} \right] \Psi^{\|}_{4;V}(v, Px)\,,
\end{align}
\begin{align}
& \big\langle V(p,\eta^*)\big|\bar{q}_1(x) g_s \widetilde{G}_{\mu\nu}(v x) \gamma_\alpha \gamma_5 q_2(-x)\big| 0 \big\rangle = - f_V^{\|} m_V P_\alpha \left[ P_\nu \eta_{\perp\mu}^\ast - P_\mu \eta_{\perp\nu}^\ast \right] \widetilde\Phi_{3;V}^{\|}(v, Px) \nn
& \hspace{5.5cm} - f_V^{\|} m_V^3 {\eta^\ast \cdot x \over P\cdot x} \left[ P_\mu g_{\alpha\nu}^{\perp} - P_\nu g_{\alpha\mu}^\perp \right] \widetilde{\Phi}^{\|}_{4;V}(v, Px) \nn
& \hspace{5.5cm} - f_V^{\|} m_V^3 {\eta^\ast \cdot x \over (P\cdot x)^2} P_\alpha \left[ P_\mu x_{\nu} - P_\nu x_{\mu} \right] \widetilde{\Psi}^{\|}_{4;V}(v, Px)\,,
\end{align}
where
\begin{align}
& P_\mu = p_\mu - {1\over 2} x_\mu {m_V^2\over P\cdot x}, \\
& \eta_\mu^\ast = {\eta^\ast \cdot x \over P\cdot x} \Big( P_\mu - {m_V^2 \over 2 P\cdot x} x_\mu \Big) + \eta_{\perp \mu}^\ast, \\
& g_{\mu\nu}^\perp = g_{\mu\nu}-\frac1{P\cdot x}(P_\mu x_\nu + P_\nu x_\mu), \\
& \Phi_{3p}(v,Px) = \int \mathcal{D}\underline{\alpha} \, e^{iP\cdot x (\alpha_1 -\alpha_2+ v\alpha_3)} \Phi_{3p}(\alpha_1,\alpha_2,\alpha_3),
\end{align}
$\underline{\alpha}$ is the set of three momentum fractions
$\underline{\alpha}=\{\alpha_1,\alpha_2,\alpha_3\}$.
 The integration measure is defined as
\begin{equation}
 \int {\cal D}\underline{\alpha} \equiv \int_0^1 d\alpha_1
  \int_0^1 d\alpha_2\int_0^1 d\alpha_3 \,\delta\left(1-\sum \alpha_i\right).
\label{eq:measure}
\end{equation}
$P$ and $x$ are two introduced light-like vectors and the dual gluon field strength tensor is defined as $\widetilde{G}_{\mu\nu} = {1\over 2} \epsilon_{\mu\nu\rho\sigma} G^{\rho\sigma}$, {with $\epsilon_{0123} = -1$}.

\section{Parameterizations for the vector meson DAs }\label{app:lcda}
Below are the parameterizations used for the vector meson DAs in our numerical analysis:
\begin{itemize}
\item  the twist-2 DA:
\vspace{-2mm}
\begin{align}
\label{eq:confexp}
\phi_{2;V}^{\parallel,\perp}(u,\mu) = 6 u \bar u \left\{ 1 + \sum_{n=1}^\infty
a_n^{\parallel,\perp}(\mu) C_n^{3/2}(\xi)\right\},
\end{align}
expressed in terms of the (non-perturbative) Gegenbauer moments $a_n^{\parallel,\perp}$ and the Gegenbauer polynomials $C_n^{3/2}$, where $\xi=2u-1$, the renormalization of $a^{\parallel,\perp}_n$ occurs multiplicatively to leading-logarithmic accuracy as
\begin{equation}
a^{\parallel,\perp}_n(\mu) = E_{n,\rm LO}^{\parallel,\perp}(\mu, \mu_0)\, a^{\parallel,\perp}_n(\mu_0).
\end{equation}
Next-to-leading order accuracy reveals a more complex scale dependence of the Gegenbauer moments, which can be summarized as: \cite{Wang:2017ijn,Agaev:2010aq}
\begin{align}
\big[ f^{\perp}_V(\mu) a^{\perp}_{n,{\rm NLO}}& (\mu)\big] =  E_{n,\rm NLO}^{\perp}  (\mu,\mu_0)
\left[f^{\perp}_V(\mu_0)  a^{\perp}_n(\mu_0)\right]\nn
&\quad+\frac{\alpha_s}{4\pi}\sum_{k=0}^{n-2}
E_{n,\rm LO}^{\perp}(\mu,\mu_0)\, d^{k}_{n}(\mu,\mu_0)\left[f^{\perp}_V(\mu_0) a^{\perp}_k(\mu_0)\right],\\
a^{\parallel}_{n,{\rm NLO}}(\mu)&=  E_{n,\rm NLO}^{\parallel} (\mu,\mu_0)
a^{\parallel}_n(\mu_0)+\frac{\alpha_s}{4\pi}\sum_{k=0}^{n-2} \,
E_{n,\rm LO}^{\parallel}(\mu,\mu_0)\, d^{k}_{n}(\mu,\mu_0)\,a^{\parallel}_k(\mu_0),
\label{eq:evoaNLO}
\end{align}
where Ref.\cite{Agaev:2010aq,Wang:2017ijn} explicitly provide the formulas for the RG factors $E^{\parallel,\perp}_{n,{\rm (N)LO}}(\mu,\mu_0)$ and off-diagonal mixing coefficients $d^k_n(\mu,\mu_0)$ within the $\overline {\text{MS}}$ scheme. The evolution of $a^{\perp}_n(\mu)$ is strongly connected to $f^{\perp}_V(\mu)$ due to the definition of the conformal operator.
\item the twist-3 two-particle DAs \cite{Ball:2007rt}:
\begin{align}
&\psi_{3;V}^\parallel(u,\mu)  =  6 u\bar u \bigg\{ 1 + \left(
\frac{1}{3}a_1^\perp(\mu) + \frac{5}{3} \kappa_{3V}^\perp(\mu)\!\right)
C_1^{3/2}(\xi)  \nn
&\hspace*{1.5cm}+ \left( \frac{1}{6}a_2^\perp(\mu)  + \frac{5}{18}
\omega_{3V}^\perp(\mu)\right) C_2^{3/2}(\xi)-
\frac{1}{20}\lambda_{3V}^\perp(\mu) C_3^{3/2}(\xi)\bigg\}\nn
&\hspace*{1.5cm}+ 3\,
\frac{m_{q_2}(\mu)+m_{q_1}(\mu)}{m_{V}}\,\frac{f_{V}^\parallel}{f_{V}^\perp(\mu)}\left\{
u\bar u \left(1 + 2 \xi a_1^\parallel(\mu) + 3 (7-5u\bar u) a_2^\parallel(\mu)\right)\right. \nn
&{}\left.\hspace*{1.5cm}+ \bar
u \ln \bar u \left(1+3 a_1^\parallel(\mu) + 6 a_2^\parallel(\mu)\right)  + u \ln u
\left(1-3 a_1^\parallel(\mu) + 6 a_2^\parallel(\mu)\right)\right\}\nn
&\hspace*{1.5cm}- 3\,
\frac{m_{q_2}(\mu)-m_{q_1}(\mu)}{m_{V}}\,\frac{f_{V}^\parallel}{f_{V}^\perp(\mu)}\left\{
u\bar u \left(9 a_1^\parallel(\mu) + 10\xi a_2^\parallel(\mu)\right)\right.\nn
&{}\left.\hspace*{1.5cm}  + \bar
u \ln \bar u \left(1+3 a_1^\parallel(\mu) + 6 a_2^\parallel(\mu)\right)- u \ln u
\left(1-3 a_1^\parallel(\mu) + 6 a_2^\parallel(\mu)\right)\right\},
\end{align}
\begin{align}
&\phi_{3;V}^\perp(u,\mu)= \frac{3}{4}\,(1+\xi^2) + \frac{3}{2}\,\xi^3
a_1^{\|}(\mu) + \left( \frac{3}{7}\,a_2^{\|}(\mu) + 5
\zeta_{3V}^{\|}(\mu)\right) (3\xi^2-1) \nn
&\hspace*{1.5cm} + \Big(5\kappa_{3V}^{\|}(\mu)- \frac{15}{16}\,\lambda_{3V}^{\|}(\mu)+
\frac{15}{8}\,\widetilde{\lambda}_{3V}^{\|}(\mu) \Big) \xi
(5\xi^2-3)\nn
&\hspace*{1.5cm}+ \Big(\frac{9}{112}\,a_2^{\|}(\mu) +
\frac{15}{32}\,\omega_{3V}^{\|}(\mu) -
\frac{15}{64}\,\widetilde\omega_{3V}^{\|}(\mu) \Big)(35\xi^4-30\xi^2+3)\nn
&{\hspace*{1.5cm}} +\frac{3}{2}\,\frac{m_{q_2}(\mu)+m_{q_1}(\mu)}{m_{V}}\,\frac{f_{V}^\perp(\mu)}{
f_{V}^\parallel} \left\{ 2 + 9 \xi a_1^\perp(\mu) + 2 (11-30 u {\bar u})\right.\nn
&{}\left.\hspace*{1.5cm} \times
    a_2^\perp(\mu)+
\Big(1-3 a_1^\perp(\mu) + 6 a_2^\perp(\mu)\Big) \ln u + \Big(1+3 a_1^\perp(\mu) + 6 a_2^\perp(\mu)\Big)
\ln {\bar u} \right\}\nn
&  {\hspace*{1.5cm}}-\frac{3}{2}\,\frac{m_{q_2}(\mu)-m_{q_1}(\mu)}{m_{V}}\,\frac{f_{V}^\perp(\mu)}{
f_{V}^\parallel} \left\{ 2\xi + 9 (1-2u {\bar u}) a_1^\perp(\mu) + 2 \xi (11-20 u {\bar u})
   \right.\nn
&{}\left.\hspace*{1.5cm} \times a_2^\perp(\mu) +
(1+3 a_1^\perp(\mu) + 6 a_2^\perp) \ln {\bar u} - (1-3 a_1^\perp(\mu) + 6 a_2^\perp(\mu))
\ln u \right\}.
\end{align}

\item the twist-4  two-particle DAs \cite{Ball:2007zt}:
\begin{align}
&\phi^\perp_{4;V}(u,\mu) =
30 u^2\bar u^2 \left\{ \left(\frac{4}{3}\, \zeta^\perp_{4V}(\mu) -
\frac{8}{3}\, \widetilde{\zeta}^\perp_{4V}(\mu) + \frac{2}{5} +
\frac{4}{35}\, a_2^\perp(\mu)\right)  \right.
\nn
&{}\hspace*{1.5cm} + \Big( \frac{3}{25}\,a_1^\perp(\mu) +
\frac{1}{3}\,\kappa_{3V}^\perp(\mu) - \frac{1}{45}\, \lambda_{3V}^\perp(\mu)
- \frac{1}{15}\,\theta_1^\perp(\mu) \nn
&\hspace*{1.5cm}+ \frac{7}{30}\,\theta_2^\perp(\mu)
+\frac{1}{5}\, \widetilde{\theta}_1^\perp(\mu) -
\frac{3}{10}\,\widetilde{\theta}_2^\perp(\mu)\Big) C_1^{5/2}(\xi)
\nn
&\hspace*{1.5cm}+\left.\left(\frac{3}{35}\,a_2^\perp(\mu) +
\frac{1}{60}\,{\omega}_{3V}^\perp(\mu) \right) C_2^{5/2}(\xi)
-\frac{4}{1575}\,\lambda_{3V}^\perp(\mu) C_3^{5/2}(\xi)\right\}\nn
&\hspace*{1.5cm}+ \left( 5 \kappa_{3V}^\perp(\mu) - a_1^\perp(\mu) -
20 \widetilde{\phi}_2^\perp(\mu)\right)
\nonumber
\end{align}
\begin{align}
&\hspace*{1.5cm}\times\left\{-4 u^3 (2-u) \ln u + 4 \bar u^3 (2-\bar u) \ln \bar u +
\frac{1}{2}\,u \bar u \xi \left(3\xi^2-11\right)\right\}\nn
&\hspace*{1.5cm}+  \left( 2 \omega_{3V}^\perp(\mu) - \frac{36}{11}\, a_2^\perp(\mu)
-\frac{252}{55}\, \langle\!\langle Q^{(1)}\rangle\!\rangle(\mu) -
\frac{140}{11}\, \langle\!\langle Q^{(3)}\rangle\!\rangle(\mu)\right)
\nn
&\hspace*{1.5cm}\times\left\{ u^3
(6 u^2-15u+10)\ln u + \bar u^3 (6 \bar u^2-15\bar u+10)\ln\bar u -
\frac{1}{8}\,u\bar u \left( 13\xi^2-21\right)\right\}.\\
&\psi^\parallel_{4;V}(u,\mu) = \psi^{\parallel,\rm T4}_{4;V}(u,\mu)
+ \psi^{\parallel,\rm WW}_{4;V}(u,\mu)\,,\\
&\psi_{4;V}^{\parallel,\rm T4}(u,\mu) =-
\frac{20}{3}\,\zeta_{4V}^\parallel(\mu) C_2^{1/2}(\xi) +
\left(10\theta_1^\parallel(\mu) - 5 \theta_2^\parallel(\mu) \right) C_3^{1/2}(\xi)\,\\
&\psi_{4;V}^{\parallel,\rm WW}(u,\mu) = 1
+ \left( 12\kappa_{4V}^\parallel(\mu) + \frac{9}{5}\,
   a_1^{\parallel}(\mu)\right) C_1^{1/2}(\xi)
+ \left(-1-\frac{2}{7}\,a_2^{\parallel}(\mu)+ \frac{40}{3}\,
   \zeta_{3V}^\parallel(\mu)\right)
  \nn
&\hspace*{1.5cm}{}\times  C_2^{1/2}(\xi)
+ \left(-\frac{9}{5}\,a_1^{\parallel}(\mu) -
   \frac{20}{3}\,\kappa_{3V}^\parallel(\mu) -
   \frac{16}{3}\,\kappa_{4V}^\parallel(\mu)\right) C_3^{1/2}(\xi)\nn
&\hspace*{1.5cm}{}
+ \left( -\frac{27}{28}\,a_2^{\parallel}(\mu) +
   \frac{5}{4}\,\zeta_{3V}^\parallel(\mu) -
   \frac{15}{8}\,\omega_{3V}^\parallel(\mu) -
   \frac{15}{16}\,\widetilde\omega_{3V}^\parallel(\mu) \right)
   C_4^{1/2}(\xi)\nn
&\hspace*{1.5cm}{} +
   6\,\frac{m_{q_2}(\mu)-m_{q_1}(\mu)}{m_{V}}\,\frac{f_{V}^\perp(\mu)}{f_{K^*}^\parallel}
   \left\{ \xi + a_1^{\perp}(\mu)\,\frac{1}{2}\,(3\xi^2-1)
    + a_2^\perp(\mu)\,\frac{1}{2}\,\xi (5\xi^2-3) \right. \nn
&\hspace*{1.5cm}\left. +
    \frac{5}{2}\, \kappa_{3V}^\perp(\mu) (3\xi^2-1) +\frac{5}{6}\, \omega_{3V}^\perp(\mu) \xi (5\xi^2-3)
    - \frac{1}{16}\,\lambda_{3V}^\perp(\mu) (35\xi^4-30\xi^2+3)\right\}.\\
&\phi^\parallel_{4;V}(u,\mu) = \phi^{\parallel,\rm T4}_{4;V}(u,\mu)
+ \phi^{\parallel,\rm WW}_{4;V}(u,\mu)\,,\\
&\phi_{4;V}^{\parallel,\rm T4}(u,\mu) = 30 u^2 {\bar u}^2 \left\{
\frac{20}{9}\,\zeta_{4V}^\parallel(\mu) + \left( - \frac{8}{15}\,
	   \theta_1^\parallel(\mu) + \frac{2}{3}\,\theta_2^\parallel(\mu)\right)
	     C_1^{5/2}(\xi)\right\} -84 \widetilde\omega_{4V}^\parallel(\mu)\nn
&\hspace*{1.5cm}{}
\times\left\{\frac{1}{8}\,u {\bar u}\,(21-13\xi^2)
+ u^3 (10-15 u + 6 u^2)\ln u
+ {\bar u}^3 (10-15 {\bar u} + 6 {\bar u}^2)\ln {\bar u}\right\} \nn
&\hspace*{1.5cm}{}+ 80\psi_2^\parallel(\mu)  \left\{ u^3 (2-u) \ln u - {\bar u}^3
(2-{\bar u}) \ln {\bar u} - \frac{1}{8}\,(3\xi^2-11)\right\},\\
&\phi_{4;K^*}^{\parallel,\rm WW}(u,\mu) =  30 u^2 {\bar u}^2 \bigg\{
\frac{4}{5}\left( 1 + \frac{1}{21}\,a_2^\parallel(\mu) +
           \frac{10}{9}\,\zeta_{3V}^\parallel(\mu)\right)\nn
& \hspace*{1.5cm}           +\left( \frac{17}{50}\,a_1^\parallel(\mu)
	   + \frac{2}{5}\,\widetilde\lambda_{3V}^\parallel(\mu) -
	   \frac{1}{5}\,\lambda_{3V}^\parallel(\mu)\right)
	     C_1^{5/2}(\xi)\nn
&\hspace*{1.5cm}{}\left. + \frac{1}{10}\,\left( \frac{9}{7}\, a_2^\parallel(\mu) +  \frac{1}{9}\, \zeta_{3V}^\parallel(\mu) +
\frac{7}{6}\,\omega_{3V}^\parallel(\mu) -\frac{3}{4}\,\widetilde\omega_{3V}^\parallel(\mu) \right)C_2^{5/2}(\xi) \right\}\nn
&\hspace*{1.5cm}{}+ 2 \left( - 2 a_2^\parallel(\mu) - \frac{14}{3}\,\zeta_{3V}^\parallel(\mu) + 3
\omega_{3V}^\parallel(\mu)\right)
\bigg\{\frac{1}{8}u {\bar u}(21-13\xi^2)\nn
&\hspace*{1.5cm}{} + u^3 (10-15 u + 6 u^2)\ln u
+ {\bar u}^3 (10-15 {\bar u} + 6 {\bar u}^2)\ln {\bar u}\bigg\} \nn
&\hspace*{1.5cm}{} + 4 \Big( a_1^\parallel(\mu) - \frac{40}{3}\kappa_{3V}^\parallel(\mu)
 \Big)\left\{ u^3 (2-u) \ln u - {\bar u}^3
(2-{\bar u}) \ln {\bar u}- \frac{1}{8}\,(3\xi^2-11)\right\}\nn
&\hspace*{1.5cm}{} + \frac{m_{q_2}(\mu)+m_{q_1}(\mu)}{m_{V}}\,\frac{f_{V}^\perp(\mu)}{f_{V}^\parallel}\,
6 u {\bar u} \left\{ 2 (3 + 16 a_2^\perp) + \frac{10}{3}( \kappa_{3V}^\perp(\mu)
- a_1^\perp(\mu)) C_1^{3/2}(\xi)\right.
\nonumber
\end{align}
\begin{align}
&\hspace*{1.5cm}\left. + \left( \frac{5}{9}\,\omega_{3V}^\perp(\mu) -
a_2^\perp(\mu) \right) C_2^{3/2}(\xi) - \frac{1}{10}\,\lambda_{3V}^\perp
C_3^{3/2}(\xi) \right\}+ 24\frac{m_{q_2}(\mu)+m_{q_1}(\mu)}{m_{V}}\nn
&\hspace*{1.5cm}{} \times\frac{f_{V}^\perp(\mu)}{f_{V}^\parallel}
\left\{ (1-3 a_1^\perp + 6 a_2^\perp(\mu)) u^2 \ln u+
(1+3 a_1^\perp(\mu) + 6 a_2^\perp(\mu)) {\bar u}^2 \ln {\bar u} \right\}
\nn
&\hspace*{1.5cm}
{}+\frac{m_{q_2}(\mu)-m_{q_1}(\mu)}{m_{V}}\,\frac{f_{V}^\perp(\mu)}{f_{V}^\parallel}\,
6 u {\bar u} \left\{ -\left( 10 \kappa_{3V}^\perp(\mu) +
\frac{82}{5}\,a_1^\perp(\mu) \right) C_1^{3/2}(\xi)\right.\nn
& \hspace*{1.5cm}{}+ 20 \left( \frac{10}{189} + \frac{1}{3}\, a_2^\perp(\mu) -
\frac{1}{21}\, \omega_{3V}^\perp(\mu) \right) C_2^{3/2}(\xi) +
\left( \frac{7}{54}\,\lambda_{3V}^\perp(\mu) + \frac{2}{5}\,a_1^\perp(\mu)
\right)
\nn
& \hspace*{1.5cm}\left.\times C_3^{3/2}(\xi) + \left( \frac{1}{5}\,a_2^\perp(\mu) - \frac{2}{315}-
\frac{1}{21}\, \omega_{3V}^\perp(\mu)\right) C_4^{3/2}(\xi)
+ \frac{2}{135}\,\lambda_{3V}^\perp(\mu) C_5^{3/2}(\xi)\right\}
\nn
&\hspace*{1.5cm}{}+ \frac{m_{q_2}(\mu)-m_{q_1}(\mu)}{m_{V}}\,\frac{f_{V}^\perp(\mu)}{f_{V}^\parallel}\,
\left\{  (5u^2-23 - 54 a_1^\perp(\mu) - 108 a_2^\perp(\mu)) \ln{\bar u}\right.
\nn
&\hspace*{1.5cm}\left.
- \left(5{\bar u}^2-23 + 54 a_1^\perp(\mu) - 108 a_2^\perp(\mu)\right) \ln u\right\}.
\end{align}
\item the twist-4  three-particle DAs \footnote{
We adopt a convention denoted as $\underline{\alpha}=\{\alpha_1,\alpha_2,\alpha_3\}=\{\alpha_{\bar q_1},\alpha_{q_2},\alpha_g\}$. It's worth mentioning that this differs from the convention used in Ref.\cite{Ball:2007zt} where the only distinction is the exchange of $\alpha_1$ for $\alpha_2$.}
\begin{align}
&\Psi_{4;V}^\parallel(\underline{\alpha}) = 120
\alpha_1\alpha_2\alpha_3 \Big[ \psi_0^\parallel +
\psi_1^\parallel(\alpha_1-\alpha_2) +
\psi_2^\parallel (3\alpha_3-1) \Big],\\
&\widetilde\Psi_{4;V}^\parallel(\underline{\alpha})= 120
\alpha_1\alpha_2\alpha_3 \Big[ \widetilde\psi_0^\parallel +
  \widetilde\psi_1^\parallel(\alpha_1-\alpha_2) +
\widetilde\psi_2^\parallel (3\alpha_3-1) \Big],\\
&\lefteqn{{\widetilde\Phi}_{4;V}^\parallel(\underline{\alpha}) = 30 \alpha_3^2\left\{ \phi^\parallel_{0}(1-\alpha_3)
     +\phi^\parallel_{1}\left[\alpha_3(1-\alpha_3)-
6\alpha_1\alpha_2\right]\right.}
\nn
&\hspace{0.5cm} \left.{} +\phi^\parallel_{2}\left[\alpha_3(1-\alpha_3)-\frac{3}{2}(\alpha_1^2+\alpha_2^2)\right]-(\alpha_1-\alpha_2)\left[ \theta^\parallel_0 +\alpha_3\theta^\parallel_1 + \frac{1}{2}\,(5 \alpha_3-3)
\theta^\parallel_2\right]\right\},
\\
&\lefteqn{ {\Phi}_{4;K^*}^\parallel(\underline{\alpha})  =
 30 \alpha_3^2\left\{ \theta^\parallel_{0}(1-\alpha_3)
          +\theta^\parallel_{1}\left[\alpha_3(1-\alpha_3)-
6\alpha_1\alpha_2\right]
\right.}\nn
&\hspace{0.5cm}\left. {}  +\theta^\parallel_{2}\left[\alpha_3(1-\alpha_3)-
\frac{3}{2}(\alpha_1^2+\alpha_2^2)\right]-
(\alpha_1-\alpha_2)\left[ \phi^\parallel_0 + \alpha_3
\phi^\parallel_1 + \frac{1}{2}\,(5 \alpha_3-3)
\phi^\parallel_2\right]\right\}.\nn
\hspace*{20pt}
\end{align}

\end{itemize}

\acknowledgments

We thank Chao Wang and Yu-Ming Wang for valuable discussions.
This work is supported by the National Natural Science Foundation of China with Grant No. 12075125 and 12247162.
H. Y. Jiang gratefully acknowledges financial support from China Scholarship Council.

\bibliographystyle{JHEP}
\bibliography{biblio}

\end{document}